\documentclass[aps,pre,twocolumn,amsmath,amssymb,nofootinbib,superscriptaddress,floatfix]{revtex4-1}

\usepackage{amsmath,amsfonts,amssymb}
\usepackage{amssymb}
\usepackage{amsthm}
\usepackage[dvipdf]{color}
\usepackage{graphicx}
\usepackage{dcolumn}
\usepackage{bm} 
\usepackage{ulem}
\usepackage{cancel}

\begin{document}

\title{Phonon Echo from Multi-Level Systems and Many-Body Interactions in Low-Temperature Glasses}

\author{Di Zhou}
\email{dizhou@bit.edu.cn}
\affiliation{Key Lab of Advanced Optoelectronic Quantum Architecture and Measurement (MOE),
School of Physics, Beijing Institute of Technology, Beijing 100081, China}
\affiliation{Department of Physics, University of Illinois at Urbana-Champaign, Urbana, Il 61801, 
United States of America}
\affiliation{Department of Physics, University of Michigan, Ann Arbor, MI 48109-1040, United States of America}
\affiliation{School of Physics, Georgia Institute of Technology, Atlanta, GA 30332, United States of America}

	

\begin{abstract}
At low temperatures, glasses exhibit distinctive properties compared to crystalline solids. A notable example is the phonon echo, a phenomenon that motivated the two-level-system (TLS) model. This model has successfully explained many universal anomalies in glasses. Here, we extend the TLS framework to a multi-level system and show that phonon echoes persist when nonlinear energy structures and disorder are included. By incorporating virtual phonon exchange, we introduce many-body interactions between these multi-level systems, leading to nonlinear eigen-energies that enhance the echo signal. Meanwhile, finite-temperature thermal fluctuations cause dephasing, resulting in a decay of echo amplitude over time. The analytical and numerical results are consistent across semi-classical and quantum regimes. Our work validates the multi-level-system model and underscores the role of many-body interactions in low-temperature glassy dynamics.
\end{abstract}

    
\maketitle

\section{Introduction}

The study of amorphous solids (glasses) at low temperatures has revealed a set of universal anomalies that starkly contrast with the behavior of crystalline materials~\cite{Pohl1971PRB, Phillips1987RPP, Varma1972Phil, Hunklinger1977Acoustic, Parshin1985PRB, Ramos2020LTP}. More than fifty years ago, a key observation, namely the linear temperature dependence of heat capacity below $10\,\mathrm{K}$ across a broad range of insulating glasses~\cite{Pohl2011RMP, Yu1991CCMP, Yu1989PRL, LASJAUNIAS1993SSC, Tietje1986ZCM, Zamponi2015PNAS, Zamponi2016PRL}, established the foundational universality of these glassy phenomena. Subsequent research identified a wealth of related anomalies, including the quadratic temperature dependence of thermal conductivity~\cite{Stefhens1976PRB, Ramos1997PRB, Buchenau1993PRL}, universally low ultrasound internal friction~\cite{Vural2011JNCS, Yu2020PRL, Ramos1992PRB, Pohl1991PRB, Classen2000PRL, Arnold1981JDPL}, anomalous shifts in sound velocity~\cite{Hunklinger1978PRL, Hunklinger1998PRL, Federle1982JPC, Piche1974PRL} and dielectric constant~\cite{Pollak1972PRL, Hunklinger1981, Peter1999PRL, Hunklinger2000PRL, Classen1994ADP}, and the saturation of the phonon mean free path~\cite{Kane1983PRB, HUNKLINGER1972PLA, SCHICKFUS1977PLA, PELLE2000131JAC}.

Among these properties, the phonon echo effect is particularly notable~\cite{Golding1979PRB, Burin2013EL}. It is typically observed at temperatures around $100 \,\text{mK}$, where the application of two hypersonic pulses (with frequency $\sim 1\,\text{GHz}$) separated by a time interval $\tau$ leads to the spontaneous emergence of a third pulse at time $2\tau$. This effect closely resembles spin echoes in nuclear magnetic resonance~\cite{Hahn1950PR, Xin2021PRB} and photon echoes in quantum optics~\cite{Abanin2017PRB, Hartmann1964PRL, Zhai2020PRA, Ilya2025PRA}, and was a key motivation for the development of the two-level-system (TLS) model of glasses~\cite{Phillips1972JLTP, Phillips1987RPP, Varma1972Phil, Jackle1972ZP, Berret1988ZPBCM, Hunklinger1975JPCSSP, Smolyakov1982SPU}. In this model, atoms or small groups of atoms tunnel between two potential minima, with a broad distribution of parameters leading to a nearly constant density of states~\cite{Phillips1987RPP, Hunklinger1986PLTP, Phillips1981, Charbonneau2016PRL}. Their random dynamics couple to long-wavelength phonon strain fields, resulting in a strain-TLS interaction formally analogous to the coupling between inhomogeneous magnetic fields and magnetic dipole moments. Consequently, the TLS model explains phonon echoes using equations of motion identical to those for spin echoes. Such echo phenomena are significant because they reveal memory effects in random, nonlinear systems~\cite{Nagel2019RMP, Torner2011RMP, ma2023PRL}. Just as spin echoes are central to nuclear magnetic resonance and photon echoes are pivotal in quantum information, phonon echoes impose strong constraints on theoretical models, such as the TLS framework, for describing glassy dynamics.

Despite the success of the TLS model in explaining low-temperature anomalies in glasses, such as phonon echo~\cite{BELTUKOV2010SSC}, the framework faces significant limitations~\cite{Zamponi2020PRL}. First, the assumption of exactly two potential minima for tunneling atoms lacks a fundamental justification~\cite{Vural2011JNCS, Yu2020PRL}. In fact, photon and elastic echoes have been explained by randomized three-level systems~\cite{Zhai2020PRA, Richard1976PRA} and by jammed solids with nonlinear Hertzian interactions~\cite{Burton2016PRE, li2012ARPL}, suggesting that the TLS picture is not the sole model for the acoustic and thermal universal anomalies in low-temperature glasses. A natural extension, namely the multi-level-system model, may also provide a more complete description of echo phenomena in glasses. Second, conventional TLS models often neglect interactions between individual two-level systems. By analogy with magnetic dipoles, which acquire long-range $1/r^{3}$ coupling via virtual photon exchange~\cite{peskin2018introduction}, TLSs coupled to phonon strain fields should likewise develop long-range many-body interactions through virtual phonon exchange. These interactions not only are essential for explaining universal phononic properties in glasses but also naturally lead to a richer effective level structure. For instance, when two TLSs interact, their coupling enlarges the joint Hilbert space, effectively forming a composite four-level system. This demonstrates how incorporating many-body interactions intrinsically motivates the extension to multi-level-system descriptions. Third, certain universal observations remain not fully understood within the TLS framework if many-body interactions are not considered~\cite{Yu2020PRL}. A notable example is the universally small internal friction in glasses~\cite{Pohl2011RMP}. Within the TLS model, it depends on several independent parameters that vary across materials. However, experimentally, its magnitude remains universally small, varying by no more than about a factor of 20 among different glasses.

Building on this foundation, a key question emerges: can the multi-level-system model, combined with mutual many-body interactions among these multi-level-systems, account for the phonon echo in low-temperature glasses, and how do such interactions govern its dynamics? While in two-level systems (TLS) the nonlinear energy structure arises from the sharp cutoff of a two-level configuration, studies of photon echoes in three-level systems ~\cite{Zhai2020PRA} and elastic echoes in jammed solids ~\cite{Burton2016PRE} confirm that nonlinear eigenvalue structures and randomness are decisive for echo generation. The multi-level model naturally generalizes this mechanism by incorporating non-equidistant and randomly distributed level spacings. Moreover, virtual phonon exchange ~\cite{Joffrin1975JDP, Busiello2010RPJ, West1993PRB, Burin2004JLTP} induces many-body couplings between multi-level-systems, analogous to magnetic dipole-dipole interactions. Herein, we show that these interactions effectively create nonlinear energy structures, thereby amplifying the phonon echo signal. Meanwhile, finite-temperature thermal fluctuations cause dephasing, leading to a decay of echo amplitude over time.


In this work, we investigate phonon echo within multi-level framework in both semi-classical and quantum regimes. In the semi-classical limit, the dynamics are described via the WKB approximation for large quantum numbers~\cite{Van1967PR}. Using the method of multiple scales-a standard perturbative technique in nonlinear mechanics~\cite{vakakis2001normal}-we analytically derive frequency shifts induced by on-site potential nonlinearities and many-body interactions. In the quantum regime, echo dynamics are characterized through third-order response functions~\cite{Yuan2023PRB} and simulated via the Liouville-von Neumann equation for the density matrix. All results of phonon echo are validated through consistent analytical and numerical calculations. Our findings confirm the essential role of multi-level systems and many-body interactions in the low-temperature dynamics of amorphous solids. 




This paper is organized as follows. In Section \uppercase\expandafter{\romannumeral2}, we introduce the multi-level system model and the virtual phonon exchange process that generates many-body interactions between multi-level systems. Section \uppercase\expandafter{\romannumeral3} investigates phonon echoes in the semi-classical limit of multi-level systems, demonstrating the emergence of many-body-induced echoes. Finally, Section \uppercase\expandafter{\romannumeral4} presents a quantum mechanical analysis of multi-level systems, with a focus on phonon echoes and dephasing effects.


\section{Multi-level-system framework with many-body interactions}

To investigate the emergence of echo signals in glasses, we first formulate a multi-level-system model. We begin by briefly reviewing the two-level-system (TLS) framework, extend it to a multi-level description, and then introduce many-body interactions between the resulting multi-level systems.

In an elastic medium, the equilibrium position of a point mass is described by its three-dimensional coordinate $\bm{x}$, and its displacement field by $\bm{u}(\bm{x})$. In the linear elastic limit, which applies to both crystalline and amorphous solids, the strain tensor is defined as
\begin{equation}
\epsilon_{ij}(\bm{x}) = \frac{1}{2}\left( \frac{\partial u_i(\bm{x})}{\partial x_j} + \frac{\partial u_j(\bm{x})}{\partial x_i} \right),
\label{eq:strain}
\end{equation}
where quadratic terms in $\bm{u}$ are neglected under the linear approximation~\cite{landau1986theory}. Here, $i,j = 1,2,3$ correspond to the Cartesian coordinates $x,y,z$, respectively.

\subsection{Brief Review of the Two-Level-System Model}


\begin{figure}[htbp]
\includegraphics[scale=0.54]{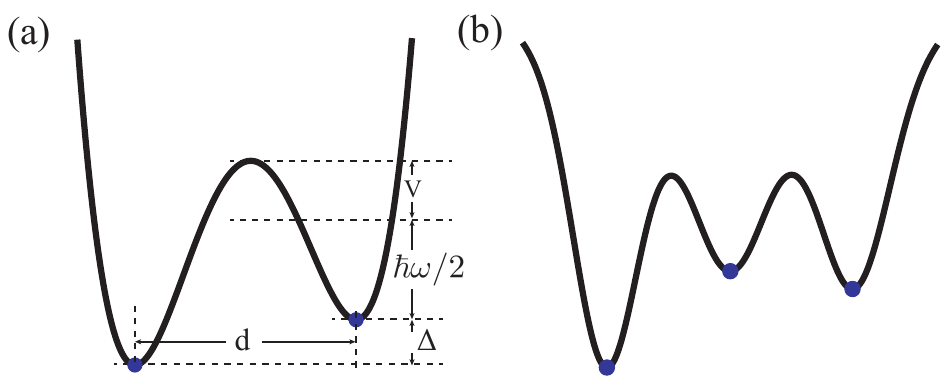}
\caption{Schematic illustration of (a) the two-level system and (b) the multi-level system models. Blue dots in both panels mark the potential minima. In (a), the ground-state energy of the harmonic potential is $\hbar\omega/2$, on the order of the Debye energy. The parameters $\Delta$, $d$, and $V$ denote the asymmetry between wells, their spatial separation, and the barrier height, respectively.
}\label{fig1}
\end{figure}

According to the two-level-system (TLS) model, TLSs are assumed to be randomly distributed throughout the amorphous solid~\cite{Varma1972Phil}. Each TLS corresponds to one or a few atoms tunneling between two potential minima~\cite{Phillips1981}, leading to an effective description of the glass system. In phonon-echo experiments, the system is driven by externally applied hypersonic phononic excitations. The Hamiltonian of the glass is therefore influenced by both the intrinsic strain field $\epsilon_{ij}(\bm{x})$, which arises from thermal and zero-point phonons, and the external driving strain field $\varepsilon_{ij}(\bm{x},t)$. The total strain acting on a TLS is thus $\epsilon_{ij}(\bm{x}) + \varepsilon_{ij}(\bm{x},t)$. The full Hamiltonian can be written as
\begin{eqnarray}\label{1}
\hat{H}_{\rm glass} = \hat{H}_{\rm el}+\sum_{s=1}^{N_{\rm TLS}}\hat{H}^{(s)}_{\rm TLS}(\epsilon_{ij}^{(s)}+\varepsilon_{ij}^{(s)}(t)),
\end{eqnarray}
where $\hat{H}_{\rm el}$ is the purely harmonic Hamiltonian in the long-wavelength linear-elastic limit. It takes the quadratic form
\begin{equation}\label{43}
\hat{H}_{\rm el} = \int_V d^3x \bigg[\sum_{i=1}^3 \frac{1}{2}\rho\dot{u}_i^2(\bm{x}) +\sum_{i,j,k,l=1}^3\frac{1}{2}C_{ijkl}\epsilon_{ij}(\bm{x})\epsilon_{kl}(\bm{x})\bigg],
\end{equation}
where $\rho$ is the mass density of the medium (amorphous or crystalline), $\dot{u}_i = \partial u_i/\partial t$ is the velocity component, and $C_{ijkl}$ is the elastic tensor~\cite{landau1986theory}, which reflects the isotropy of amorphous systems~\cite{Vural2011JNCS}.



The second term in Eq.~(\ref{1}) represents the TLS Hamiltonian, with $N_{\mathrm{TLS}}$ denoting the total number of TLSs in the amorphous system. The superscript $(s)$ labels a TLS located at position $\bm{x}_s$, and $\epsilon_{ij}^{(s)} + \varepsilon_{ij}^{(s)}(t)$ indicates the dependence of the TLS Hamiltonian on the total strain field at that location. Within linear response theory~\cite{Kubo1962JPSJ}, the TLS Hamiltonian can be expanded with respect to the strain field:
\begin{equation}\label{4}
\hat{H}_{\rm TLS}^{(s)}\big(\epsilon^{(s)}_{ij} + \varepsilon^{(s)}_{ij}(t)\big)
= \hat{H}_{\rm TLS}^{(s)}
+ \sum_{i,j=1}^3 \bigl[\epsilon_{ij}^{(s)} + \varepsilon_{ij}^{(s)}(t)\bigr] \, \hat{\mu}_{ij}^{(s)},
\end{equation}
where $\hat{H}_{\rm TLS}^{(s)}$ is the strain-free TLS Hamiltonian. Due to its two-level structure, it can be expressed as a $2\times2$ matrix in its eigenbasis:
\begin{equation}
\hat{H}_{\rm TLS}^{(s)} = \frac{1}{2} \, E^{(s)} \, \sigma_z,
\end{equation}
with eigenvalues $\pm E^{(s)}/2$ (see Fig.~\ref{fig1}(a)). Here, $E^{(s)} = (\Delta^{(s)2} + \Delta_0^{(s)2})^{1/2}$, where $\Delta^{(s)}$ is the asymmetry between the two wells and $\Delta_0^{(s)} = \hbar\omega e^{-\lambda}$. The parameter $\hbar\omega/2$ represents the ground-state energy of the harmonic approximation near a potential minimum. The tunneling parameter is $\lambda = d(2m_{\text{TLS}}V/\hbar^2)^{1/2}$, in which $d$ is the spatial separation of the minima, $V$ is the barrier height, and $m_{\mathrm{TLS}}$ is the effective mass of the tunneling particle. The Pauli matrices $\sigma_x, \sigma_y, \sigma_z$ provide the algebraic structure of the TLS Hamiltonian. Owing to the inherent disorder in glasses, the level splittings $E^{(s)}$ follow a broad distribution, giving rise to an approximately constant density of states. This uniform distribution is responsible for the linear temperature dependence of the heat capacity observed in insulating glasses~\cite{Varma1972Phil}. The coupling term in Eq.~(\ref{4}) describes TLS-strain interactions~\cite{Joffrin1975JDP}. The TLS-strain coupling tensor $\hat{\mu}_{ij}^{(s)}$ is defined as the first-order variation of the Hamiltonian with respect to strain:
\begin{eqnarray}\label{5}
\hat{\mu}_{ij}^{(s)} = \frac{\delta\hat{H}_{\rm TLS}^{(s)}}{\delta\epsilon_{ij}^{(s)}}
= D_{ij}^{(s)} \sigma_z + M_{ij}^{(s)} \sigma_x,
\end{eqnarray}
where $D_{ij}^{(s)}$ and $M_{ij}^{(s)}$ are matrix elements that depend on the microscopic details of the TLS at $\bm{x}_s$. These elements vary randomly across different TLS positions. In the standard TLS model, the tensorial character of the phonon strain field is often neglected, and the coupling constants are replaced by their orientation-averaged values~\cite{Hunklinger1986PLTP}. Collecting the contributions, the glass Hamiltonian originally introduced in Eq.~(\ref{1}) can be written as
\begin{eqnarray}\label{25}
\hat{H}_{\rm glass}
 & = & \hat{H}_{\rm el}
+ \sum_{s} \biggl( \hat{H}_{\rm TLS}^{(s)} + \sum_{ij} \epsilon_{ij}^{(s)} \hat{\mu}_{ij}^{(s)} \biggr)\nonumber \\
& {} &
+ \sum_{s} \sum_{ij} \varepsilon_{ij}^{(s)}(t) \, \hat{\mu}_{ij}^{(s)}.
\end{eqnarray}

In theoretical treatments of both spin echo and phonon echo, it is commonly assumed that the intrinsic coupling between the strain field and the TLS systems, namely the term $\epsilon_{ij}^{(s)}\hat{T}_{ij}^{(s)}$ in Eq.~(\ref{25}), is neglected~\cite{Hunklinger1986PLTP}. Under this approximation, the glass Hamiltonian, in analogy with spin echo theory, reduces to
\begin{eqnarray}\label{53}
\hat{H}_{\rm glass} \approx \hat{H}_{\rm el} + 
\sum_{s}\hat{H}_{\rm TLS}^{(s)}  + \sum_s \sum_{ij} \varepsilon_{ij}^{(s)}(t)\hat{\mu}^{(s)}_{ij}.
\end{eqnarray}
Following this simplified Hamiltonian, the two-level-system (TLS) Hamiltonian, a ${2} \times {2}$ matrix algebraically equivalent to spin operators~\cite{Hahn1950PR}, provides a natural framework for analyzing phonon echo phenomena. In direct analogy with spin-magnetic field coupling, the TLS-strain interaction, represented by the last term in Eq.~(\ref{53}), generates pseudo-spin dynamics formally identical to those in spin echo systems. The intrinsic randomness of the TLS Hamiltonian further acts as a spatially inhomogeneous spin-magnetic field coupling, producing ensembles of pseudo-spins that precess at different rates within the glass, giving rise to the phonon echo signal. 

To depict the echo signal in glasses, we define the expectation value of the pseudo-spin operator for each TLS:
\begin{eqnarray}\label{10}
\langle \hat{\mu}_{ij}^{(s)} \rangle =  
\operatorname{Tr}\!\left(\hat{\rho}^{(s)}_I(t)\,\hat{\mu}_{ij,I}^{(s)}(t)\right),
\end{eqnarray}
where $\hat{\rho}^{(s)}_I$ denotes the density matrix of the $s$th TLS (located at position $\bm{x}_s$) in the interaction picture at inverse temperature $\beta = 1/k_B T$, and $\hat{\mu}_{ij,I}^{(s)}(t)$ denotes the pseudo-spin operator in the same picture. When external strain pulses are applied, the first pulse rotates the pseudo-spin orientation, corresponding to the expectation value of the TLS Pauli matrices as in Eq.~(\ref{10}), by $\pi/2$ into the transverse plane. The second pulse subsequently applies a $\pi$ rotation, reversing the motion of the faster precessing pseudo-spins and rephasing them with the slower ones. This time-reversal mechanism leads to an echo, because the ensemble expectation value of the acoustic pseudo-spin operators combines coherently at $t = 2\tau$ to produce a macroscopic signal. Consequently, the total pseudo-spin polarization of the glassy system,
\begin{eqnarray}\label{48}
\langle \hat{\mu}_{ij} \rangle = \sum_{s=1}^{N_{\rm TLS}} 
\langle \hat{\mu}_{ij}^{(s)} \rangle,
\end{eqnarray}
rephases and yields a detectable echo. The TLS model therefore explains phonon echo phenomena through its direct formal analogy with the standard spin-echo description.

Within this simplified glass Hamiltonian, as shown in Eq.~\eqref{53}, the phonon echo emerges naturally. However, retaining the full glass Hamiltonian in Eq.~\eqref{25}, namely the glass Hamiltonian with the intrinsic coupling term $\epsilon_{ij}^{(s)}\hat{T}_{ij}^{(s)}$
 between the strain field and TLSs, induces effective many-body interactions among spatially separated TLSs, analogous to remote magnetic dipole coupling. We will explicitly derive this interaction, incorporate it into the multi-level-system model in Sec.~\uppercase\expandafter{\romannumeral2}(C), and subsequently examine its influence on the phonon echo within the multi-level framework in Secs.~\uppercase\expandafter{\romannumeral3} and~\uppercase\expandafter{\romannumeral4}.




\subsection{Extending to Multi-Level-System Model}

The TLS model indicates that phonon echo emerges from the combined effect of glassy randomness and the sharp cutoff in the energy eigenvalues inherent to the two-level nature. It is therefore the nonlinearity of the level spacing, rather than the mere existence of two levels, that plays the decisive role. Building upon the conventional two-level-system framework, we extend to a multi-level-system Hamiltonian with non-equidistant energy-level spacing, which can likewise induce phonon echo in glassy systems.

Consider a block of amorphous solid that contains a total of $N_{\mathrm{MLS}}$ multi-level-systems. The Hamiltonian is
\begin{eqnarray}\label{28}
\hat{H}_{\rm glass} = \hat{H}_{\rm el} + \sum_{s=1}^{N_{\rm MLS}} \hat{H}_{\rm MLS}^{(s)}(\epsilon_{ij}^{(s)}+\varepsilon_{ij}^{(s)}(t)),
\end{eqnarray}
where $\hat{H}_{\mathrm{el}}$ is the long-wavelength phonon Hamiltonian [Eq.~(\ref{43})], and $\hat{H}_{\mathrm{MLS}}^{(s)}(\epsilon_{ij}^{(s)}+\varepsilon_{ij}^{(s)}(t))$ describes atoms tunneling among multiple minima of the potential well at $\bm{x}_s$, as schematically illustrated in Fig.~\ref{fig1}(b). Analogous to the TLS Hamiltonian, the multi-level-system Hamiltonian expands in powers of the strain field:
\begin{equation}\label{2}
\hat{H}_{\rm MLS}^{(s)}(\epsilon_{ij}^{(s)}+\varepsilon_{ij}^{(s)}(t)) = \hat{H}_{\rm MLS}^{(s)} + \sum_{i,j=1}^3 \left[\epsilon_{ij}^{(s)} +\varepsilon_{ij}^{(s)}(t)\right]\hat{T}_{ij}^{(s)} ,
\end{equation}
with
\begin{eqnarray}\label{3}
\hat{T}_{ij}^{(s)} = {\delta \hat{H}_{\rm MLS}}/{\delta \epsilon_{ij}^{(s)}},
\end{eqnarray}
paralleling the TLS coupling matrix in Eq.~(\ref{5}). The operator $\hat{T}_{ij}^{(s)}$ has the same dimensionality as $\hat{H}_{\mathrm{MLS}}^{(s)}$ and quantifies strain-induced shifts of multi-level-system eigenvalues. Thus, Eqs.~(\ref{2}) and (\ref{3}) establish the multi-level extension of the TLS model in glasses.

In conventional elasticity theory~\cite{landau1986theory}, $\hat{T}_{ij}^{(s)}$ coincides with the stress tensor, bridging classical continuum mechanics with the quantum multi-level-system description. Therefore, we use the terms ``stress tensor'' and ``strain-multi-level-system coupling operator'' interchangeably. With an applied external strain field $\varepsilon_{ij}^{(s)}(t)$, the complete Hamiltonian for the multi-level-system model becomes
\begin{eqnarray}\label{51}
\hat{H}_{\rm glass} & = & \hat{H}_{\rm el} + \sum_{s}\bigg(\hat{H}_{\rm MLS}^{(s)} +\sum_{ij}\epsilon_{ij}^{(s)}\hat{T}^{(s)}_{ij}\bigg) \nonumber \\
& {} & 
+\sum_s\sum_{ij}\varepsilon_{ij}^{(s)}(t)\hat{T}^{(s)}_{ij}.
\end{eqnarray}

Analogous to the standard treatment of spin and phonon echoes, where the coupling between intrinsic strain fields and TLSs is neglected, we may also omit the term $\epsilon_{ij}^{(s)}\hat{T}_{ij}^{(s)}$ in the multi-level-system model. This yields an approximate glass Hamiltonian that is already sufficient to derive the phonon echo:
\begin{equation}\label{54}
\hat{H}_{\rm glass} \approx \hat{H}_{\rm el} + \sum_{s}\hat{H}_{\rm MLS}^{(s)} 
+\sum_s\sum_{ij}\varepsilon_{ij}^{(s)}(t)\hat{T}^{(s)}_{ij}.
\end{equation}
As in the TLS model, where the phonon echo is extracted from the ensemble average of the pseudo-spin (Pauli) operators, we define the corresponding ensemble average of the stress tensor operator $\hat{T}_{ij}^{(s)}$ for the multi-level system:
\begin{equation}\label{49}
\langle \hat{T}_{ij} \rangle(t) = \sum_{s=1}^{N_{\mathrm{MLS}}} \operatorname{Tr}\!\left(\hat{\rho}^{(s)}_I(t)\,\hat{T}_{ij,I}^{(s)}(t)\right),
\end{equation}
where $\hat{\rho}^{(s)}_I(t)$ is the density matrix of the multi-level system located at $\bm{x}_s$ in the interaction picture, and $\hat{T}_{ij,I}^{(s)}(t)$ is the corresponding stress tensor operator in the same picture. This quantity will later serve, in both the semi-classical and quantum analyses, to reveal the phonon echo within the multi-level-system framework.

The approximated Hamiltonian shown in Eq.~\eqref{54} is already capable of demonstrating the phonon echo. However, retaining the full strain-stress coupling term $\epsilon_{ij}^{(s)}\hat{T}_{ij}^{(s)}$ in the multi-level-system Hamiltonian, as in Eq.~\eqref{51}, naturally introduces a weak many-body interaction between spatially separated multi-level systems, analogous to magnetic dipole-dipole interactions. In the following sections, we first use the approximated Hamiltonian in Eq.~\eqref{54} to investigate the multi-level extension of phonon echo in Secs.~\uppercase\expandafter{\romannumeral3}(A) and~\uppercase\expandafter{\romannumeral4}(A), and subsequently analyze the influence of the many-body interaction on the echo in Secs.~\uppercase\expandafter{\romannumeral3}(B) and~\uppercase\expandafter{\romannumeral4}(B).

\subsection{Many-Body Interaction Between Multi-Level Systems}

In the conventional TLS description of glassy dynamics~\cite{Golding1979PRB, Phillips1987RPP}, the Hamiltonian shares the algebraic structure of spin operators. In spin systems, magnetic dipole moments interact remotely via a $1/r^3$ coupling, suggesting that TLSs can likewise couple over long distances. Extending this framework to multi-level systems, we derive the corresponding many-body interaction in the framework of multi-level-system model. 

This mechanism can be understood from two complementary perspectives: (i) A multi-level system at position $\bm{x}$ perturbs the surrounding atomic configuration, generating a local strain field. A second system at $\bm{x}'$, with atoms tunneling in its own potential wells, experiences this deformation, leading to strain-mediated stress-stress coupling and thereby an elastic many-body interaction. (ii) Because multi-level systems intrinsically couple to phonon strain fields, an excited system at $\bm{x}$ can emit a phonon that propagates through the amorphous medium and is absorbed by another system at $\bm{x}'$. Such virtual phonon exchange~\cite{Yu2017PRA, peskin2018introduction} induces an effective long-range interaction between the two systems.


Starting from the glass Hamiltonian in Eq.~\eqref{51}, we absorb the intrinsic strain-stress coupling into the phonon degrees of freedom by completing the square in the elastic strain fields~\cite{Joffrin1975JDP}. This yields an equivalent Hamiltonian
\begin{equation}\label{29}
\hat{H}_{\rm glass} = \hat{H}_{\rm el}(\hat{T}_{ij})+ \sum_s \hat{H}_{\rm MLS}^{(s)}+ \sum_{s}\sum_{ij}\varepsilon_{ij}^{(s)}(t)\hat{T}_{ij}^{(s)}+ \hat{V},
\end{equation}
where $\hat{H}_{\rm el}(\hat{T}_{ij})$ is the phonon energy with equilibrium positions shifted by the intrinsic coupling, and the mutual many-body interaction term takes the form
\begin{eqnarray}\label{13}
 \hat{V} = \sum_{ss'}\sum_{ijkl}\Lambda_{ijkl}^{(ss')}\hat{T}_{ij}^{(s)}\hat{T}_{kl}^{(s')}.
\end{eqnarray}
Here $s$ and $s'$ label multi-level systems, and $i,j,k,l=1,2,3$ denote Cartesian components. The coefficient $\Lambda_{ijkl}^{(ss')}$ encodes the remote interaction between stress tensors at positions $\bm{x}_s$ and $\bm{x}_{s'}$. This coefficient was first derived in Ref.~\cite{Joffrin1975JDP}, later corrected in Ref.~\cite{zhou2019universal}, and is reproduced in Appendix~A as
\widetext
\begin{eqnarray}\label{14}
 & {} & \Lambda_{ijkl}^{(ss')} = -\frac{1}{32\pi\rho c_t^2}\sum_{\substack{(a,b,c,d)=(i,j,k,l),\\ (i,j,l,k),(j,i,l,k),(j,i,k,l)}}\left[\frac{\delta_{bd}-3n_b n_d}{|\bm{x}_s-\bm{x}_{s'}|^3}(1-\delta_{ss'})+\frac{4\pi}{3L^3}\delta_{bd}\delta_{ss'}\right]\delta_{ac} +\frac{1-{c_t^2}/{c_l^2}}{2\rho c_t^2}  \nonumber \\ 
 & {} & \quad \cdot  \left[\frac{(1-\delta_{ss'})}{8\pi |\bm{x}_s-\bm{x}_{s'}|^3}+\frac{4\delta_{ss'}}{15L^3}\right] 
 \Bigg(\sum_{\substack{(a,b,c,d)=(i,j,k,l),\\ (i,k,j,l),(j,k,i,l)}}\delta_{ab}\delta_{cd} -3\sum_{\substack{ (a,b,c,d) = (i,j,k,l), (i,k,j,l),\\ (i,l,j,k),(j,k,i,l), (j,l,i,k), (k,l,i,j) }} n_a n_b\delta_{cd} +15n_in_jn_kn_l\Bigg).\qquad 
 \end{eqnarray}
\endwidetext
Here $\hat{n}=(\bm{x}_s-\bm{x}_{s'})/|\bm{x}_s-\bm{x}_{s'}|$ is the unit vector connecting the two systems, with $n_i$ its Cartesian components; $c_l$ and $c_t$ are the longitudinal and transverse sound velocities, respectively; $L$ is the average spacing between nearest-neighbor multi-level systems ($L\gg a$, with $a\sim 3$\AA\ the atomic distance); and $\delta_{ab}$ is the Kronecker delta. Despite its apparent complexity, $\Lambda_{ijkl}^{(ss')}$ is essentially a fourth-order tensor generalization of the familiar dipole-dipole coefficient. Its algebraic decay, $\Lambda_{ijkl}^{(ss')}\sim 1/|\bm{x}_s-\bm{x}_{s'}|^3$, follows directly from the Fourier transform of elastic fields in three dimensions, exactly analogous to magnetic dipole-dipole interactions.

Incorporating both the many-body interaction and an externally applied time-dependent strain field $\varepsilon_{ij}^{(s)}(t)$, the effective Hamiltonian for the multi-level systems is given by Eq.~\eqref{29}. Using this time-dependent Hamiltonian, we analyze how many-body interactions affect phonon-echo phenomena in both the semi-classical and quantum regimes. The semi-classical approach provides mathematical simplicity and physical intuition, whereas the finite-temperature quantum framework yields a complete microscopic description. Both approaches lead to consistent results, which we verify analytically and through numerical simulations.

\section{Phonon echo in the semi-classical framework}


We construct a semi-classical analog of the multi-level-system model using the WKB approximation, which maps the quantum multi-level system in the large-quantum-number limit to a nonlinear oscillator with one degree of freedom. For a multi-level system located at $\bm{x}_s$, the Hamiltonian $\hat{H}_{\mathrm{MLS}}^{(s)}$ is described by a generalized coordinate $q_s$, an abstract \textit{phase variable} that tracks cyclic quantum evolution, analogous to the angular coordinate of a wave packet moving along a closed phase-space orbit. Its time derivative $\dot{q}_s$ gives the instantaneous phase velocity. The effective mass $m$, determined by the curvature of the energy spectrum with respect to the quantum number, quantifies the stiffness of the level structure: narrow (wide) level spacing corresponds to a light (heavy) mass, yielding fast (slow) phase evolution under perturbations. The semi-classical dynamics then correspond to a fictitious particle of mass $m$ moving in a one-dimensional nonlinear potential $V^{(s)}(q_s)$ derived from $\hat{H}_{\mathrm{MLS}}^{(s)}$, with kinetic energy $\tfrac{1}{2} m\dot{q}_s^2$. The potential $V^{(s)}(q_s)$ is designed via the WKB quantization condition 
\begin{equation}\label{56}
\int \sqrt{2m\big[E_n^{(s)}-V^{(s)}(q_s)\big]}\,dq_s = \pi\hbar\Bigl(n+\tfrac12\Bigr),\qquad n \in \mathbb{Z},
\end{equation}
where $E_n^{(s)}$ is the eigenvalue of $\hat{H}_{\mathrm{MLS}}^{(s)}$.

The effective semi-classical potential $V^{(s)}$ is constructed by applying inversion techniques of Eq.~\eqref{56} to energy-dependent potentials~\cite{PhysRevD.109.096014}, which include both the tensorial nature of external strain fields and intrinsic stress tensors. After averaging over all strain-field orientations, the potential reduces to a scalar form; this corresponds to neglecting the tensorial character of the phonon strain field and replacing the coupling matrix elements by their orientation-averaged values, as is conventional in the TLS model~\cite{Hunklinger1986PLTP}. Below, we use the generalized coordinate $q_s$ and the derived potential landscape $V^{(s)}(q_s)$ to study phonon echo in the semi-classical framework, focusing on the weakly nonlinear and zero-temperature limits.

\subsection{Phonon echo in a semi-classical framework without many-body interactions}


This section examines phonon echo in the weakly nonlinear, zero-temperature semi-classical regime, explicitly \textit{excluding} many-body interactions. In this limit, the semi-classical multi-level systems are treated as independent particles. Our focus is on the semi-classical description of on-site multi-level systems and their echo response.

The semi-classical potential is naturally expressed as a polynomial in $q_s$, reflecting the non-equidistant energy spectra of multi-level quantum systems. A purely harmonic potential would yield equally spaced levels, whereas anharmonic terms capture deviations from this uniform spacing. Nonlinear contributions to the semi-classical potential generate frequency shifts in the oscillations of nonlinear normal modes~\cite{vakakis2001normal, Leamy2019PRE, Zhou2022NC}. These shifts are central to the emergence of the semi-classical phonon echo. In what follows, we compute the frequency shift of nonlinear normal modes as the oscillation amplitude increases, employing the perturbative method of multiple scales. We then demonstrate that the echo signal arises directly from this nonlinearity-induced frequency shift.

The general semi-classical potential involves an infinite polynomial expansion. In the weakly nonlinear regime, however, the quartic term provides the leading correction. Thus, our analysis is restricted to fourth order:
\begin{equation}
V^{(s)}(q_s) = \tfrac{1}{2}k_{2}^{(s)} q_s^2 + \tfrac{1}{3}k_3 q_s^3 + \tfrac{1}{4}k_4 q_s^4,
\end{equation}
where the harmonic coefficient $k_{2}^{(s)}$ varies across sites to represent the randomized dynamics of multi-level systems, and $k_3$ and $k_4$ denote the cubic and quartic contributions.


The semi-classical potential governs the dynamics of each nonlinear oscillator $s$. At zero temperature, the motion follows the Newtonian equation
\begin{equation}
m\ddot{q}_s = -k_{2}^{(s)} q_s - k_3 q_s^2 - k_4 q_s^3,
\end{equation}
where $m$ is the effective semi-classical mass set by the curvature of the multi-level energy spectrum. In the weakly nonlinear regime $k_3 A,\, k_4 A^2 \ll k_{2,s}$ ($A$ being the mode amplitude), we apply the method of multiple scales~\cite{Rosa2023NJP, Ruzzene2018PRE, Zhou2024NJP, Romeo2012}, a standard perturbative technique for weakly nonlinear systems. As outlined in Appendix~B, we introduce a book-keeping parameter in front of the nonlinear terms, expand both the time scale and $q_s$ in multiple scales, and match the equations of motion order by order while removing secular terms. This procedure reveals two main effects: (i) the cubic term $k_3$ shifts the equilibrium position away from zero, with a displacement proportional to the mode amplitude; (ii) the quartic term $k_4$ induces a frequency shift
\begin{equation}\label{7}
\omega^{(s)}(A) = \omega^{(s)} + \frac{3k_4}{8m\omega^{(s)}} A^2,
\end{equation}
where $\omega^{(s)} = (k_{2}^{(s)}/m)^{1/2}$ is the linear natural frequency of oscillator $s$. The quadratic coefficient $k_{2}^{(s)}$ varies randomly among sites, making $\omega^{(s)}$ site-dependent and thereby introducing intrinsic randomness into the ensemble of non-interacting nonlinear oscillators.

To probe the phonon echo, we apply two impulsive kicks at $t=0$ and $t=\tau$, which instantaneously modify the oscillator velocity. In its orientation-averaged scalar form, the semi-classical driving force is
\begin{equation}\label{6}
F^{(s)}(t) = I_1^{(s)} \delta(t) + I_2^{(s)} \delta(t-\tau),
\end{equation}
where $I_1^{(s)}$ and $I_2^{(s)}$ are the impulse magnitudes.

\begin{figure}[htbp]
\includegraphics[scale=0.54]{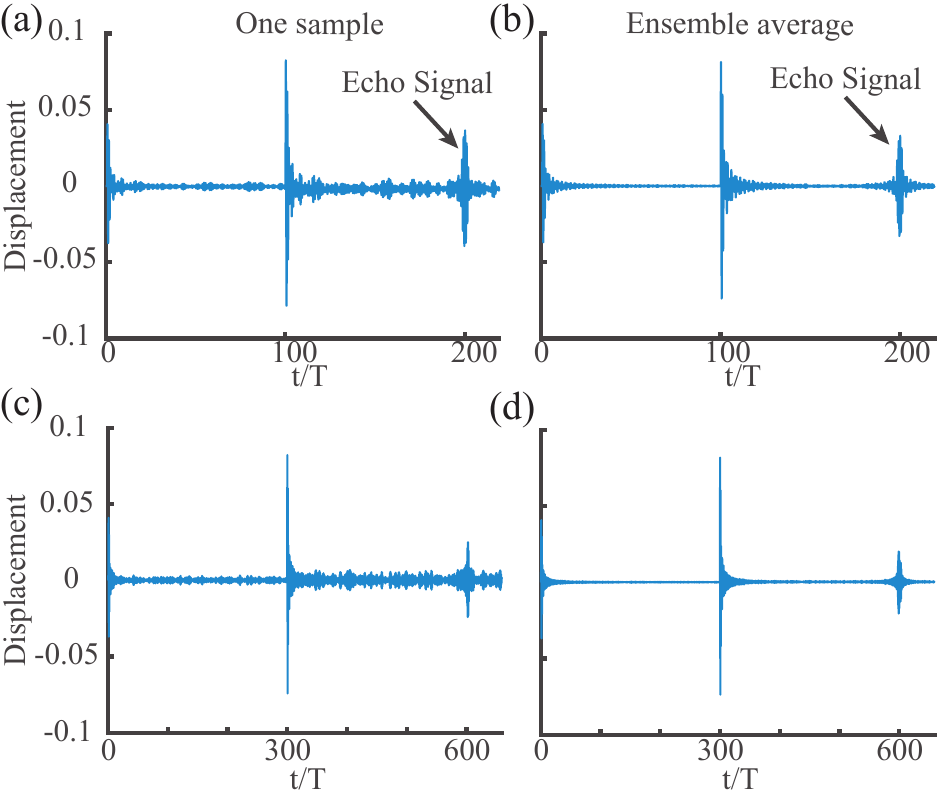}
\caption{Numerical simulations of the phonon echo within the semi-classical framework. 
The Newtonian dynamics are integrated using the fourth-order Runge--Kutta algorithm. 
A total of $10^3$ randomly generated, non-interacting, and nonlinear oscillators are simulated with parameters: 
oscillator mass $m=1$, mean quadratic stiffness $\bar{k}_{2,s}=1$ with standard deviation ${\rm std}(k_{2,s})=1$, 
cubic stiffness $k_3\in[-1,1]$ uniformly distributed, and quartic stiffness $k_4=1$. 
The first pulse is applied at $\tau_1=0.5=0.08T$, where $T=2\pi(m/\bar{k}_{2,s})^{1/2}$ is the oscillator period, 
with duration $\Delta\tau_1=0.2=0.032T$ and amplitude $I_1/\Delta\tau_1=0.5$. 
The second pulse is applied at $\tau_2=100T$, with duration $\Delta\tau_2=0.2=0.032T$ and amplitude $I_2/\Delta\tau_2=1$. 
The pulse separation is $\tau=\tau_2-\tau_1=100T$, predicting an echo signal at $\tau_1+2\tau=200T$. 
The numerical time step is set to $dt=0.1$. 
(a) Summation of the displacement--time relation for a single set of $10^3$ oscillators. 
(b) Ensemble-averaged dynamics obtained from $N=100$ independent realizations of randomly generated oscillator groups. 
(c) Same simulation as in (a), but with the second external pulse applied at $\tau_2=300T$. 
(d) Same ensemble averaging as in (b), with the second external pulse applied at $\tau_2=300T$.
 }\label{fig2}
\end{figure}

After the first pulse ($t=0$), oscillator $s$ oscillates with amplitude $A_1^{(s)}$ and frequency $\omega^{(s)}(A_1^{(s)})$ given by Eq.~\eqref{7}. For $0<t<\tau$, the nonlinear normal mode approximates $q_s(t) = A_1^{(s)} \sin\omega^{(s)}(A_1^{(s)})t$ in the \textit{weakly nonlinear regime}~\cite{Zhou2020PRB}. At $t=\tau$, the second kick changes the amplitude to $A_2^{(s)}$, and oscillations thereafter occur at $\omega^{(s)}(A_2^{(s)})$. As derived in Appendix~B, for $t>\tau$ the displacement splits into two parts:
\begin{eqnarray}\label{8}
q_s(t>\tau) = q_{s,{\rm ne}}(t) + q_{s,{\rm e}}(t).
\end{eqnarray}
The first term, $q_{s,\mathrm{ne}}$, is the \textit{non-echo} contribution present in both linear and nonlinear oscillators; it does not produce the echo. Summing over the amorphous ensemble yields
\begin{eqnarray}\label{19}
\langle q_{\rm ne}\rangle(t)
 & = & \sum_{s=1}^{N_{\rm MLS}} \bigg[A_1^{(s)}\sin\omega^{(s)} t \nonumber \\
& {} & + (A_1^{(s)}-A_2^{(s)})\sin\omega^{(s)}(t-\tau)\bigg],
\end{eqnarray}
which, assuming a uniform distribution of oscillator frequencies (as in the standard TLS model), decays as $\sim 1/t$ in time. This decay is illustrated in the Newtonian simulation results for the ensemble-averaged displacement shown in Figs.~\ref{fig2}(b,d), where theory and simulation show excellent agreement. At the echo time $t=2\tau \gg 2\pi/\omega^{(s)}$, this contribution becomes negligible compared to the echo signal.

The second term in Eq.~(\ref{8}), $q_{s,{\rm e}}$, provides the non-vanishing echo contribution. Performing the summation over the semi-classical displacement $q_s$ yields the semi-classical echo signal
\begin{equation}\label{9}
\langle q_{\rm e}\rangle(t)
= \sum_{s=1}^{N_{\rm MLS}}
\frac{3k_4 A_1^{(s)}}{16 m\omega^{(s)2}}  (A_1^{(s)2}-A_2^{(s)2})\sin\omega^{(s)}(t-2\tau).
\end{equation}
This result has several notable features.  
(i) The signal becomes macroscopic around $t=2\tau$. The change in amplitude from the second pulse mixes the frequencies $\omega^{(s)}(A_1^{(s)})$ and $\omega^{(s)}(A_2^{(s)})$, generating a difference mode $\omega^{(s)}(A_1^{(s)})-\omega^{(s)}(A_2^{(s)})$ with phase $\omega^{(s)}(t-2\tau)$. This represents the semi-classical continuous extension of the nonlinear energy-level structure beyond the discrete two-level case.  
(ii) Because the natural frequencies $\omega^{(s)}$ vary randomly across sites, the echo decays rapidly for $|t-2\tau|\gtrsim 10\times 2\pi/\omega^{(s)}$, as seen in Fig.~\ref{fig2}. This indicates the essential role of randomness, because identical oscillators would always stay coherent, which contradicts with the experimental observations that in glasses the echo signal only occurs at $t=2\tau$. This memory effect distinguishes amorphous systems from crystals and illustrates how nonlinearity and randomness generate memory in disordered and biophysical systems~\cite{ZhouHaiJun2005PRL, Wu2025NC, Zhou2018PRL, Tang2024PRL}. (iii) The echo amplitude scales with the quartic nonlinearity $k_4$, following $A_1^{(s)}(A_1^{(s)2}-A_2^{(s)2})$. This dependence arises from the change in frequency shift before and after the second pulse, underscoring the central role of nonlinearity-induced frequency modulation. Moreover, this term is proportional to $I_1^{(s)} I_2^{(s)2}$, consistent with conventional photon and spin echoes~\cite{Hahn1950PR, Zhai2020PRA}. (iv) Equation~\eqref{9} shows the echo is a \textit{third-order} perturbative process, in agreement with spin echo~\cite{PhysRev.141.499, Cooke1982JPCSSP} and photon echo~\cite{cho2019JOSA, Bishop2004PRB, Patton2005PRL}. 

Figure~\ref{fig2} presents numerical simulations validating the theoretical predictions of phonon echo in the semi-classical framework. In Figs.~\ref{fig2}(a,c), a single set of $10^3$ non-interacting nonlinear oscillators is driven by two external pulses. The sum of their displacements spontaneously yields a macroscopic echo at $t=2\tau$. We further sum over $N=100$ independent oscillator displacements, which suppresses random spikes outside the echo window. The resulting ensemble-summed signals, as shown in Figs.~\ref{fig2}(b,d), show excellent agreement with our analytical results.

\begin{figure}[htbp]
\includegraphics[scale=0.525]{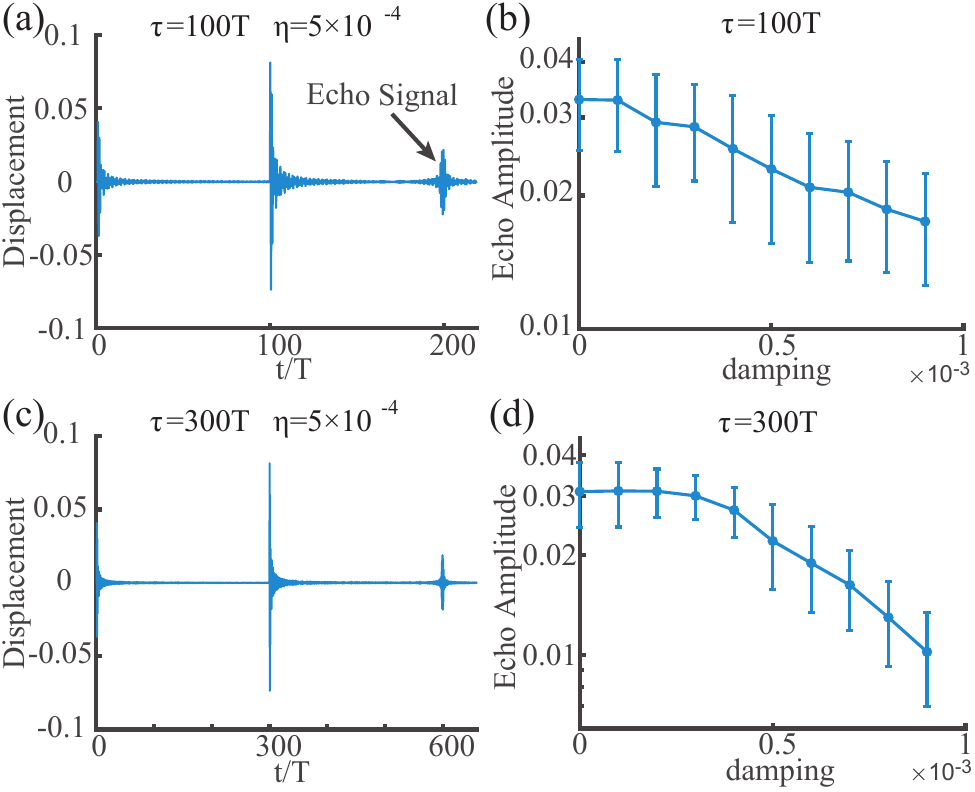}
\caption{Influence of damping on the semi-classical phonon echo.
(a) Echo signal using the parameters from Fig.~\ref{fig2} with pulse separation $\tau = 100T$ and damping coefficient $\eta=5\times 10^{-4}$.
(b) Echo amplitude as a function of $\eta$ ($0 \leq \eta \leq 9\times 10^{-4}$). Dots and error bars denote the mean and standard deviation from $N=100$ samples.
(c) Same as (a) but with $\tau = 300T$.
(d) Same as (b) but with $\tau = 300T$.}\label{fig3}
\end{figure}

We further consider the coupling of the multi-level system to an environment, which mediates energy exchange between them. In the semi-classical limit, where the large particle number yields a macroscopic coherent state (such as the collective phonon echo signal), and assuming Ohmic system-environment coupling and Markovian dynamics, the environmental influence can be approximated by a viscous damping term $f_s = -\eta \dot{q}_s$ acting on the $s$-th semi-classical degree of freedom. This damping introduces dephasing in the ensemble-averaged echo signal. As illustrated in Fig.~\ref{fig3}, the echo amplitude decays exponentially with increasing damping coefficient $\eta$.

From the above analysis, it is evident that a purely harmonic oscillator cannot produce a phonon echo signal. This is numerically verified in Fig.~\ref{fig4}(a), thereby confirming that nonlinearity is essential for the emergence of the echo. We will discuss this result in Sec.~\uppercase\expandafter{\romannumeral3}(B).

Although the method of multiple scales has long been employed to analyze nonlinear mode oscillations, this work presents its first application to the echo phenomenon in semi-classical systems. This highlights the versatility and effectiveness of the method of multiple scales in nonlinear mechanics.

\subsection{Phonon echo in a semi-classical framework with many-body interaction}

\begin{figure}[htbp]
\includegraphics[scale=0.54]{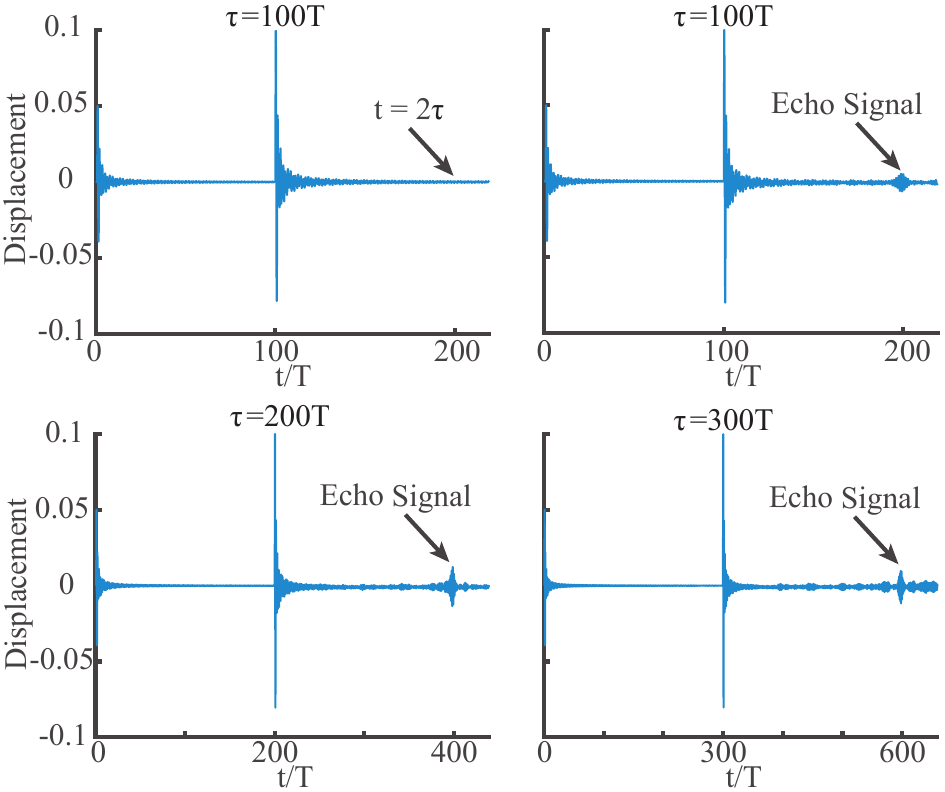}
\caption{Numerical results of the phonon echo within the semi-classical framework, induced solely by many-body interactions. A total of $10^3$ randomly generated, interacting nonlinear oscillators are simulated, with ensemble averaging over $N=100$ samples. The oscillators are placed on a $10\times 10\times 10$ cubic lattice, each labeled by $(n_1, n_2, n_3)$ with $1\le n_1, n_2, n_3\le 10$. Parameters are as in Fig.~\ref{fig2}, but with $k_3=k_4=0$ to ensure each individual oscillator is purely linear. The mutual interaction coefficients satisfy $V_{nn'}^{(ss')}=0$ for $n\neq 2$ or $n'\neq 2$. In (a) $V_{22}^{(ss')}=0$ (non-interacting case); in (b), (c), and (d) $V_{22}^{(ss')}=2\xi$, with $\xi$ uniformly distributed in $[0,1]$ over each interacting pair $s,s'$. Interactions are restricted to nearest neighbors: an oscillator at $(n_1,n_2,n_3)$ couples only to the six neighbors at $(n_1\pm1,n_2,n_3)$, $(n_1,n_2\pm1,n_3)$, and $(n_1,n_2,n_3\pm1)$. This simplification does not alter the qualitative echo behavior. The pulse separation $\tau$ is $100T$, $100T$, $200T$, and $300T$ for panels (a)-(d), respectively.
}\label{fig4}
\end{figure}

The preceding analysis shows that a purely harmonic oscillator, whose eigenfrequency is amplitude-independent, cannot exhibit an echo, as shown in Fig.~\ref{fig4}(a). In contrast, many-body interactions in the semi-classical limit can generate effective nonlinearities beyond the purely harmonic potential. This mechanism enables semi-classical echo signals to emerge even when each local oscillator alone is clear of echo signal. In what follows, we study purely harmonic oscillators with many-body interactions and analyze the resulting echo signal.


The semi-classical limit of the many-body interaction in Eq.~\eqref{13} can be obtained as follows. In quantum mechanics, the displacement operator is expressed via ladder operators, which act off-diagonally in the energy eigenbasis. Likewise, the stress tensor operator $\hat{T}_{ij}^{(s)}$ defined in Eq.~\eqref{3} couples different oscillator levels and is therefore also off-diagonal in that basis. Thus, the stress-tensor operators can be written as polynomials of the displacement operator. This relation follows from evaluating the expectation value of $\hat{T}_{ij}^{(s)}$ in the wave function of the multi-level system at $\bm{x}_s$:
\begin{equation}\label{50}
\langle \psi^{(s)} | \hat{T}_{ij}^{(s)} | \psi^{(s)} \rangle
= \sum_{n \ge 0} a_{ij,n}^{(s)}\, q_s^{\,n},
\end{equation}
where the coefficients $a_{ij,n}^{(s)}$ vary from site to site and are constrained to be real numbers, reflecting the Hermitian nature of the quantum stress-tensor operators. If the system is invariant under inversion about $q_s=0$, odd-$n$ terms vanish; otherwise, linear and higher odd terms may appear.

Substituting the polynomial expansion into the many-body interaction yields an effective semi-classical coupling potential in the generalized coordinates $\{q_s\}$:
\begin{equation}\label{15}
V = \sum_{s\neq s'} \sum_{n,n'} \frac{V_{nn'}^{(ss')}}{2}\, q_s^{n} q_{s'}^{n'} ,
\end{equation}
where the interaction strength is
\begin{equation}
V_{nn'}^{(ss')} \equiv \sum_{ijkl} 
\frac{\Lambda_{ijkl}^{(ss')} a_{ij,n}^{(s)} a_{kl,n'}^{(s')}}{|\bm{x}_s-\bm{x}_{s'}|^3}.
\end{equation}
This potential contains cross terms between $q_s$ and $q_{s'}$, thereby introducing nonlinear coupling into the semi-classical equations of motion. The corresponding generalized restoring force is $Q_s = -\partial V/\partial q_s$, so that the equation of motion for particle $s$ becomes
\begin{equation}\label{16}
m \ddot{q}_s = - k_{2,s} q_s - k_3 q_s^2 - k_4 q_s^3 
- \sum_{s' \neq s} \sum_{n,n'} n\, V_{nn'}^{(ss')} \, q_s^{\,n-1} q_{s'}^{\,n'} .
\end{equation}
For purely harmonic oscillators, the on-site nonlinear terms are set to zero, which isolates the echo signal as arising solely from many-body induced nonlinearity rather than intrinsic anharmonicity. We retain these parameters in the derivation to keep the result general; the purely harmonic case is recovered by taking $k_3 = k_4 = 0$. Applying the method of multiple scales (Appendix~B) to a set of mode amplitudes $\{A\} = (A^{(1)}, A^{(2)}, \ldots, A^{(N_{\mathrm{MLS}})})$, the leading contribution to the amplitude-induced frequency shift, denoted as $\delta\omega^{(s)}(t) = \omega^{(s)}(t) - \omega^{(s)}$, comes from the $n=n'=2$ term in $V_{nn'}^{(ss')}$, giving
\begin{equation}\label{17}
\delta\omega^{(s)}(t) = \frac{3k_4}{8m\omega^{(s)}} A^{(s)2}(t) 
+ \frac{1}{2m\omega^{(s)}} \sum_{s'\neq s} V_{22}^{(ss')} A^{(s')2}(t).
\end{equation}
Here $A^{(s')}$ is the mode amplitude of the $s'$-th oscillator, and the natural frequency $\omega^{(s)} = (k_{2,s}/m)^{1/2}$ varies across sites due to the intrinsic randomness of glassy systems. This result shows that, even with vanishing on-site anharmonicity ($k_3=k_4=0$), many-body interactions generate an effective nonlinearity that shifts the oscillation frequency through coupling to remote oscillators. The corresponding numerical verifications are exhibited in Figs.~\ref{fig4}(b-d).


\section{Phonon Echo in Quantum Mechanical Multi-Level Systems at Finite Temperatures}

Building on the simplified zero-temperature analysis of the semi-classical phonon echo, we now present a full quantum description at finite temperature. Within the quantum multi-level-system framework, both the echo signal and its associated dephasing arise naturally.

\subsection{Quantum Echo in Multi-Level Systems without Many-Body Interactions}

\begin{figure}[htbp]
\includegraphics[scale=0.54]{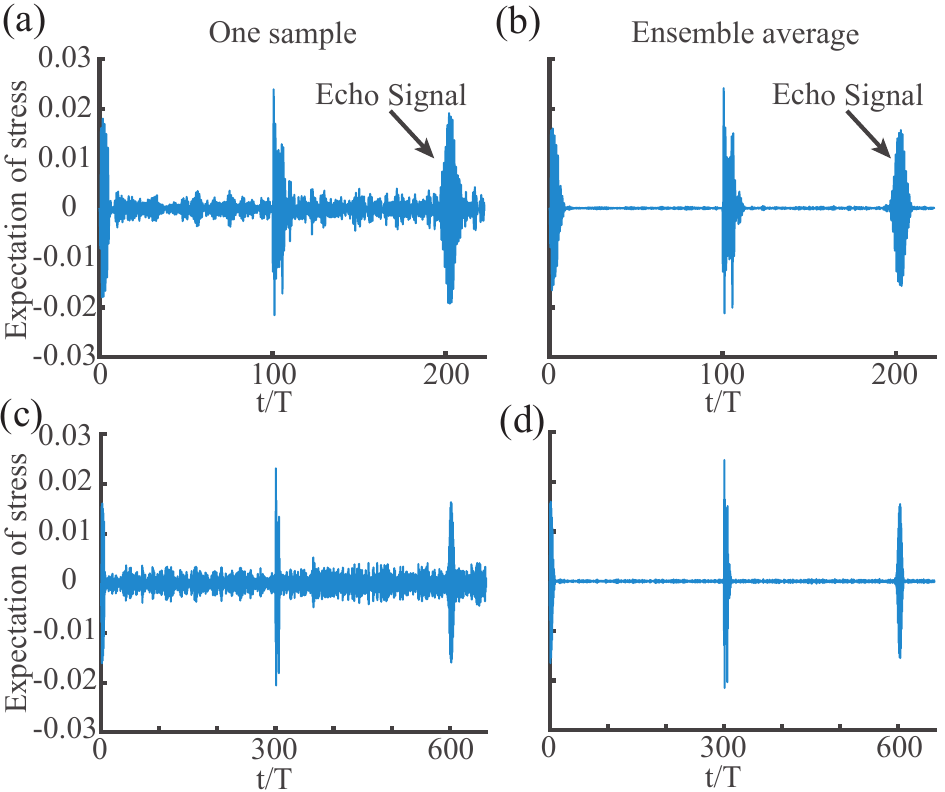}
\caption{Numerical simulation of the phonon echo from the multi-level-system model in the quantum regime. The Liouville-von Neumann dynamics are integrated using a fourth-order Runge-Kutta algorithm with $\hbar=1$, $k_{\mathrm{B}}=1$, and temperature $T=0.1$. A total of $10^3$ randomly generated, non-interacting quantum multi-level systems are placed on a $10\times10\times10$ lattice. Each system is described by a $5\times5$ Hamiltonian (a five-level system) with eigenvalues $\epsilon = 0,0.3,1.3,3,5.1$. The orientation-averaged strain-field strengths are $\epsilon_1=0.1$ for the first pulse and $\epsilon_2=0.2$ for the second pulse. The coupling stress tensor to the external strain field is $\hat{T}_{\mathrm{ext}}^{(s)}+\delta\hat{T}_{\mathrm{ext}}^{(s)}$, where the tensorial character has been averaged over orientations. The zeroth-order part $\hat{T}_{\mathrm{ext}}^{(s)}$ is a $5\times5$ Hermitian matrix with elements 
$\hat{T}_{\mathrm{ext},nn'}^{(s)} = 0.2\sqrt{{n}/{2}}\,\delta_{n',n+1}$ and $\hat{T}_{\mathrm{ext},nn'}^{(s)} = \hat{T}_{\mathrm{ext},n'n}^{(s)}$. The perturbation $\delta\hat{T}_{\mathrm{ext}}^{(s)}$ is a Hermitian random $5\times5$ matrix whose elements are uniformly distributed in $[-0.5,0.5]$. (a) Pulse separation $\tau=100T$; total stress-tensor expectation value summed over $10^3$ systems.  
(b) Ensemble average of the signal in (a) over $N=100$ independent realizations.  
(c) Same as (a) but with $\tau=300T$.  
(d) Same as (b) but with $\tau=300T$. 
   }\label{fig5}
\end{figure}

We begin with a collection of multi-level systems having randomly distributed energy eigenvalues. As shown in Eq.~\eqref{49}, the echo signal is governed by the density matrix $\hat{\rho}^{(s)}$ of the $s$-th multi-level system at position $\bm{x}_s$. Under an external hypersonic strain field $\varepsilon_{ij}^{(s)}(t)$, the system experiences a time-dependent perturbation $\varepsilon_{ij}^{(s)}(t)\hat{T}_{ij}^{(s)}$ [see Eq.~\eqref{54}]. The density matrix then evolves according to the Liouville-von Neumann equation in the interaction picture:
\begin{equation}\label{39}
\mathrm{i}\,\partial_t \hat{\rho}_I^{(s)}(t) 
= \sum_{ij} \varepsilon_{ij}^{(s)}(t)\,
\bigl[\hat{T}_{ij,I}^{(s)}(t),\,\hat{\rho}_I^{(s)}(t)\bigr],
\end{equation}
where the interaction-picture operator is defined by $\hat{A}_I(t)=e^{\mathrm{i}\hat{H}_{\mathrm{MLS}}^{(s)}t}\hat{A}e^{-\mathrm{i}\hat{H}_{\mathrm{MLS}}^{(s)}t}$. Both $\hat{\rho}_I^{(s)}(t)$ and $\hat{T}_{ij,I}^{(s)}(t)$ are expressed in this picture; here $\mathrm{i}=\sqrt{-1}$ and $\hbar$ is set to unity. The stress tensor, defined in Eqs.~\eqref{2} and \eqref{3}, quantifies the coupling between multi-level systems and the phonon strain field. The external strain field is modeled as two impulsive pulses:
\begin{equation}\label{40}
\varepsilon_{ij}^{(s)}(t) = I^{(s)}_{1,ij}\,\delta(t) + I^{(s)}_{2,ij}\,\delta(t-\tau).
\end{equation}
At the initial time $t=0$, the density matrix is given by the Boltzmann distribution
\begin{equation}
\hat{\rho}_I^{(s)}(0) = \sum_i p_i^{(s)} \, |i^{(s)}\rangle\langle i^{(s)}|,
\end{equation}
with $p_i^{(s)} = \exp(-\beta E_i^{(s)})/\mathcal{Z}^{(s)}$ the occupation probability of the $i$-th eigenstate at temperature $T$, $\beta = 1/k_{\mathrm{B}}T$, and $\mathcal{Z}^{(s)}$ the partition function of the $s$-th system. In the following, we compute the time evolution of the density matrix under the combined influence of the multi-level Hamiltonian and the external perturbation, which ultimately yields the time-dependent expectation value of the stress-tensor operators and hence the echo signal.

To this end, we perturbatively expand the expectation value of the stress tensor in powers of the externally applied strain field:
\begin{equation}\label{22}
\langle \hat{T}_{ij}^{(s)}\rangle (t) 
\overset{\mathrm{def}}{=} \sum_{n=0}^\infty \langle \hat{T}_{ij}^{(s)}\rangle_n (t),
\end{equation}
where the $n$th-order contribution is
\begin{equation}\label{40}
\langle \hat{T}_{ij}^{(s)}\rangle_n (t) = 
\operatorname{Tr}\!\left(\hat{\rho}^{(s)}_{I,n}(t)\,\hat{T}_{ij,I}^{(s)}(t)\right),
\end{equation}
with $\hat{\rho}^{(s)}_{I,n}(t)$ the $n$th-order expansion of the density matrix:
\begin{equation}\label{23}
\begin{aligned}
\hat{\rho}^{(s)}_{I,n}(t) &= 
 \left(-\mathrm{i}\right)^n  
 \int_{-\infty}^t dt_1 \int_{-\infty}^{t_1} dt_2 \cdots \int_{-\infty}^{t_{n-1}} dt_n \\
 &\quad\bigl[\hat{H}_I'^{(s)}(t_1),\bigl[\hat{H}_I'^{(s)}(t_2),\ldots,
 \bigl[\hat{H}_I'^{(s)}(t_n),  \\
 &\quad \sum_i p_i^{(s)} |i^{(s)}\rangle\langle i^{(s)}| \bigr]\cdots\bigr]\bigr],
\end{aligned}
\end{equation}
where $\hat{H}_I'^{(s)}(t)$ is the external time-dependent perturbation in the interaction picture. 

We now analyze each term in Eq.~\eqref{22} and its role in the phonon echo response. In particular, we will show that the third-order term $\langle \hat{T}_{ij}^{(s)}\rangle_3(t)$ is responsible for generating the phonon echo signal. This result agrees with the third-order nonlinear response predicted by the semi-classical theory of phonon echo.

Analogous to the isotropic mechanical and dielectric response of amorphous systems~\cite{Berthier2023ModernCS, PhysRevLett.127.215501, PhysRevLett.94.205501, PhysRevLett.126.118004, PhysRevLett.124.225502, PhysRevLett.123.105701}, and as noted in Eq.~\eqref{50}, the isotropy of glasses causes all even-order contributions to the stress-tensor expectation value to vanish. This behavior resembles the absence of second-harmonic generation in inversion-symmetric nonlinear structures~\cite{RevModPhys.48.1, PhysRevLett.7.118, PhysRevB.100.220501, PhysRevLett.124.013901, Tang2023FOP, Tempelman2021PRB, PhysRevLett.126.113901, PhysRevLett.111.263901}. In particular,
\begin{equation}\label{44}
\langle \hat{T}_{ij}^{(s)}\rangle_0 = \langle \hat{T}_{ij}^{(s)}\rangle_2 = 0.
\end{equation}
The detailed derivation and proof are given in Appendix~C.

We now compute the first-order expansion of the stress-tensor expectation value (see Appendix~C for details):
\begin{equation}\label{24}
\begin{aligned}
\langle \hat{T}^{(s)}_{ij}\rangle_1 
&= -\frac{\mathrm{i}}{\hbar}\sum_{kl}\sum_{ab} p_a^{(s)} 
\Bigl[\bigl(I^{(1)}_{kl} e^{\mathrm{i}\omega_{ab}^{(s)} t} 
+ I^{(2)}_{kl} e^{\mathrm{i}\omega_{ab}^{(s)}(t-\tau)}\bigr) \\
&\qquad\cdot 
\langle a^{(s)}|\hat{T}_{ij}^{(s)}|b^{(s)}\rangle 
\langle b^{(s)}|\hat{T}_{kl}^{(s)}|a^{(s)}\rangle - \mathrm{c.c.}\Bigr],
\end{aligned}
\end{equation}
where $\omega_{ab}^{(s)} = (E_a^{(s)} - E_b^{(s)})/\hbar$ is the energy gap between eigenstates $|a^{(s)}\rangle$ and $|b^{(s)}\rangle$, and $\hat{T}_{ij}^{(s)}$ is the stress-tensor operator in the Schr\"{o}dinger picture. This expression follows from the standard Kubo formula~\cite{Shalchi2011PRE}, which is equivalent to the resonant susceptibility. The result contains two oscillatory contributions: one centered at time $t$ (response to the pulse at $t=0$) and another centered at $t-\tau$ (response to the pulse at $t=\tau$). These contributions correspond to Eq.~\eqref{19} in the semi-classical limit, where all terms vanish at $t=2\tau$ and thus produce no echo. Likewise, the fully quantum expression in Eq.~\eqref{24} yields an expectation value that also vanishes at $t=2\tau$ after disorder and ensemble averaging.

The third-order expansion of the stress-tensor expectation value contributes directly to the phonon echo signal. As derived in Appendix~C, the third-order term reads
\begin{equation}\label{35}
\begin{aligned}
\langle \hat{T}_{ij}^{(s)}\rangle_3 &= 
\sum_{klmnpq} 
2\left(\frac{\mathrm{i}}{\hbar}\right)^3 
\sum_{ab} p_a^{(s)} I_{kl}^{(2)} I_{mn}^{(2)} I_{pq}^{(1)} \\
&\quad\cdot\bigg[ 
e^{\mathrm{i}\omega_{ab}^{(s)}(t-2\tau)} 
\langle a^{(s)}|\hat{T}_{pq}^{(s)}|b^{(s)}\rangle 
\langle b^{(s)}|\hat{T}_{mn}^{(s)}|a^{(s)}\rangle \\
&\qquad\cdot
\langle a^{(s)}|\hat{T}_{ij}^{(s)}|b^{(s)}\rangle 
\langle b^{(s)}|\hat{T}_{kl}^{(s)}|a^{(s)}\rangle 
- \mathrm{c.c.} \bigg],
\end{aligned}
\end{equation}
where ``c.c.'' denotes the complex conjugate. At $t=2\tau$, the phase factors $e^{\mathrm{i}\omega_{ab}^{(s)}(t-2\tau)}$ reduce to unity for all multi-level systems; consequently, the expectation values of the stress tensors add coherently, producing a macroscopic echo signal. The emergent echo response of the glassy system is therefore obtained by summing over all multi-level systems:
\begin{equation}\label{42}
\langle \hat{T}_{ij}\rangle_{\mathrm{e}} 
= \sum_{s=1}^{N_{\mathrm{MLS}}} \langle \hat{T}_{ij}^{(s)}\rangle_3.
\end{equation}
As shown in Fig.~\ref{fig5}, the non-equidistant energy spacing of the multi-level system naturally gives rise to the phonon echo signal.

This echo term, as the third-order perturbation of the stress-tensor expectation value, corresponds to a three-step quantum process coupling the lower eigenstate $|a^{(s)}\rangle$ and the upper eigenstate $|b^{(s)}\rangle$. The first pulse at $t=0$ excites a coherence between $|a^{(s)}\rangle$ and $|b^{(s)}\rangle$, creating the off-diagonal density matrix element $\rho_{ba}^{(1)} \propto I_1 e^{-\mathrm{i}\omega_{ab}^{(s)}t}$. During the delay $\tau$, the phase evolves, accumulating a factor $e^{-\mathrm{i}\omega_{ab}^{(s)}\tau}$. The second pulse at $t=\tau$ produces two key effects: (i) It converts the existing coherence into a population difference that retains the phase information accumulated before $\tau$. In the perturbative expansion, this step effectively reverses the sign of the accumulated phase, transforming $e^{-\mathrm{i}\omega_{ab}^{(s)}\tau}$ into $e^{+\mathrm{i}\omega_{ab}^{(s)}\tau}$ (phase conjugation). The amplitude of this contribution is proportional to $I_2^2$.  
(ii) It then regenerates a new coherence from this population difference. The subsequent free evolution from $\tau$ to $t$ introduces a phase factor $e^{-\mathrm{i}\omega_{ab}^{(s)}(t-\tau)}$. After these steps, the total phase factor becomes $e^{-\mathrm{i}\omega_{ab}^{(s)}(t-2\tau)}$. At $t=2\tau$, this phase equals unity for all molecules, irrespective of their individual detuning. As a result, the coherences re-phase and generate a macroscopic signal. The echo amplitude scales as $I_1 I_2^2$, confirming that it represents a third-order response with respect to the external driving fields.

Finally, we address the absence of the echo signal in a purely harmonic quantum oscillator, paralleling its absence in the purely harmonic semi-classical case. For a harmonic quantum oscillator, the level spacing is constant. While one might anticipate nonzero second- and third-order contributions to $\langle \hat{T}_{ij} \rangle$ that could potentially yield an echo, the harmonic oscillator is a \textit{linear system}. Its response is fully described by linear response theory, and higher-order nonlinear susceptibilities vanish due to the structure of the coordinate coupling. The vanishment of the higher order terms is justified as follows. For the second-order term, transitions among three quantum states $|n\rangle$, $|n+1\rangle$, and $|n+2\rangle$, with energy spacings $\omega_{n+1,n}=\omega$ and $\omega_{n+2,n}=2\omega$, could in principle produce a phase factor $e^{\mathrm{i}\omega(t-2\tau)}$ in the density matrix, suggesting an echo at $t=2\tau$. However, under the linear driving $\hat{H}'(t) = -F(t)\hat{x}$ with the ladder-operator representation $\hat{x} = \sqrt{{\hbar}/{2m\omega_0}}(\hat{a}^\dagger+\hat{a})$, only transitions $|n\rangle \to |n\pm1\rangle$ are allowed; the direct transition $|n\rangle \to |n\pm2\rangle$ is forbidden. A process $|n\rangle \to |n\pm2\rangle$ would require quadratic driving $\sim x^2 F$, which merely renormalizes the spring constant: ${p^2}/{2m} + {kx^2}/{2} \to {p^2}/{2m} + {(k+F)x^2}/{2}$. Hence, no genuine echo arises at second order. The third-order term involves a fourth-order correlation function with amplitude proportional to $F^3$. Because the displacement in a harmonic oscillator is linear in bosonic operators, Wick's theorem reduces quartic correlations to products of two-point functions, causing odd-order correlations to vanish. The commutator of displacement operators, $[\hat{x}(t),\hat{x}(t')] = ({\mathrm{i}\hbar}/{m\omega_0})\sin[\omega_0(t'-t)]$, is a c-number; consequently, higher nested commutators vanish, and the third-order stress-tensor contribution is identically zero. In summary, a purely harmonic oscillator cannot produce an echo signal.

\subsection{Quantum Echo and Dephasing Time Induced by Many-Body Interactions}

We now consider the phonon echo signal in a multi-level-system model where mutual many-body interactions between the multi-level systems are incorporated. The Hamiltonian of the entire system is governed by Eq.~(\ref{29}). An exact treatment of such an interacting many-body system is hindered by the exponential growth of the Hilbert space dimension, a manifestation of the curse of dimensionality~\cite{PhysRevLett.108.258701}, which renders both analytical solutions and full numerical simulations intractable. To make theoretical progress, we restrict our analysis to the weak-coupling regime, where the interaction strength is small compared to the typical energy-level spacings of the unperturbed multi-level-system Hamiltonian, i.e.,
\begin{eqnarray}
\left|\langle n'|\hat{V}|n\rangle\right| \ll \left|E_n^{(0)}-E_{n'}^{(0)}\right|,
\end{eqnarray}
for states $n$ and $n'$ connected by the matrix element of the interaction operator $\hat{V}$. Here, $E_n^{(0)}$ denotes the unperturbed energy of the entire system, and $|n\rangle$ denotes the corresponding unperturbed eigenstate. These unperturbed eigenstates can be well approximated by direct products of single-block eigenstates, representing the zeroth-order limit of the perturbing interaction:
\begin{eqnarray}\label{55}
|n\rangle = \bigotimes_{s=1}^{N_{\rm MLS}} |n^{(s)}\rangle,
\end{eqnarray}
where $|n^{(s)}\rangle$ denotes the $n^{(s)}$-th eigenstate of the local Hamiltonian $\hat{H}_{\rm MLS}^{(s)}$ at position $\bm{x}_s$. Thus, $|n\rangle$ represents the $n$-th eigenstate of the \textit{unperturbed} multi-level-system Hamiltonian of the entire system. This product-state basis provides a computationally tractable starting point for perturbative analyses, including the cumulant expansion employed below to evaluate dephasing effects.

Under the weak-coupling approximation, the eigenstates of the system are well approximated by the unperturbed product states given in Eq.~(\ref{55}). Consequently, at zeroth order in the many-body interaction $\hat{V}$, the phonon echo signal is simply obtained by substituting these product states into Eqs.~(\ref{35}) and (\ref{42}), yielding a coherent signal that does not account for interaction-induced dephasing. To capture the leading-order effects of the many-body interactions, one could in principle compute the first-order correction to the wave functions and evaluate their contribution to the stress-tensor expectation value via Eq.~(\ref{35}). However, such a first-order correction in $\hat{V}$ primarily modifies the amplitude of the echo signal while preserving its coherent nature.

More importantly, many-body interactions induce an additional effect beyond this amplitude correction: dephasing, which emerges from the accumulation of random relative phases among different quantum states due to interaction-induced energy fluctuations. This dephasing effect, which leads to the decay of the echo signal over time, is not captured at first order in $\hat{V}$. Instead, it requires a treatment that accounts for the second-order energy shifts or the statistical accumulation of phase differences, often formulated through a cumulant expansion of the time-evolution operator. Dephasing arises from the second-order cumulant of the phase fluctuations induced by many-body interactions, which scales as $V^2$ and leads to exponential decay of coherence.

To capture the dephasing effect induced by the many-body interactions, we analyze the temporal evolution of the density matrix $\hat{\rho}(t)$. Its matrix element governs the stress tensor and, consequently, the phonon echo signal. In the absence of external pulses, the dynamics is generated solely by the many-body interaction $\hat{V}$. For a perturbative treatment, we adopt the interaction picture with respect to the unperturbed Hamiltonian $\sum_s \hat{H}_{\text{MLS}}^{(s)}$. The density matrix in the interaction picture, $\hat{\rho}_I(t) = e^{\mathrm{i}\sum_s \hat{H}_{\rm MLS}^{(s)} t} \hat{\rho}(t) e^{-\mathrm{i} \sum_s \hat{H}_{\rm MLS}^{(s)} t}$, evolves according to the Liouville-von Neumann equation:
\begin{eqnarray}\label{41}
\mathrm{i}\partial_t \hat{\rho}_I(t) = [\hat{V}_I(t), \hat{\rho}_I(t)], \qquad t \neq 0,\tau,
\end{eqnarray}
where $\hat{V}_I(t) = e^{\mathrm{i}\sum_s\hat{H}_{\rm MLS}^{(s)} t}\hat{V} e^{-\mathrm{i}\sum_s\hat{H}_{\rm MLS}^{(s)} t}$ is the many-body interaction in the interaction picture. Equation~(\ref{41}) is exact. In the weak many-body interaction regime, we solve it perturbatively to obtain an analytic expression for the dephasing of the echo.

We assume that at $t=0$ (immediately before the application of the first pulse), the system is in a state of thermal equilibrium with respect to the unperturbed Hamiltonian $\sum_s \hat{H}_{\mathrm{MLS}}^{(s)}$. Given the weakness of the many-body interaction $\hat{V}$ compared to the local level spacings, the initial density matrix is well approximated by the factorized state:
\begin{equation}
\hat{\rho}_I(t=0) \approx  \bigotimes_{s=1}^{N_{\mathrm{MLS}}} \frac{e^{-\beta \hat{H}_{\mathrm{MLS}}^{(s)}}}{\mathcal{Z}^{(s)}} \equiv \bigotimes_{s=1}^{N_{\mathrm{MLS}}} \hat{\rho}_I^{(s)}(t=0),
\end{equation}
where $\beta = 1/(k_B T)$, $\mathcal{Z}^{(s)}$ is the corresponding partition function for multi-level-system at position $\bm{x}_s$, and $\hat{\rho}_I^{(s)}(t=0)$ is the local thermal density matrix of the multi-level system. This factorized initial condition not only reflects the physical assumption of local thermalization prior to the echo sequence but also provides a computationally tractable starting point for the subsequent perturbative analysis of the dephasing dynamics.

The phonon echo signal, given by the expectation value of the stress tensor of all multi-level-systems, namely $\sum_s\hat{T}_{ij}^{(s)}$, is subject to dephasing due to many-body interactions, with temperature effects encoded in the initial thermal density matrix. The decay of the echo amplitude is directly governed by the time evolution of the off-diagonal elements (coherences) of the density matrix in the interaction picture, $\rho_{I,ab}(t) = \langle a | \hat{\rho}_I(t) | b \rangle$, where $|a\rangle$ and $|b\rangle$ denote unperturbed eigenstates of the multi-level Hamiltonian. From the Liouville-von Neumann equation, the evolution of a specific matrix element is obtained as:
\begin{eqnarray}\label{36}
& {} & \frac{d\rho_{I,ab}(t)}{dt} = \nonumber \\
& {} & -\mathrm{i} \sum_c \Big[ 
\langle a | \hat{V}_I(t) | c \rangle  \rho_{I,cb}(t) 
-
\rho_{I,ac}(t)  \langle c | \hat{V}_I(t) | b \rangle 
\Big],\qquad 
\end{eqnarray}
where the sum runs over the complete set of unperturbed eigenstates $\{|c\rangle\}$. This equation describes how the many-body interaction $\hat{V}_I(t)$ couples different coherence elements, thereby inducing dephasing and eventual decay of the echo signal.

We now introduce a series of approximations that extract the dephasing dynamics from Eq.~(\ref{36}). Specifically: (i) We focus on a specific pair of unperturbed eigenstates $|a\rangle$ and $|b\rangle$ and assume that, to leading order, the evolution of the coherence $\rho_{I,ab}(t)$ is not significantly coupled to other coherences $\rho_{I,cd}(t)$ with $(c,d) \neq (a,b)$. In other words, terms in Eq.~(\ref{36}) that involve matrix elements $\rho_{I,ac}$ or $\rho_{I,cb}$ for $c \neq a,b$ are neglected. (ii) The diagonal elements (populations) of the density matrix evolve on a time scale much longer than the echo time $2\tau$ of interest. Hence, over the relevant time window, the populations $p_i(t) = \rho_{I,ii}(t)$ can be treated as approximately constant and equal to their initial thermal values. (iii) The leading dephasing mechanism arises from the fluctuations of the diagonal matrix elements of the interaction, $\langle a|\hat{V}_I(t)|a\rangle$ and $\langle b|\hat{V}_I(t)|b\rangle$. These fluctuations modulate the instantaneous transition frequency between states $|a\rangle$ and $|b\rangle$, thereby causing phase dispersion. Off-diagonal elements of $\hat{V}$ contribute only at second order in perturbation theory, e.g., via virtual transitions that induce energy shifts $\delta E_a = \sum_{c\neq a} |\langle a|\hat{V}|c\rangle|^2/(E_a^{(0)}-E_c^{(0)}) \sim \mathcal{O}(V^2)$. While these shifts renormalize the mean frequency, they do not lead to dephasing unless they also fluctuate. (iv) We adopt a secular approximation~\cite{Tempelman2021PRB} (or rotating-wave approximation~\cite{PhysRevLett.98.013601}) under which rapidly oscillating terms associated with off-diagonal couplings to states $|c\rangle$ ($c \neq a,b$) average to zero on the time scale of the coherence evolution. This is justified when the typical energy differences $|E_a^{(0)} - E_c^{(0)}|$ are large compared to the inverse dephasing time.

Incorporating the four approximations outlined above, we obtain a simplified equation of motion for the coherence $\rho_{I,ab}(t)$ (see Appendix~E for a detailed derivation):
\begin{eqnarray}\label{37}
\rho_{I,ab}(t) = \rho_{I,ab}(0) \exp\left[\mathrm{i}\int_0^t \delta\omega_{ab}(t')\,dt'\right],
\end{eqnarray}
where the instantaneous frequency difference is given by
\begin{eqnarray}
\delta\omega_{ab}(t) = \langle b|\hat{V}_I(t)|b\rangle - \langle a|\hat{V}_I(t)|a\rangle.
\end{eqnarray}
A technical subtlety arises: if $|a\rangle$ and $|b\rangle$ are eigenstates of the unperturbed Hamiltonian $\sum_s \hat{H}_{\rm MLS}^{(s)}$, the matrix elements $\langle a|\hat{V}_I(t)|a\rangle$ and $\langle b|\hat{V}_I(t)|b\rangle$ would indeed be time-independent. However, in the physical situation we consider, the system is subject to thermal fluctuations and mutual couplings among the multi-level systems. As a result, the environment of a given coherence $(a,b)$ evolves stochastically, rendering $\delta\omega_{ab}(t)$ a random variable. We therefore model $\delta\omega_{ab}(t)$ as a stationary Gaussian stochastic process, a description justified by the central limit theorem when the frequency shift results from a large number of independent fluctuating sources. Its statistical properties are characterized by the correlation function
\begin{eqnarray}\label{38}
\langle \delta\omega_{ab}(t) \delta\omega_{ab}(t') \rangle = 2\Gamma^2\, e^{-|t-t'|/\tau_c},
\end{eqnarray}
where $\Gamma$ sets the typical fluctuation amplitude and $\tau_c$ is the correlation time of the environment. Performing an ensemble average over the stochastic realizations and employing a second-order cumulant expansion (valid in the weak-coupling regime where $\hat{V}$ can be treated perturbatively), we obtain the echo signal of the stress tensor:
\begin{eqnarray}\label{47}
\langle \hat{T}_{ij}\rangle_{\rm e}(t) =
e^{-\frac{2\Gamma^2}{\tau_c}t}\,
\sum_s\langle \hat{T}_{ij}^{(s)}\rangle_{\rm e}(t).
\end{eqnarray}
Here $\langle \hat{T}_{ij}^{(s)}\rangle_{\rm e}(t)$ is the echo contribution from the $s$-th multi-level system in the absence of mutual interactions, shown in Eqs.~(\ref{35}) and (\ref{42}). The exponential factor captures the dephasing induced by the many-body interactions; in the non-interacting limit ($\Gamma \to 0$), this factor reduces to unity and no decay occurs.

\begin{figure}[htbp]
\includegraphics[scale=0.54]{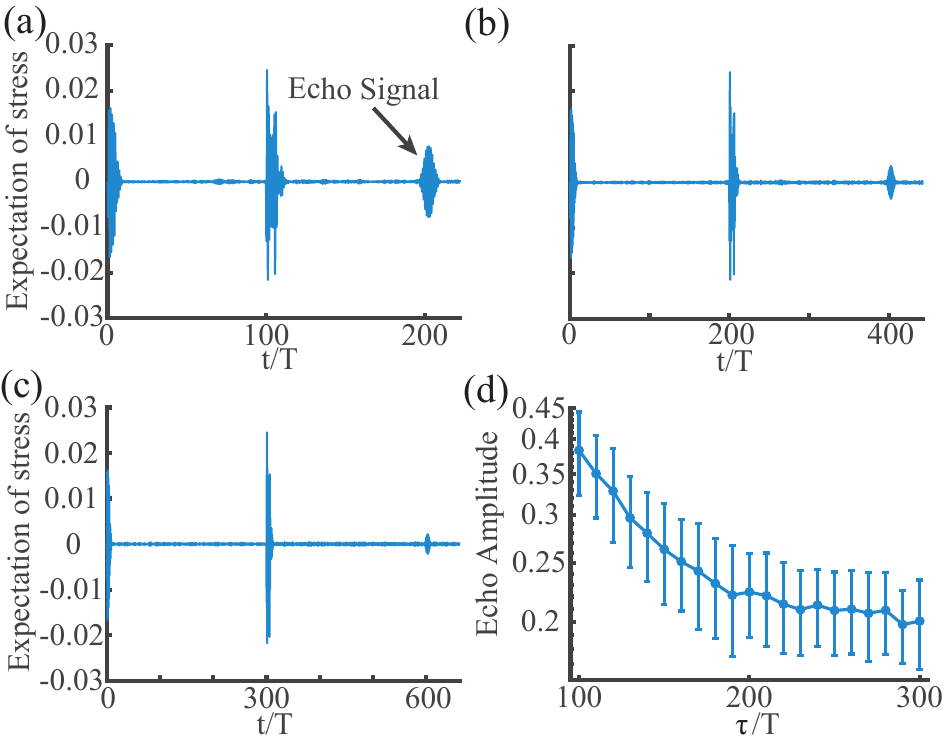}
\caption{Phonon echo induced by multi-level systems with many-body interactions. All parameters of the individual multi-level-system model are adopted from Fig.~\ref{fig5}. For simplicity, the many-body interaction is restricted to nearest neighbors, which does not alter 
the qualitative behavior of the echo signal compared to the full interaction. The strength of the orientation-averaged many-body interaction is $\Lambda=0.1$. The orientation-averaged stress tensor for the many-body interaction yields a $5\times 5$ Hermitian matrix, $\hat{T}_{{\rm int}}^{(s)} + \delta \hat{T}_{{\rm int}}^{(s)}$, where the zeroth-order term is $\hat{T}_{{\rm int},nn'}^{(s)} = \tfrac{0.1}{2}  n(n-1) \delta_{nn'}$. The perturbation, $\delta \hat{T}_{{\rm int}}^{(s)}$, consists of Hermitian random $5\times 5$ matrices with elements uniformly distributed in $[-0.5, 0.5]$. The phonon echo is examined for time intervals between the two external pulses: (a) $\tau=100T$, (b) $\tau=200T$, and (c) $\tau=300T$. 
Panel (d) shows the dephasing of echo amplitude versus the time interval between pulses, where dots and error bars represent the mean value and standard deviation of the echo amplitude across $N=100$ randomly generated samples of the interacting multi-level systems.}\label{fig6}
\end{figure}


To further verify the analytic results for the dephasing mechanism, we conduct a self-consistent mean-field simulation that directly incorporates the nonlinear couplings among the multi-level systems. The self-consistent feedback between sites generates an effective dephasing environment for each individual multi-level system. As shown in Fig.~\ref{fig6}, the mean-field simulation also yields a decay behavior of the echo amplitude. This consistency indicates that the essential dephasing physics is captured by the stochastic dynamics emerging from the many-body nonlinear couplings. Although the error bars overlap between neighboring data points, the mean values show a consistent decreasing trend with increasing $\tau$.



\section{Discussions}

In this work, we present two complementary approaches, namely the semi-classical and quantum dynamics, for studying phonon echo phenomena, extending the conventional two-level-system framework to a multi-level-system description. By considering virtual phonon exchange processes, we derive the full glass Hamiltonian that incorporates many-body interactions between stress tensors, reminiscent of dipole-dipole interactions in magnetic materials. These interactions exhibit a characteristic long-range $1/r^3$ dependence.

When two external signals are applied, phonon echoes emerge from the interplay of two key effects: the nonlinear structure of the energy eigenvalues and the spatial randomness of the oscillators. The nonlinear eigenvalue structure induces frequency shifts associated with changes in oscillation amplitude, while randomness drives decoherence, yielding a coherent macroscopic state only at the echo time. Many-body interactions further enhance the echo amplitude, even when individual oscillators remain purely linear, thereby amplifying the collective signal. Moreover, the combined influence of finite temperature and many-body coupling introduces dephasing, leading to a reduction in echo amplitude as the time interval between input pulses increases.

Future investigations may explore the role of multi-level systems and many-body interactions in phonon saturation effects~\cite{PELLE2000131JAC}, as well as the coupling between phonon strain fields and topological defects in glasses~\cite{Wu2025NC, paulose2015topological}. Extending these concepts from fully disordered structures to less disordered systems, such as disordered solids and quasicrystals~\cite{Zhou2019PRX, PhysRevB.34.596, PhysRevB.34.617, PhysRevLett.53.2477, PhysRevB.101.115413}, represents a promising direction for future research. Constructing artificial mechanical and elastic systems with embedded active elements~\cite{PhysRevLett.125.118001, Sone2019PRL, Zhou2020PRR, PhysRevLett.126.198001, Ma2025PRB, souslov2017topological} may open new avenues for exploring the interaction between non-Hermiticity~\cite{PhysRevLett.121.026808, PhysRevLett.121.086803} and echo phenomena, which could prove important in unraveling activity-induced memory effects in man-made structures.



\section{Acknowledgment}

Di Zhou would like to acknowledge insightful discussion to Anthony J. Leggett, Shinsei Ryu, Rudro Rana Biswas, and Xiao Chen. This work is supported by the National Science Foundation of China Grant No. 12374157.

\appendix

\section{Deriving many-body interactions between multi-level systems}

We define the displacement field of an atom at position $\bm{x}$ as $\bm{u}(\bm{x})$, 
where $u_i(\bm{x})$ denotes its $i$-th Cartesian component, with $i=1,2,3$ corresponding to the $x$, $y$, and $z$ directions. 
Consider a unit volume $V=L^3$ (with side length $L$) in the amorphous solid, containing atoms that oscillate around their force-balanced equilibrium positions. 
The displacement field is Fourier transformed as
\begin{equation}
u_{i}(\bm{x}) \overset{\rm def}{=} \frac{1}{\sqrt{V}} 
\sum_{\bm{q}} u_{i}(\bm{q}) \, e^{-\mathrm{i}\bm{q}\cdot\bm{x}},
\end{equation}
with the inverse Fourier transformation 
\begin{equation}
u_{i}(\bm{q}) = \frac{1}{\sqrt{V}} \int_V d^3x\,  u_{i}(\bm{x}) \, e^{\mathrm{i}\bm{q}\cdot\bm{x}},
\end{equation}
where $\bm{q}$ is the wavevector in reciprocal space. 
The corresponding elastic strain field, within the linear elastic regime, is given by
\begin{equation}
\epsilon_{ij}(\bm{x}) = \frac{-\mathrm{i}}{2\sqrt{V}}\sum_{\bm{q}} e^{-\mathrm{i}\bm{q}\cdot\bm{x}} 
\left[q_j u_{i}(\bm{q}) + q_i u_j(\bm{q}) \right].
\end{equation}

It is convenient to decompose the reciprocal-space elastic strain fields into longitudinal and transverse components. The displacement field can be expanded in the phonon polarization basis as
\begin{equation}
\bm{u}(\bm{q}) = \sum_{\mu = l, t_1, t_2} u_\mu(\bm{q}) \, \bm{\mathrm{e}}_\mu(\bm{q}),
\end{equation}
where $u_\mu(\bm{q})$ denotes the amplitude of the $\mu$-th polarization mode. Here, $\mu = l$ corresponds to the longitudinal polarization, while $\mu = t_1, t_2$ represent the two transverse polarizations, and $\bm{\mathrm{e}}_\mu(\bm{q})$ are the corresponding unit polarization vectors~\cite{Joffrin1975JDP}. The longitudinal polarization vector is given by
\begin{equation}
\bm{\mathrm{e}}_{l}(\bm{q}) = \frac{\bm{q}}{|\bm{q}|},
\qquad
\mathrm{e}_{l,i}(\bm{q}) = \frac{q_i}{|\bm{q}|}, \quad i = 1,2,3,
\end{equation}
with $i$ indexing the Cartesian coordinates $(x,y,z)$. The transverse polarization vectors satisfy the orthogonality conditions
\begin{equation}\label{A4}
\bm{\mathrm{e}}_{t_1}(\bm{q}) \cdot \bm{q} =
\bm{\mathrm{e}}_{t_2}(\bm{q}) \cdot \bm{q} =
\bm{\mathrm{e}}_{t_1}(\bm{q}) \cdot \bm{\mathrm{e}}_{t_2}(\bm{q}) = 0,
\end{equation}
and obey the completeness relation
\begin{equation}
\sum_{\mu = l,t_1,t_2} \mathrm{e}_{\mu,i}(\bm{q}) \, \mathrm{e}_{\mu,j}(\bm{q})
= \delta_{ij}.
\end{equation}
Here $\mathrm{e}_{\mu,i}(\bm{q})$ denotes the $i$-th Cartesian component of the polarization vector for mode $\mu$. These vectors yield the condition 
\begin{equation}
|\bm{\mathrm{e}}_\mu(\bm{q})|\equiv 1, \qquad \mu = l, t_1, t_2.
\end{equation}
Following the above definitions, it is still evident that the choice of the transverse polarizations is arbitrary. Thus, we choose 
the transverse polarization vectors as, for example,
\begin{equation}
\bm{\mathrm{e}}_{t_1}(\bm{q}) = \frac{\bm{z} \times \bm{q}}{|\bm{z} \times \bm{q}|},
\quad
\bm{\mathrm{e}}_{t_2}(\bm{q}) = \frac{\bm{q} \times (\bm{z} \times \bm{q})}{|\bm{q} \times (\bm{z} \times \bm{q})|},
\end{equation}
for $\bm{q}$ not parallel to the $z$-axis. By decomposing the displacement field into longitudinal and transverse components, namely 
$\bm{u}_l(\bm{q}) = u_l(\bm{q})\textbf{e}_l(\bm{q})$ and 
$\bm{u}_{t_1,t_2}(\bm{q}) = u_{t_1,t_2}(\bm{q})\,\textbf{e}_{t_1,t_2}(\bm{q})$, 
we find that: $\nabla \times \bm{u}_l(\bm{x}) = 0$, $\nabla \cdot \bm{u}_l(\bm{x}) \neq 0$, $\nabla \times \bm{u}_{t_1,t_2}(\bm{x}) \neq 0$, and $\nabla \cdot \bm{u}_{t_1,t_2}(\bm{x}) = 0$. This confirms the expected physical behavior: longitudinal waves are compressional 
($\nabla\cdot\bm{u}_l\neq 0$) but irrotational ($\nabla\times\bm{u}_l=0$), 
while transverse waves are shear waves that are divergence-free ($\nabla\cdot\bm{u}_{t_1,t_2}=0$) 
but have non-zero curl ($\nabla\times\bm{u}_{t_1,t_2}\neq 0$).

We now derive the elastic Hamiltonian of the amorphous solid based on continuum elasticity. 
Starting from the standard form in real space (Eq.~(\ref{43})) and applying the Fourier 
transformation, we obtain
\begin{equation}
\hat{H}_{\rm el} = \sum_{\bm{q}}\sum_{i=1}^3 \frac{1}{2}\rho |\dot{u}_i(\bm{q})|^2 
+ \sum_{\bm{q}}\sum_{ijkl} \frac{1}{2}C_{ijkl}\epsilon_{ij}(\bm{q})\epsilon_{kl}(-\bm{q}),
\end{equation}
Decomposing the displacement field into phonon polarization modes, we find
\begin{equation}\label{A12}
\hat{H}_{\rm el} = \sum_{\bm{q},\mu} \left[\frac{|\pi_\mu(\bm{q})|^2}{2\rho}
+ \frac{1}{2}\rho \omega_{\bm{q}\mu}^2 |u_\mu(\bm{q})|^2\right],
\end{equation}
where $\pi_\mu(\bm{q}) = \rho \dot{u}_\mu(\bm{q})$ is the canonical momentum, and 
the phonon frequency $\omega_{\bm{q}\mu}$ in the long-wavelength limit is determined by
\begin{equation}
\rho \omega_{\bm{q}\mu}^2 = \sum_{ijkl} C_{ijkl} {\rm e}_{\mu,i}(\bm{q})  {\rm e}_{\mu,k}(\bm{q}) q_j q_l.
\end{equation}

We now proceed from Eq.~(\ref{A12}) to derive the mutual many-body interaction between multi-level systems. 
In the multi-level-system model, the phonon strain field couples linearly to the multi-level systems, enabling phonon scattering processes in which the two-level systems can absorb or emit phonon energy. 
As discussed in the main text, we denote the $s$th multi-level system located at position $\bm{x}_s$. 
The minimal coupling term is linear in the strain fields and represents the most relevant contribution within the framework of linear elasticity, 
\begin{equation}
\hat{H}_{\rm coup}^{(s)} = 
\sum_{ij} \epsilon_{ij}^{(s)}  \hat{T}_{ij}^{(s)},
\end{equation}
where $\epsilon_{ij}^{(s)}$ denotes the strain field coupled to the $s$th multi-level system, and $\hat{T}_{ij}^{(s)}$ is the stress tensor operator describing the minimal coupling matrix between the $s$th multi-level system and the phonon strain field. 
This operator serves as the coupling coefficient between the multi-level systems and the phonon strain field. 

Following the analogy with the derivation of the magnetic dipole-dipole interaction, we now derive the many-body interaction arising from correlations of stress tensors. 
To this end, we combine a total of $N_{\rm MLS}=N^3$ multi-level systems into a super block of amorphous solid with side length $NL$, and derive the many-body interaction between stress tensor operators. 
Using the reality condition of the displacement field $\bm{u}(\bm{x})$, the coupling Hamiltonian can be written as
\begin{eqnarray}\label{A5}
\hat{H}_{\rm coup} 
& = & \sum_{s=1}^{N^3}\sum_{ij}\epsilon_{ij}^{(s)}\hat{T}_{ij}^{(s)}\nonumber \\
& = & \frac{1}{4\sqrt{V}}\sum_s \sum_{ij} \hat{T}_{ij}^{(s)}
\sum_{\bm{q},\mu}\left[\mathrm{i}u_{\mu}(\bm{q})e^{\mathrm{i}\bm{q}\cdot \bm{x}_s}+{\rm c.c.}\right]\nonumber \\
& {} & 
\left[q_j\mathrm{e}_{\mu, i}(\bm{q})+q_i\mathrm{e}_{\mu, j}(\bm{q})\right],
\end{eqnarray}
where $\hat{H}_{\rm coup}$ denotes the stress-strain coupling (multi-level system to phonon strain field coupling) for the super block glass Hamiltonian of side length $NL$, containing a total of $N^3$ multi-level systems. 
Here ``c.c.'' denotes the complex conjugate of the preceding term. 
Since the minimal coupling is linear in the displacement fields, it can be absorbed into the elastic Hamiltonian, which is quadratic in the displacement fields, by completing the square. 
As a result, the sum of the elastic Hamiltonian and the stress-strain coupling takes the alternative form
\begin{equation}\label{A6}
\hat{H}_{\rm el}+\hat{H}_{\rm coup}
=\hat{H}_{\rm el}(\hat{T}_{ij})+\hat{V},
\end{equation}
where the ``stress-tensor shifted phonon Hamiltonian'' $\hat{H}_{\rm el}(\hat{T}_{ij})$ is obtained by absorbing the stress-strain coupling $\hat{H}_{\rm coup}$ (linear in the strain field) and completing the square:
\begin{equation}\label{A13}
\hat{H}_{\rm el}(\hat{T}_{ij})
=\sum_{\bm{q},\mu}\left(
\frac{|\pi_{\mu}(\bm{q})|^2}{2\rho}
+\frac{\rho\omega^2_{\bm{q}\mu}}{2}\big|u_{\mu}(\bm{q})-u_{\mu}^{(0)}(\bm{q})\big|^2
\right).
\end{equation}
In Eq.~(\ref{A13}), the ``equilibrium position'' $u_{\mu}^{(0)}(\bm{q})$ is given by
\begin{equation}\label{A7}
u_{\mu}^{(0)}(\bm{q})
=\sum_s \sum_{ij}\frac{\mathrm{i} e^{-\mathrm{i}\bm{q}\cdot \bm{x}_s}}{2\sqrt{V}\rho\omega_{\bm{q}\mu}^2}
\left[q_j \mathrm{e}_{\mu, i}(\bm{q})+q_i \mathrm{e}_{\mu, j}(\bm{q})\right]
\hat{T}_{ij}^{(s)} .
\end{equation}

Based on the expression of the equilibrium position, we now compute the many-body interaction $\hat{V}$, which can be expressed as
\begin{eqnarray}\label{A11}
\hat{V}
& = & -\sum_{\bm{q},\mu}
\frac{\rho\omega^2_{\bm{q}\mu}}{2}\big|u_{\mu}^{(0)}(\bm{q})\big|^2
\nonumber \\
& = &  -\sum_{\bm{q},\mu}\frac{\big[q_j\mathrm{e}_{\mu, i}(\bm{q})+q_i\mathrm{e}_{\mu, j}(\bm{q})\big]
\big[q_k\mathrm{e}_{\mu, l}(\bm{q})+q_l\mathrm{e}_{\mu, k}(\bm{q})\big]}{8\rho V\omega^2_{\bm{q}\mu}}\nonumber \\
& {} & 
\sum_{ss'}\sum_{ijkl} \hat{T}_{ij}^{(s)} \hat{T}_{kl}^{(s')}
\cos\!\left[\bm{q}\cdot (\bm{x}_s-\bm{x}_{s'})\right].
\end{eqnarray}
We obtain the many-body interaction as expressed in Eq.~(\ref{13}). 
The coefficient $\Lambda_{ijkl}$ is obtained from Eq.~(\ref{A11}) by integrating over the wavevector $\bm{q}$, leading to the coefficient presented in Eq.~(\ref{14}). 

Finally, it is notable that the coefficient of the many-body interaction between multi-level systems satisfies the symmetry property
\begin{equation}\label{B70}
\Lambda_{ijkl}^{(ss')} 
= \Lambda_{klij}^{(ss')}
= \Lambda_{ijkl}^{(s's)}.
\end{equation}

\section{Method of multiple-scale derivation of frequency shift in nonlinear oscillators}

In this section, we introduce the standard multiple-scale method in nonlinear mechanics, which yields the semi-classical phonon echo results presented in Eqs.~(\ref{7}) and (\ref{17}). 
The starting point is the nonlinear equation of motion for oscillator $s$,
\begin{equation}
m\ddot{q}_s = -k_{2,s}q_s - k_3 q_s^2 - k_4 q_s^3 
 - \sum_{s'\neq s} \sum_{n,n'} n V_{nn'}^{(ss')} q_s^{\,n-1} q_{s'}^{\,n'} .
\end{equation}

To apply the multiple-scale method, we introduce a book-keeping parameter $\epsilon \ll 1$ in front of the nonlinear terms, i.e., 
$k_3 \to \epsilon k_3$, $k_4 \to \epsilon k_4$, and $V_{nn'}^{(ss')} \to \epsilon V_{nn'}^{(ss')}$. 
The time variable is then expanded into multiple scales:
\begin{equation}
T_0 = t, \quad T_1 = \epsilon t, \quad \ldots, \quad T_n = \epsilon^{\,n-1} t, \quad \ldots
\end{equation}
which yields the derivative expansion
\begin{equation}
\frac{d}{dt} = D_0 + \epsilon D_1 + \cdots, \quad
\text{with } D_j \equiv \frac{\partial}{\partial T_j}.
\end{equation}
The displacement is expanded accordingly as
\begin{equation}
q_s(t) = q_{s,0}(T_0,T_1,\ldots) + \epsilon q_{s,1}(T_0,T_1,\ldots) + \cdots .
\end{equation}

Matching orders in $\epsilon$, the zeroth-order equation of motion is
\begin{equation}
m D_0^2 q_{s,0} + k_{2,s} q_{s,0} = 0,
\end{equation}
which admits the harmonic oscillator solution
\begin{equation}
q_{s,0} = A_s(T_1) e^{\mathrm{i}\omega^{(s)} T_0 + \mathrm{i}\theta_s(T_1)} + \text{c.c.},
\end{equation}
where $A_s(T_1)$ is the slowly varying amplitude, $\omega^{(s)} = (k_{2,s}/m)^{1/2}$ the natural frequency, and $\theta_s(T_1)$ the slowly varying phase. At $\mathcal{O}(\epsilon)$, the equation of motion becomes
\begin{align}
m D_0^2 q_{s,1} + k_{2,s} q_{s,1} 
&= -2m D_0 D_1 q_{s,0} - k_3 q_{s,0}^2 - k_4 q_{s,0}^3 \nonumber \\
&\quad - \sum_{s'\neq s} \sum_{n,n'} n V_{nn'}^{(ss')} q_{s,0}^{\,n-1} q_{s',0}^{\,n'} .
\end{align}
This equation describes the driving force generated by the zeroth-order motion and its feedback to the first-order displacement.

Among the nonlinear terms, only the $(n,n')=(2,2)$ contribution survives in the secular (resonant) part. 
Eliminating secular growth requires the phase $\theta_s(T_1)$ to satisfy
\begin{equation}
\delta\omega^{(s)} \equiv D_1 \theta_s = 
\frac{3k_4}{8m\omega^{(s)}} A_s^2 
+ \frac{1}{2m\omega^{(s)}} \sum_{s'\neq s} V_{22}^{(ss')} A_{s'}^2 .
\end{equation}
Consequently, the effective oscillation frequency becomes
\begin{equation}
\omega^{(s)}_{\text{eff}}(\{A_s\}) = \omega^{(s)} + \delta\omega^{(s)},
\end{equation}
which corresponds to the frequency shift in Eqs.~(\ref{7}) and (\ref{17}), arising from both on-site and interaction-induced nonlinearities.  

Finally, we note that the quadratic term $-k_3 q_{s,0}^2$ induces a static displacement 
$\delta q_s = -k_3 A_s^2 / (2k_{2,s})$, which shifts the equilibrium position. 
This static displacement does not affect the frequency shift and is irrelevant for the phonon echo signal. 
Redefining $q_s \to q_s - \delta q_s$, the oscillator motion remains approximately sinusoidal in time, with the weak amplitude-dependent frequency shift predicted by the multiple-scale analysis.

\section{Third-order perturbative expansion for quantum echo in a multi-level system without many-body interaction}

To derive the phonon echo in the quantum regime, we employ the density matrix formalism, which naturally incorporates thermal effects. At $t=0$, the system is assumed to be thermalized, with  $\hat{\rho}(t=0) = \sum_i p_i^{(s)} |i^{(s)}\rangle \langle i^{(s)}| $, where $p_i^{(s)} = e^{-\beta E_i^{(s)}}/\mathcal{Z}^{(s)}$ is the Boltzmann weight at temperature $T$, $\beta = 1/k_B T$ is the inverse temperature, and $E_i^{(s)}$ denotes the eigenenergy of the $s$th multi-level system. In the interaction picture, the density operator is denoted by $\hat{\rho}^{(s)}_I(t)$, and its temporal evolution is given by  
\begin{eqnarray}\label{1.2}
\hat{\rho}^{(s)}_I(t) = \hat{U}^{(s)}(t,0)\, \hat{\rho}^{(s)}_I(0)\, \hat{U}^{(s)\dag}(t,0),
\end{eqnarray}
where the time-evolution operator reads  
\begin{eqnarray}
\hat{U}^{(s)}(t,0) = \mathcal{T} \exp\left[-\frac{\mathrm{i}}{\hbar}\int_0^t \hat{H}_I'^{(s)}(t')\, dt'\right],
\end{eqnarray}
with $\hat{H}_I'^{(s)} = \sum_{ij} \epsilon_{ij}^{(s)}(t)\, \hat{T}_{ij,I}^{(s)}(t)$, 
the externally applied time-dependent perturbation Hamiltonian in the interaction picture. Here, $\hat{T}_{ij,I}^{(s)}(t)$ is the stress tensor operator of the multi-level system located at position $\bm{x}_s$, and $\mathcal{T}$ denotes the time-ordering operator.

Thus, by expanding the time-evolution operator in powers of the external perturbation, we obtain a perturbative series for the expectation value of the stress tensor:
\begin{equation}
\langle \hat{T}_{ij}^{(s)}\rangle (t) 
= \operatorname{Tr}\left[\hat{\rho}_I^{(s)}(t)\hat{T}_{ij,I}^{(s)}(t)\right]
= \sum_{n=0}^\infty \operatorname{Tr}\left[\hat{\rho}_{I,n}^{(s)}(t)\hat{T}_{ij,I}^{(s)}(t)\right],
\end{equation}
where $\hat{\rho}_{I,n}^{(s)}(t)$ denotes the $n$th-order contribution to the density matrix in the interaction picture.

The zeroth-order contribution vanishes for centrosymmetric systems in thermal equilibrium:
\begin{eqnarray}
\langle \hat{T}_{ij}^{(s)}\rangle_0 
& = & \operatorname{Tr}\left[\hat{\rho}_{I,0}^{(s)}(t)\hat{T}_{ij,I}^{(s)}(t)\right] \nonumber \\
 & = & \operatorname{Tr}\left[\hat{\rho}_{0}^{(s)} e^{i\hat{H}_0^{(s)} t/\hbar}\hat{T}^{(s)}_{ij}e^{-i\hat{H}^{(s)}_0 t/\hbar}\right] = 0,
\end{eqnarray}
which directly leads to Eq.~(\ref{44}) in the main text.

The first-order expansion of the density matrix corresponds to the Kubo formula, i.e. the resonance susceptibility that typically arises in the low-temperature, high-frequency regime. The first-order expansion of density matrix is obtained from Eq.~(\ref{23}) by asking $n=1$, from which we obtain the corresponding order of the stress tensor expectation value is
\begin{eqnarray}
& {} & \langle \hat{T}_{ij}^{(s)}\rangle_1 
=\nonumber \\
& {} & 
-\frac{\mathrm{i}}{\hbar}\sum_{kl}\int_{-\infty}^t  dt_1 \, \epsilon^{(s)}_{kl}(t_1) 
\operatorname{Tr}\!\left(\hat{\rho}_{I,0}^{(s)} \left[\hat{T}_{ij,I}^{(s)}(t), \hat{T}_{kl,I}^{(s)}(t_1)\right]\right)\nonumber \\
\end{eqnarray}
Transforming the operators back to the Schr\"{o}dinger picture, two choices arise: (i) $t_1=0$, corresponding to the first triggering signal, and (ii) $t_1=\tau$, corresponding to the second excitation. These two contributions combine to yield Eq.~(\ref{24}) in the main text.


The second-order expansion of the density matrix is derived from Eq.~(\ref{23}) by asking $n=2$, from which the second-order contribution to the stress tensor is
\begin{eqnarray}\label{33}
& {} & \langle \hat{T}^{(s)}_{ij}\rangle_2 = \nonumber \\
& {} & 
\frac{-1}{2\hbar^2} \int_{-\infty}^t dt_1 \int_{-\infty}^t dt_2 \sum_{abc} \langle a^{(s)}|\hat{T}^{(s)}_{ij}(t)|b^{(s)}\rangle \nonumber \\
& {} & 
\bigg[(p_a^{(s)}-p_c^{(s)})\langle b^{(s)}|\hat{H}_I'^{(s)}(t_1)|c^{(s)}\rangle \langle c^{(s)}|\hat{H}_I'^{(s)}(t_2)|a^{(s)}\rangle \nonumber \\
& {} & 
 - (p_c^{(s)}-p_b^{(s)})\langle b^{(s)}|\hat{H}_I'^{(s)}(t_1)|c^{(s)}\rangle \langle c^{(s)}|\hat{H}_I'^{(s)}(t_2)|a^{(s)}\rangle\bigg]\nonumber \\
\end{eqnarray}
In Eq.~(\ref{33}), we impose the $\delta$-function constraints of the triggering signals. Four choices arise: (i) $t_1=t_2=0$, (ii) $t_1=0$, $t_2=\tau$, (iii) $t_1=\tau$, $t_2=0$, and (iv) $t_1=t_2=\tau$. These yield
\begin{eqnarray}\label{34}
& {} & \langle \hat{T}^{(s)}_{ij}\rangle_2 = \nonumber \\
& {} & 
\frac{-1}{2\hbar^2}  \sum_{abc} 
\sum_{klpq} (p_a^{(s)}-p_c^{(s)}) \nonumber \\
& {} & 
\bigg[\bigg(e^{\mathrm{i}\omega_{ab}^{(s)} t} I_{kl}^{(1)} I_{pq}^{(1)} + e^{\mathrm{i}\omega_{ca}^{(s)} t+\mathrm{i}\omega_{ab}^{(s)} \tau} I_{kl}^{(1)} I_{pq}^{(2)} \nonumber \\
& {} & +e^{\mathrm{i}\omega_{ab}^{(s)} t+\mathrm{i}\omega_{bc}^{(s)}\tau} I_{kl}^{(2)}I_{pq}^{(1)}+e^{\mathrm{i}\omega_{ab}^{(s)}(t-\tau)} I_{kl}^{(2)} I_{pq}^{(2)} \bigg) \nonumber \\
& {} & \langle a^{(s)}|\hat{T}^{(s)}_{ij}|b^{(s)}\rangle \langle b^{(s)}|\hat{T}_{kl}^{(s)}|c^{(s)}\rangle \langle c^{(s)}|\hat{T}_{pq}^{(s)}|a^{(s)}\rangle
+ {\rm c.c.}\bigg].\qquad 
\end{eqnarray}
Here $\omega_{ab} = (E_a^{(s)}-E_b^{(s)})/\hbar$ denotes the energy gap between eigenstates $|a^{(s)}\rangle$ and $|b^{(s)}\rangle$. The four terms in Eq.~(\ref{34}) behave differently: the first oscillates sinusoidally in $t$ and cannot produce an echo at $t=2\tau$; the second and third contain apparent factors of $\omega_{ab}^{(s)}(t-2\tau)$ under special coincidences such as $\omega_{ca}^{(s)}=-2\omega_{ab}^{(s)}$ or $\omega_{bc}^{(s)}=-2\omega_{ab}^{(s)}$, but these conditions are not generically satisfied in a random ensemble of multi-level systems and are quickly suppressed by dephasing; the fourth term carries a phase factor $\omega_{ab}^{(s)}(t-\tau)$, which also fails to yield an echo at $t=2\tau$. We therefore conclude that the second-order contribution $\langle \hat{T}_{ij}^{(s)}\rangle_2$ cannot generate an echo signal. When summed over all multi-level systems, the randomness of the eigenfrequencies leads to phase cancellation, and the isotropic nature of glasses ensures that the thermal equilibrium stress vanishes, as

The third-order expansion of the density matrix is given by Eq.~(\ref{23}) with $n=3$. Thus, we obtain the term responsible for the echo contribution:

\widetext
\begin{eqnarray}\label{45}
\langle \hat{T}_{ij}^{(s)}\rangle_3 (t)
& = &  \sum_{klmnpq}\int_0^\infty dt_3 \int_0^\infty dt_2 \int_0^\infty dt_1 \, 
\epsilon^{(s)}_{kl}(t-t_3) \epsilon^{(s)}_{mn}(t-t_3-t_2) \epsilon^{(s)}_{pq}(t-t_3-t_2-t_1) \left(\frac{\mathrm{i}}{\hbar}\right)^3 \sum_{abcd}p_a^{(s)} \nonumber \\
& {} & 
\bigg[\left(
e^{\mathrm{i}\omega^{(s)}_{ab}t_3}\langle a^{(s)}| \hat{T}_{ij}^{(s)}|b^{(s)}\rangle \langle b^{(s)}| \hat{T}_{kl}^{(s)}|c^{(s)}\rangle 
- {\rm c.c.}(a\leftrightarrows c)
\right)
e^{\mathrm{i}(\omega_{ad}^{(s)}t_1+\omega_{ac}^{(s)}t_2)}
\langle c^{(s)}| \hat{T}_{mn}^{(s)} |d^{(s)}\rangle \langle d^{(s)}|\hat{T}_{pq}^{(s)}|a^{(s)}\rangle\nonumber \\
& {} & + \left(e^{\mathrm{i}\omega_{cd}^{(s)}t_3}  \langle b^{(s)}| \hat{T}_{kl}^{(s)}|c^{(s)}\rangle \langle c^{(s)}| \hat{T}_{ij}^{(s)} |d^{(s)}\rangle - {\rm c.c.}(b\leftrightarrows d)\right) 
e^{\mathrm{i}(\omega_{ad}^{(s)}t_1+\omega_{bd}^{(s)}t_2)} 
\langle a^{(s)}| \hat{T}_{mn}^{(s)} |b^{(s)}\rangle \langle d^{(s)}| \hat{T}_{pq}^{(s)}|a^{(s)}\rangle\nonumber \\
& {} & 
+ \left(e^{\mathrm{i}\omega^{(s)}_{cd}t_3} \langle b^{(s)}| \hat{T}^{(s)}_{kl}|c^{(s)}\rangle \langle c^{(s)}| \hat{T}^{(s)}_{ij}|d^{(s)}\rangle 
- {\rm c.c.}(b\leftrightarrows d)\right) 
e^{\mathrm{i}(\omega^{(s)}_{ba}t_1+\omega^{(s)}_{bd}t_2)}
\langle a^{(s)}| \hat{T}^{(s)}_{pq}|b^{(s)}\rangle \langle d^{(s)}| \hat{T}^{(s)}_{mn} |a^{(s)}\rangle\nonumber \\
& {} & + \left(e^{\mathrm{i}\omega^{(s)}_{cd}t_3} 
  \langle c^{(s)}| \hat{T}^{(s)}_{ij}|d^{(s)}\rangle \langle d^{(s)}| \hat{T}_{kl}^{(s)}|a^{(s)}\rangle  
- {\rm c.c.}(a\leftrightarrows c)
\right)
e^{\mathrm{i}(\omega_{ba}^{(s)}t_1+\omega_{ca}^{(s)}t_2)}
\langle a^{(s)}|\hat{T}^{(s)}_{pq}|b^{(s)}\rangle \langle b^{(s)}| \hat{T}^{(s)}_{mn} |c^{(s)}\rangle\bigg]\nonumber \\
\end{eqnarray}
\endwidetext

In the result of Eq.~(\ref{45}), the input triggering signals are represented by $\delta$-functions in the time domain. There are several possible choices for the parameters $(t_1, t_2, t_3)$. However, due to the inherent randomness and nonlinearity of the multi-level system, the energy gaps such as $\omega_{ab}^{(s)}$, $\omega_{bc}^{(s)}$, $\omega_{cd}^{(s)}$, and $\omega_{da}^{(s)}$ fluctuate strongly. One cannot rely on the accidental coincidence that these gaps occasionally satisfy the constraint $\omega_{ab}^{(s)} = -\omega_{bc}^{(s)}$ to produce an echo signal, since such coincidence is merely fortuitous when $|c^{(s)}\rangle \neq |a^{(s)}\rangle$. If instead $|c^{(s)}\rangle = |a^{(s)}\rangle$, the transition process involves only the pair of levels $|a^{(s)}\rangle$ and $|b^{(s)}\rangle$, rather than three distinct states. In this case, the constraint $\omega_{ab}^{(s)} = -\omega_{bc}^{(s)}$ is automatically satisfied, as $|c^{(s)}\rangle$ and $|a^{(s)}\rangle$ are identical. Even with random fluctuations in the energy levels, this condition remains valid. Under this selection rule for deriving the echo signal, we find that in the external excitations, which are $\delta$-functions, the only non-vanishing choice is
\begin{eqnarray}
& {} & (t_1, t_2, t_3) = (\tau, 0, t-\tau):\nonumber \\
& {} & 
\epsilon^{(s)}_{kl}(t-t_3) = \epsilon^{(s)}_{kl}(\tau)=I_{kl}^{(2)}, \nonumber \\
& {} & 
\epsilon^{(s)}_{mn}(t-t_3-t_2)=\epsilon^{(s)}_{mn}(\tau)=I_{mn}^{(2)}, \nonumber \\
& {} & 
\epsilon^{(s)}_{pq}(t-t_3-t_2-t_1)=\epsilon^{(s)}_{pq}(0)=I_{pq}^{(1)}.
\end{eqnarray}
At first glance, another choice $(t_1, t_2, t_3)=(0,\tau,t-\tau)$ might also appear to generate an echo. However, this requires $\omega_{ba}^{(s)}=-\omega_{cb}^{(s)}$ with $|c^{(s)}\rangle \neq |a^{(s)}\rangle$. Such a condition may hold in specific multi-level systems, but it cannot be satisfied universally across amorphous materials, where level spacings are random. In general, this coincidence cannot occur for all multi-level systems. Therefore, the second seemingly plausible choice, $(t_1, t_2, t_3)=(0,\tau,t-\tau)$, is not a robust option for glass systems.

Based on the preceding discussion, the echo signal is robust only when it arises from a pair of energy levels $|a^{(s)}\rangle$ and $|b^{(s)}\rangle$, which can be regarded as an effective two-level system, even though the full system is multi-level. With this simplification, Eq.~(\ref{45}) reduces to  
\begin{eqnarray}\label{46}
& {} & \langle \hat{T}_{ij}^{(s)}\rangle_3 (t) =\nonumber \\
 & {} & 2\left(\frac{\mathrm{i}}{\hbar}\right)^3 \sum_{klmnpq} I_{pq}^{(1)}I_{mn}^{(2)}I_{kl}^{(2)} 
 \sum_{ab}p_a^{(s)} e^{\mathrm{i}\omega_{ab}^{(s)} (t-\tau)}\nonumber \\
 & {} & \bigg[ \left(e^{\mathrm{i}\omega^{(s)}_{ab} \tau}  \langle a^{(s)}| \hat{T}_{mn}^{(s)} |b^{(s)}\rangle \langle b^{(s)}|\hat{T}_{pq}^{(s)}|a^{(s)}\rangle 
+ {\rm c.c.}\right) \nonumber \\
& {} & \langle a^{(s)}| \hat{T}_{ij}^{(s)}|b^{(s)}\rangle \langle b^{(s)}| \hat{T}_{kl}^{(s)}|a^{(s)}\rangle -{\rm c.c.}
\bigg].
\end{eqnarray}
In Eq.~(\ref{46}), the result can be separated into two contributions. The first exhibits a phase factor of $\omega_{ab}^{(s)}t$, which differs from the echo signal, whose characteristic phase factor is $\sim \omega(t-2\tau)$ and appears at $t=2\tau$. The second contribution carries a phase factor of $\omega_{ab}^{(s)}(t-2\tau)$, consistent with the temporal profile of the echo signal, expressed in Eq.~(\ref{35}). At this stage, the phonon echo emerges from the third-order expansion of the stress tensor. Furthermore, Eq.~(\ref{46}) shows that the echo amplitude is proportional to $I^{(1)}I^{(2)2}$, reminiscent of the spin echo in two-level systems.

\section{Temporal evolution of the matrix element of the density matrix in the presence of many-body interactions}

To derive the temporal evolution of the matrix element in the density matrix, we perform the time derivative on the matrix element $\rho_{ab}$:
\begin{eqnarray}
\frac{d\rho_{I,ab}}{dt} 
& = &
\operatorname{Tr}\left(-\mathrm{i}[\hat{V}_I(t),\hat{\rho}_I(t)] |b\rangle \langle a|\right)\nonumber \\
& = & 
-\mathrm{i}\sum_c \bigg( \langle a |\hat{V}_I(t)|c\rangle \langle c |\hat{\rho}_I(t)|b\rangle \nonumber \\
& {} & 
- \langle a |\hat{\rho}_I(t)|c\rangle \langle c |\hat{V}_I(t)|b\rangle \bigg),
\end{eqnarray}
We then split the summation over the quantum state 
$c$ to the cases to three cases, namely $c=a$, $c=b$, and $c\neq a,b$. By doing so, the temporal evolution of the density matrix element reads
\begin{eqnarray}
\frac{d\rho_{I, ab}}{dt} 
& = & \mathrm{i}  \delta\omega_{ab}(t) \rho_{I, ab}(t)
-\mathrm{i} \sum_{c \neq a}\langle a|\hat{V}_I(t)|c\rangle \rho_{cb}(t)\nonumber \\
 & {} & +\mathrm{i} \sum_{c\neq b} \rho_{ac}(t)\langle c|\hat{V}_I(t)|b\rangle .
\end{eqnarray}
Thus, for for $c\neq a,b$, the density matrix element evolves as 
\begin{eqnarray}
\frac{d\rho_{I,cb}}{dt} 
& = & \mathrm{i}  \delta\omega_{cb}(t) \rho_{I,cb}(t)
-\mathrm{i} \sum_{d \neq c}\langle c|\hat{V}_I(t)|d\rangle \rho_{I,db}(t)\nonumber \\
& {} & 
+\mathrm{i} \sum_{d\neq b} \rho_{I,cd}(t)\langle d|\hat{V}_I(t)|b\rangle \nonumber \\
& \approx &
\mathrm{i} \bigg[\sum_{d\neq b} \rho_{I,cd}(t) \langle d|\hat{V}_I(t)|b\rangle
- \langle c|\hat{V}_I(t)|a\rangle \rho_{I,ab}(t)\bigg]
 \nonumber \\
& \approx & 
-\mathrm{i} \langle c|\hat{V}_I(t)|a\rangle \rho_{I,ab}(t)\nonumber \\
 & = &  
-\mathrm{i}e^{\mathrm{i}\omega_{cb}t} \langle c|\hat{V}(t)|a\rangle \rho_{I,ab}(t)
\end{eqnarray}
leading to the fast-oscillatory behavior in the density matrix element, 
\begin{eqnarray}
\rho_{I,cb}(t) \approx -\mathrm{i} \int_0^t e^{-\mathrm{i}\omega_{cb}(t-t')}\langle c|\hat{V}|d\rangle (t')\rho_{I,db}(t')dt'.
\end{eqnarray}
Since this is highly oscillatory term that averages to zero very quickly. Next, substituting this term into the eqnarray of motion for the density matrix $\rho$, which leads to Eq.~(\ref{37}).

The dephasing effect can be derived by computing the average of the density matrix element for a group of oscillators, and make use of the second order cumulant approximation,
\begin{eqnarray}
\langle \rho_{I, ab}(t)\rangle = 
\rho_{I, ab}(0) \langle e^{-\mathrm{i}\phi}\rangle = \rho_{I, ab}(0) e^{-\frac{1}{2}\langle \phi^2\rangle}
\end{eqnarray}
where the expectation value of the phase variable reads
\begin{equation}
\langle \phi^2\rangle = \int_0^{2\tau}dt_1\int_0^{2\tau}dt_2\,\,f(t_1)f(t_2)\,\, \langle \delta\omega_{ab}(t_1)\delta\omega_{ab}(t_2)\rangle,
\end{equation}
where $f(t)$ function is the time-dependent distribution function of the random variable in $\delta\omega_{ab}(t)$. We further assume that the average of the operators $\delta\omega_{ab}(t)$ for different time yield Eq.~(\ref{38}), which is the typical mathematical format for the correlation function in disordered systems. Integrating over the time domain and assuming that the distribution function is constant, we obtain 
\begin{eqnarray}
\frac{1}{2}\langle \phi^2\rangle = 2\Gamma^2\left(\frac{t}{\tau_c}-1+e^{-\frac{t}{\tau_c}}\right)
\end{eqnarray}
where $\tau_c$ is the characteristic time scale for the exponentially-decaying correlation function in the random variables. The above result reflects two limits, namely the short time regime for $t\ll \tau_c$, and long-time regime for $t\gg \tau_c$. Since the echo experiment typically demands $\tau \gg 100 \times 2\pi/\omega_{1,0}$, where $\omega_{1,0}$ is the eigenfrequency between the first excitation and ground state, we have the long-time regime, and then we have 
\begin{eqnarray}
\langle \rho_{I, ab}(t)\rangle = \rho_{I, ab}(0) \exp\left(-\frac{2\Gamma^2}{\tau_c}t\right)
\end{eqnarray}
leading to Eq.~(\ref{47}) in the main text.


\begin{thebibliography}{119}%
\makeatletter
\providecommand \@ifxundefined [1]{%
 \@ifx{#1\undefined}
}%
\providecommand \@ifnum [1]{%
 \ifnum #1\expandafter \@firstoftwo
 \else \expandafter \@secondoftwo
 \fi
}%
\providecommand \@ifx [1]{%
 \ifx #1\expandafter \@firstoftwo
 \else \expandafter \@secondoftwo
 \fi
}%
\providecommand \natexlab [1]{#1}%
\providecommand \enquote  [1]{``#1''}%
\providecommand \bibnamefont  [1]{#1}%
\providecommand \bibfnamefont [1]{#1}%
\providecommand \citenamefont [1]{#1}%
\providecommand \href@noop [0]{\@secondoftwo}%
\providecommand \href [0]{\begingroup \@sanitize@url \@href}%
\providecommand \@href[1]{\@@startlink{#1}\@@href}%
\providecommand \@@href[1]{\endgroup#1\@@endlink}%
\providecommand \@sanitize@url [0]{\catcode `\\12\catcode `\$12\catcode `\&12\catcode `\#12\catcode `\^12\catcode `\_12\catcode `\%12\relax}%
\providecommand \@@startlink[1]{}%
\providecommand \@@endlink[0]{}%
\providecommand \url  [0]{\begingroup\@sanitize@url \@url }%
\providecommand \@url [1]{\endgroup\@href {#1}{\urlprefix }}%
\providecommand \urlprefix  [0]{URL }%
\providecommand \Eprint [0]{\href }%
\providecommand \doibase [0]{http://dx.doi.org/}%
\providecommand \selectlanguage [0]{\@gobble}%
\providecommand \bibinfo  [0]{\@secondoftwo}%
\providecommand \bibfield  [0]{\@secondoftwo}%
\providecommand \translation [1]{[#1]}%
\providecommand \BibitemOpen [0]{}%
\providecommand \bibitemStop [0]{}%
\providecommand \bibitemNoStop [0]{.\EOS\space}%
\providecommand \EOS [0]{\spacefactor3000\relax}%
\providecommand \BibitemShut  [1]{\csname bibitem#1\endcsname}%
\let\auto@bib@innerbib\@empty
\bibitem [{\citenamefont {Zeller}\ and\ \citenamefont {Pohl}(1971)}]{Pohl1971PRB}%
  \BibitemOpen
  \bibfield  {author} {\bibinfo {author} {\bibfnamefont {R.~C.}\ \bibnamefont {Zeller}}\ and\ \bibinfo {author} {\bibfnamefont {R.~O.}\ \bibnamefont {Pohl}},\ }\href {\doibase 10.1103/PhysRevB.4.2029} {\bibfield  {journal} {\bibinfo  {journal} {Phys. Rev. B}\ }\textbf {\bibinfo {volume} {4}},\ \bibinfo {pages} {2029} (\bibinfo {year} {1971})}\BibitemShut {NoStop}%
\bibitem [{\citenamefont {Phillips}(1987)}]{Phillips1987RPP}%
  \BibitemOpen
  \bibfield  {author} {\bibinfo {author} {\bibfnamefont {W.~A.}\ \bibnamefont {Phillips}},\ }\href {\doibase 10.1088/0034-4885/50/12/003} {\bibfield  {journal} {\bibinfo  {journal} {Reports on Progress in Physics}\ }\textbf {\bibinfo {volume} {50}},\ \bibinfo {pages} {1657} (\bibinfo {year} {1987})}\BibitemShut {NoStop}%
\bibitem [{\citenamefont {Anderson}\ \emph {et~al.}(1972)\citenamefont {Anderson}, \citenamefont {Halperin},\ and\ \citenamefont {Varma}}]{Varma1972Phil}%
  \BibitemOpen
  \bibfield  {author} {\bibinfo {author} {\bibfnamefont {P.}~\bibnamefont {Anderson}}, \bibinfo {author} {\bibfnamefont {B.}~\bibnamefont {Halperin}}, \ and\ \bibinfo {author} {\bibfnamefont {C.~M.}\ \bibnamefont {Varma}},\ }\href {\doibase https://doi.org/10.1080/14786437208229210} {\bibfield  {journal} {\bibinfo  {journal} {Philos. Mag.}\ }\textbf {\bibinfo {volume} {25}} (\bibinfo {year} {1972}),\ https://doi.org/10.1080/14786437208229210}\BibitemShut {NoStop}%
\bibitem [{\citenamefont {Hunklinger}(1977)}]{Hunklinger1977Acoustic}%
  \BibitemOpen
  \bibfield  {author} {\bibinfo {author} {\bibfnamefont {S.}~\bibnamefont {Hunklinger}},\ }\enquote {\bibinfo {title} {Acoustic and dielectric properties of glasses at low temperatures},}\ \ (\bibinfo  {publisher} {Springer Berlin Heidelberg},\ \bibinfo {address} {Berlin, Heidelberg},\ \bibinfo {year} {1977})\ pp.\ \bibinfo {pages} {1--11}\BibitemShut {NoStop}%
\bibitem [{\citenamefont {Galperin}\ \emph {et~al.}(1985)\citenamefont {Galperin}, \citenamefont {Gurevich},\ and\ \citenamefont {Parshin}}]{Parshin1985PRB}%
  \BibitemOpen
  \bibfield  {author} {\bibinfo {author} {\bibfnamefont {Y.~M.}\ \bibnamefont {Galperin}}, \bibinfo {author} {\bibfnamefont {V.~L.}\ \bibnamefont {Gurevich}}, \ and\ \bibinfo {author} {\bibfnamefont {D.~A.}\ \bibnamefont {Parshin}},\ }\href {\doibase 10.1103/PhysRevB.32.6873} {\bibfield  {journal} {\bibinfo  {journal} {Phys. Rev. B}\ }\textbf {\bibinfo {volume} {32}},\ \bibinfo {pages} {6873} (\bibinfo {year} {1985})}\BibitemShut {NoStop}%
\bibitem [{\citenamefont {Ramos}(2020)}]{Ramos2020LTP}%
  \BibitemOpen
  \bibfield  {author} {\bibinfo {author} {\bibfnamefont {M.~A.}\ \bibnamefont {Ramos}},\ }\href {\doibase 10.1063/10.0000572} {\bibfield  {journal} {\bibinfo  {journal} {Low Temperature Physics}\ }\textbf {\bibinfo {volume} {46}},\ \bibinfo {pages} {104} (\bibinfo {year} {2020})}\BibitemShut {NoStop}%
\bibitem [{\citenamefont {Pohl}\ \emph {et~al.}(2002)\citenamefont {Pohl}, \citenamefont {Liu},\ and\ \citenamefont {Thompson}}]{Pohl2011RMP}%
  \BibitemOpen
  \bibfield  {author} {\bibinfo {author} {\bibfnamefont {R.~O.}\ \bibnamefont {Pohl}}, \bibinfo {author} {\bibfnamefont {X.}~\bibnamefont {Liu}}, \ and\ \bibinfo {author} {\bibfnamefont {E.}~\bibnamefont {Thompson}},\ }\href {\doibase 10.1103/RevModPhys.74.991} {\bibfield  {journal} {\bibinfo  {journal} {Rev. Mod. Phys.}\ }\textbf {\bibinfo {volume} {74}},\ \bibinfo {pages} {991} (\bibinfo {year} {2002})}\BibitemShut {NoStop}%
\bibitem [{\citenamefont {Clare}\ and\ \citenamefont {J.}(1988)}]{Yu1991CCMP}%
  \BibitemOpen
  \bibfield  {author} {\bibinfo {author} {\bibfnamefont {Y.}~\bibnamefont {Clare}}\ and\ \bibinfo {author} {\bibfnamefont {L.~A.}\ \bibnamefont {J.}},\ }\href@noop {} {\bibfield  {journal} {\bibinfo  {journal} {Comments on Condensed Matter Physics}\ }\textbf {\bibinfo {volume} {14}},\ \bibinfo {pages} {231} (\bibinfo {year} {1988})}\BibitemShut {NoStop}%
\bibitem [{\citenamefont {Yu}(1989)}]{Yu1989PRL}%
  \BibitemOpen
  \bibfield  {author} {\bibinfo {author} {\bibfnamefont {C.~C.}\ \bibnamefont {Yu}},\ }\href {\doibase 10.1103/PhysRevLett.63.1160} {\bibfield  {journal} {\bibinfo  {journal} {Phys. Rev. Lett.}\ }\textbf {\bibinfo {volume} {63}},\ \bibinfo {pages} {1160} (\bibinfo {year} {1989})}\BibitemShut {NoStop}%
\bibitem [{\citenamefont {Lasjaunias}\ \emph {et~al.}(1993)\citenamefont {Lasjaunias}, \citenamefont {Ravex}, \citenamefont {Vandorpe},\ and\ \citenamefont {Hunklinger}}]{LASJAUNIAS1993SSC}%
  \BibitemOpen
  \bibfield  {author} {\bibinfo {author} {\bibfnamefont {J.}~\bibnamefont {Lasjaunias}}, \bibinfo {author} {\bibfnamefont {A.}~\bibnamefont {Ravex}}, \bibinfo {author} {\bibfnamefont {M.}~\bibnamefont {Vandorpe}}, \ and\ \bibinfo {author} {\bibfnamefont {S.}~\bibnamefont {Hunklinger}},\ }\href {\doibase https://doi.org/10.1016/0038-1098(93)90288-X} {\bibfield  {journal} {\bibinfo  {journal} {Solid State Communications}\ }\textbf {\bibinfo {volume} {88}},\ \bibinfo {pages} {1023} (\bibinfo {year} {1993})},\ \bibinfo {note} {special Issue A Celebratory Issue to Commemorate 30 Years of Solid State Communications}\BibitemShut {NoStop}%
\bibitem [{\citenamefont {Tietje}\ \emph {et~al.}(1986)\citenamefont {Tietje}, \citenamefont {von Schickfus},\ and\ \citenamefont {Gmelin}}]{Tietje1986ZCM}%
  \BibitemOpen
  \bibfield  {author} {\bibinfo {author} {\bibfnamefont {H.}~\bibnamefont {Tietje}}, \bibinfo {author} {\bibfnamefont {M.}~\bibnamefont {von Schickfus}}, \ and\ \bibinfo {author} {\bibfnamefont {E.}~\bibnamefont {Gmelin}},\ }\href {\doibase 10.1007/BF01313693} {\bibfield  {journal} {\bibinfo  {journal} {Zeitschrift für Physik B Condensed Matter}\ }\textbf {\bibinfo {volume} {95}} (\bibinfo {year} {1986}),\ 10.1007/BF01313693}\BibitemShut {NoStop}%
\bibitem [{\citenamefont {Franz}\ \emph {et~al.}(2015)\citenamefont {Franz}, \citenamefont {Parisi}, \citenamefont {Urbani},\ and\ \citenamefont {Zamponi}}]{Zamponi2015PNAS}%
  \BibitemOpen
  \bibfield  {author} {\bibinfo {author} {\bibfnamefont {S.}~\bibnamefont {Franz}}, \bibinfo {author} {\bibfnamefont {G.}~\bibnamefont {Parisi}}, \bibinfo {author} {\bibfnamefont {P.}~\bibnamefont {Urbani}}, \ and\ \bibinfo {author} {\bibfnamefont {F.}~\bibnamefont {Zamponi}},\ }\href {\doibase 10.1073/pnas.1511134112} {\bibfield  {journal} {\bibinfo  {journal} {Proceedings of the National Academy of Sciences}\ }\textbf {\bibinfo {volume} {112}},\ \bibinfo {pages} {14539} (\bibinfo {year} {2015})},\ \Eprint {http://arxiv.org/abs/https://www.pnas.org/doi/pdf/10.1073/pnas.1511134112} {https://www.pnas.org/doi/pdf/10.1073/pnas.1511134112} \BibitemShut {NoStop}%
\bibitem [{\citenamefont {Charbonneau}\ \emph {et~al.}(2016{\natexlab{a}})\citenamefont {Charbonneau}, \citenamefont {Corwin}, \citenamefont {Parisi}, \citenamefont {Poncet},\ and\ \citenamefont {Zamponi}}]{Zamponi2016PRL}%
  \BibitemOpen
  \bibfield  {author} {\bibinfo {author} {\bibfnamefont {P.}~\bibnamefont {Charbonneau}}, \bibinfo {author} {\bibfnamefont {E.~I.}\ \bibnamefont {Corwin}}, \bibinfo {author} {\bibfnamefont {G.}~\bibnamefont {Parisi}}, \bibinfo {author} {\bibfnamefont {A.}~\bibnamefont {Poncet}}, \ and\ \bibinfo {author} {\bibfnamefont {F.}~\bibnamefont {Zamponi}},\ }\href {\doibase 10.1103/PhysRevLett.117.045503} {\bibfield  {journal} {\bibinfo  {journal} {Phys. Rev. Lett.}\ }\textbf {\bibinfo {volume} {117}},\ \bibinfo {pages} {045503} (\bibinfo {year} {2016}{\natexlab{a}})}\BibitemShut {NoStop}%
\bibitem [{\citenamefont {Stephens}(1976)}]{Stefhens1976PRB}%
  \BibitemOpen
  \bibfield  {author} {\bibinfo {author} {\bibfnamefont {R.~B.}\ \bibnamefont {Stephens}},\ }\href {\doibase 10.1103/PhysRevB.13.852} {\bibfield  {journal} {\bibinfo  {journal} {Phys. Rev. B}\ }\textbf {\bibinfo {volume} {13}},\ \bibinfo {pages} {852} (\bibinfo {year} {1976})}\BibitemShut {NoStop}%
\bibitem [{\citenamefont {Ramos}\ and\ \citenamefont {Buchenau}(1997)}]{Ramos1997PRB}%
  \BibitemOpen
  \bibfield  {author} {\bibinfo {author} {\bibfnamefont {M.~A.}\ \bibnamefont {Ramos}}\ and\ \bibinfo {author} {\bibfnamefont {U.}~\bibnamefont {Buchenau}},\ }\href {\doibase 10.1103/PhysRevB.55.5749} {\bibfield  {journal} {\bibinfo  {journal} {Phys. Rev. B}\ }\textbf {\bibinfo {volume} {55}},\ \bibinfo {pages} {5749} (\bibinfo {year} {1997})}\BibitemShut {NoStop}%
\bibitem [{\citenamefont {Gil}\ \emph {et~al.}(1993)\citenamefont {Gil}, \citenamefont {Ramos}, \citenamefont {Bringer},\ and\ \citenamefont {Buchenau}}]{Buchenau1993PRL}%
  \BibitemOpen
  \bibfield  {author} {\bibinfo {author} {\bibfnamefont {L.}~\bibnamefont {Gil}}, \bibinfo {author} {\bibfnamefont {M.~A.}\ \bibnamefont {Ramos}}, \bibinfo {author} {\bibfnamefont {A.}~\bibnamefont {Bringer}}, \ and\ \bibinfo {author} {\bibfnamefont {U.}~\bibnamefont {Buchenau}},\ }\href {\doibase 10.1103/PhysRevLett.70.182} {\bibfield  {journal} {\bibinfo  {journal} {Phys. Rev. Lett.}\ }\textbf {\bibinfo {volume} {70}},\ \bibinfo {pages} {182} (\bibinfo {year} {1993})}\BibitemShut {NoStop}%
\bibitem [{\citenamefont {Vural}\ and\ \citenamefont {Leggett}(2011)}]{Vural2011JNCS}%
  \BibitemOpen
  \bibfield  {author} {\bibinfo {author} {\bibfnamefont {D.~C.}\ \bibnamefont {Vural}}\ and\ \bibinfo {author} {\bibfnamefont {A.~J.}\ \bibnamefont {Leggett}},\ }\href {\doibase https://doi.org/10.1016/j.jnoncrysol.2011.06.035} {\bibfield  {journal} {\bibinfo  {journal} {Journal of Non-Crystalline Solids}\ }\textbf {\bibinfo {volume} {357}},\ \bibinfo {pages} {3528} (\bibinfo {year} {2011})}\BibitemShut {NoStop}%
\bibitem [{\citenamefont {Carruzzo}\ and\ \citenamefont {Yu}(2020)}]{Yu2020PRL}%
  \BibitemOpen
  \bibfield  {author} {\bibinfo {author} {\bibfnamefont {H.~M.}\ \bibnamefont {Carruzzo}}\ and\ \bibinfo {author} {\bibfnamefont {C.~C.}\ \bibnamefont {Yu}},\ }\href {\doibase 10.1103/PhysRevLett.124.075902} {\bibfield  {journal} {\bibinfo  {journal} {Phys. Rev. Lett.}\ }\textbf {\bibinfo {volume} {124}},\ \bibinfo {pages} {075902} (\bibinfo {year} {2020})}\BibitemShut {NoStop}%
\bibitem [{\citenamefont {Buchenau}\ \emph {et~al.}(1992)\citenamefont {Buchenau}, \citenamefont {Galperin}, \citenamefont {Gurevich}, \citenamefont {Parshin}, \citenamefont {Ramos},\ and\ \citenamefont {Schober}}]{Ramos1992PRB}%
  \BibitemOpen
  \bibfield  {author} {\bibinfo {author} {\bibfnamefont {U.}~\bibnamefont {Buchenau}}, \bibinfo {author} {\bibfnamefont {Y.~M.}\ \bibnamefont {Galperin}}, \bibinfo {author} {\bibfnamefont {V.~L.}\ \bibnamefont {Gurevich}}, \bibinfo {author} {\bibfnamefont {D.~A.}\ \bibnamefont {Parshin}}, \bibinfo {author} {\bibfnamefont {M.~A.}\ \bibnamefont {Ramos}}, \ and\ \bibinfo {author} {\bibfnamefont {H.~R.}\ \bibnamefont {Schober}},\ }\href {\doibase 10.1103/PhysRevB.46.2798} {\bibfield  {journal} {\bibinfo  {journal} {Phys. Rev. B}\ }\textbf {\bibinfo {volume} {46}},\ \bibinfo {pages} {2798} (\bibinfo {year} {1992})}\BibitemShut {NoStop}%
\bibitem [{\citenamefont {Raychaudhuri}\ and\ \citenamefont {Pohl}(1991)}]{Pohl1991PRB}%
  \BibitemOpen
  \bibfield  {author} {\bibinfo {author} {\bibfnamefont {A.~K.}\ \bibnamefont {Raychaudhuri}}\ and\ \bibinfo {author} {\bibfnamefont {R.~O.}\ \bibnamefont {Pohl}},\ }\href {\doibase 10.1103/PhysRevB.44.12233} {\bibfield  {journal} {\bibinfo  {journal} {Phys. Rev. B}\ }\textbf {\bibinfo {volume} {44}},\ \bibinfo {pages} {12233} (\bibinfo {year} {1991})}\BibitemShut {NoStop}%
\bibitem [{\citenamefont {Classen}\ \emph {et~al.}(2000)\citenamefont {Classen}, \citenamefont {Burkert}, \citenamefont {Enss},\ and\ \citenamefont {Hunklinger}}]{Classen2000PRL}%
  \BibitemOpen
  \bibfield  {author} {\bibinfo {author} {\bibfnamefont {J.}~\bibnamefont {Classen}}, \bibinfo {author} {\bibfnamefont {T.}~\bibnamefont {Burkert}}, \bibinfo {author} {\bibfnamefont {C.}~\bibnamefont {Enss}}, \ and\ \bibinfo {author} {\bibfnamefont {S.}~\bibnamefont {Hunklinger}},\ }\href {\doibase 10.1103/PhysRevLett.84.2176} {\bibfield  {journal} {\bibinfo  {journal} {Phys. Rev. Lett.}\ }\textbf {\bibinfo {volume} {84}},\ \bibinfo {pages} {2176} (\bibinfo {year} {2000})}\BibitemShut {NoStop}%
\bibitem [{\citenamefont {Arnold}\ \emph {et~al.}(1981)\citenamefont {Arnold}, \citenamefont {Doussineau}, \citenamefont {Fr{\'e}nois},\ and\ \citenamefont {Levelut}}]{Arnold1981JDPL}%
  \BibitemOpen
  \bibfield  {author} {\bibinfo {author} {\bibfnamefont {W.}~\bibnamefont {Arnold}}, \bibinfo {author} {\bibfnamefont {P.}~\bibnamefont {Doussineau}}, \bibinfo {author} {\bibfnamefont {C.}~\bibnamefont {Fr{\'e}nois}}, \ and\ \bibinfo {author} {\bibfnamefont {A.}~\bibnamefont {Levelut}},\ }\href {\doibase 10.1051/jphyslet:019810042013028900} {\bibfield  {journal} {\bibinfo  {journal} {Journal de Physique Lettres}\ }\textbf {\bibinfo {volume} {42}},\ \bibinfo {pages} {289} (\bibinfo {year} {1981})}\BibitemShut {NoStop}%
\bibitem [{\citenamefont {Strom}\ \emph {et~al.}(1978)\citenamefont {Strom}, \citenamefont {von Schickfus},\ and\ \citenamefont {Hunklinger}}]{Hunklinger1978PRL}%
  \BibitemOpen
  \bibfield  {author} {\bibinfo {author} {\bibfnamefont {U.}~\bibnamefont {Strom}}, \bibinfo {author} {\bibfnamefont {M.}~\bibnamefont {von Schickfus}}, \ and\ \bibinfo {author} {\bibfnamefont {S.}~\bibnamefont {Hunklinger}},\ }\href {\doibase 10.1103/PhysRevLett.41.910} {\bibfield  {journal} {\bibinfo  {journal} {Phys. Rev. Lett.}\ }\textbf {\bibinfo {volume} {41}},\ \bibinfo {pages} {910} (\bibinfo {year} {1978})}\BibitemShut {NoStop}%
\bibitem [{\citenamefont {Strehlow}\ \emph {et~al.}(1998)\citenamefont {Strehlow}, \citenamefont {Enss},\ and\ \citenamefont {Hunklinger}}]{Hunklinger1998PRL}%
  \BibitemOpen
  \bibfield  {author} {\bibinfo {author} {\bibfnamefont {P.}~\bibnamefont {Strehlow}}, \bibinfo {author} {\bibfnamefont {C.}~\bibnamefont {Enss}}, \ and\ \bibinfo {author} {\bibfnamefont {S.}~\bibnamefont {Hunklinger}},\ }\href {\doibase 10.1103/PhysRevLett.80.5361} {\bibfield  {journal} {\bibinfo  {journal} {Phys. Rev. Lett.}\ }\textbf {\bibinfo {volume} {80}},\ \bibinfo {pages} {5361} (\bibinfo {year} {1998})}\BibitemShut {NoStop}%
\bibitem [{\citenamefont {Federle}\ and\ \citenamefont {Hunklinger}(1982)}]{Federle1982JPC}%
  \BibitemOpen
  \bibfield  {author} {\bibinfo {author} {\bibfnamefont {G.}~\bibnamefont {Federle}}\ and\ \bibinfo {author} {\bibfnamefont {S.}~\bibnamefont {Hunklinger}},\ }\href {\doibase 10.1051/jphyscol:1982999} {\bibfield  {journal} {\bibinfo  {journal} {Journal de Physique Colloques}\ }\textbf {\bibinfo {volume} {43}},\ \bibinfo {pages} {C9} (\bibinfo {year} {1982})}\BibitemShut {NoStop}%
\bibitem [{\citenamefont {Pich{\'e}}\ \emph {et~al.}(1974)\citenamefont {Pich{\'e}}, \citenamefont {Maynard}, \citenamefont {Hunklinger},\ and\ \citenamefont {J{\"a}ckle}}]{Piche1974PRL}%
  \BibitemOpen
  \bibfield  {author} {\bibinfo {author} {\bibfnamefont {L.}~\bibnamefont {Pich{\'e}}}, \bibinfo {author} {\bibfnamefont {R.}~\bibnamefont {Maynard}}, \bibinfo {author} {\bibfnamefont {S.}~\bibnamefont {Hunklinger}}, \ and\ \bibinfo {author} {\bibfnamefont {J.}~\bibnamefont {J{\"a}ckle}},\ }\href {\doibase 10.1103/PhysRevLett.32.1426} {\bibfield  {journal} {\bibinfo  {journal} {Physical Review Letters}\ }\textbf {\bibinfo {volume} {32}},\ \bibinfo {pages} {1426} (\bibinfo {year} {1974})}\BibitemShut {NoStop}%
\bibitem [{\citenamefont {Pollak}\ and\ \citenamefont {Pike}(1972)}]{Pollak1972PRL}%
  \BibitemOpen
  \bibfield  {author} {\bibinfo {author} {\bibfnamefont {M.}~\bibnamefont {Pollak}}\ and\ \bibinfo {author} {\bibfnamefont {G.~E.}\ \bibnamefont {Pike}},\ }\href {\doibase 10.1103/PhysRevLett.28.1449} {\bibfield  {journal} {\bibinfo  {journal} {Phys. Rev. Lett.}\ }\textbf {\bibinfo {volume} {28}},\ \bibinfo {pages} {1449} (\bibinfo {year} {1972})}\BibitemShut {NoStop}%
\bibitem [{\citenamefont {Hunklinger}\ and\ \citenamefont {v.~Schickfus}(1981)}]{Hunklinger1981}%
  \BibitemOpen
  \bibfield  {author} {\bibinfo {author} {\bibfnamefont {S.}~\bibnamefont {Hunklinger}}\ and\ \bibinfo {author} {\bibfnamefont {M.}~\bibnamefont {v.~Schickfus}},\ }\enquote {\bibinfo {title} {Acoustic and dielectric properties of glasses at low temperatures},}\ in\ \href {\doibase 10.1007/978-3-642-81534-8_6} {\emph {\bibinfo {booktitle} {Amorphous Solids: Low-Temperature Properties}}},\ \bibinfo {editor} {edited by\ \bibinfo {editor} {\bibfnamefont {W.~A.}\ \bibnamefont {Phillips}}}\ (\bibinfo  {publisher} {Springer Berlin Heidelberg},\ \bibinfo {address} {Berlin, Heidelberg},\ \bibinfo {year} {1981})\ pp.\ \bibinfo {pages} {81--105}\BibitemShut {NoStop}%
\bibitem [{\citenamefont {Kettemann}\ \emph {et~al.}(1999)\citenamefont {Kettemann}, \citenamefont {Fulde},\ and\ \citenamefont {Strehlow}}]{Peter1999PRL}%
  \BibitemOpen
  \bibfield  {author} {\bibinfo {author} {\bibfnamefont {S.}~\bibnamefont {Kettemann}}, \bibinfo {author} {\bibfnamefont {P.}~\bibnamefont {Fulde}}, \ and\ \bibinfo {author} {\bibfnamefont {P.}~\bibnamefont {Strehlow}},\ }\href {\doibase 10.1103/PhysRevLett.83.4325} {\bibfield  {journal} {\bibinfo  {journal} {Phys. Rev. Lett.}\ }\textbf {\bibinfo {volume} {83}},\ \bibinfo {pages} {4325} (\bibinfo {year} {1999})}\BibitemShut {NoStop}%
\bibitem [{\citenamefont {Strehlow}\ \emph {et~al.}(2000)\citenamefont {Strehlow}, \citenamefont {Wohlfahrt}, \citenamefont {Jansen}, \citenamefont {Haueisen}, \citenamefont {Weiss}, \citenamefont {Enss},\ and\ \citenamefont {Hunklinger}}]{Hunklinger2000PRL}%
  \BibitemOpen
  \bibfield  {author} {\bibinfo {author} {\bibfnamefont {P.}~\bibnamefont {Strehlow}}, \bibinfo {author} {\bibfnamefont {M.}~\bibnamefont {Wohlfahrt}}, \bibinfo {author} {\bibfnamefont {A.~G.~M.}\ \bibnamefont {Jansen}}, \bibinfo {author} {\bibfnamefont {R.}~\bibnamefont {Haueisen}}, \bibinfo {author} {\bibfnamefont {G.}~\bibnamefont {Weiss}}, \bibinfo {author} {\bibfnamefont {C.}~\bibnamefont {Enss}}, \ and\ \bibinfo {author} {\bibfnamefont {S.}~\bibnamefont {Hunklinger}},\ }\href {\doibase 10.1103/PhysRevLett.84.1938} {\bibfield  {journal} {\bibinfo  {journal} {Phys. Rev. Lett.}\ }\textbf {\bibinfo {volume} {84}},\ \bibinfo {pages} {1938} (\bibinfo {year} {2000})}\BibitemShut {NoStop}%
\bibitem [{\citenamefont {Classen}\ \emph {et~al.}(1994)\citenamefont {Classen}, \citenamefont {Enss}, \citenamefont {Bechinger}, \citenamefont {Weiss},\ and\ \citenamefont {Hunklinger}}]{Classen1994ADP}%
  \BibitemOpen
  \bibfield  {author} {\bibinfo {author} {\bibfnamefont {J.}~\bibnamefont {Classen}}, \bibinfo {author} {\bibfnamefont {C.}~\bibnamefont {Enss}}, \bibinfo {author} {\bibfnamefont {C.}~\bibnamefont {Bechinger}}, \bibinfo {author} {\bibfnamefont {G.}~\bibnamefont {Weiss}}, \ and\ \bibinfo {author} {\bibfnamefont {S.}~\bibnamefont {Hunklinger}},\ }\href {\doibase 10.1002/andp.19945060314} {\bibfield  {journal} {\bibinfo  {journal} {Annalen der Physik}\ }\textbf {\bibinfo {volume} {506}},\ \bibinfo {pages} {315} (\bibinfo {year} {1994})}\BibitemShut {NoStop}%
\bibitem [{\citenamefont {Graebner}\ \emph {et~al.}(1983)\citenamefont {Graebner}, \citenamefont {Allen}, \citenamefont {Golding},\ and\ \citenamefont {Kane}}]{Kane1983PRB}%
  \BibitemOpen
  \bibfield  {author} {\bibinfo {author} {\bibfnamefont {J.~E.}\ \bibnamefont {Graebner}}, \bibinfo {author} {\bibfnamefont {L.~C.}\ \bibnamefont {Allen}}, \bibinfo {author} {\bibfnamefont {B.}~\bibnamefont {Golding}}, \ and\ \bibinfo {author} {\bibfnamefont {A.~B.}\ \bibnamefont {Kane}},\ }\href {\doibase 10.1103/PhysRevB.27.3697} {\bibfield  {journal} {\bibinfo  {journal} {Phys. Rev. B}\ }\textbf {\bibinfo {volume} {27}},\ \bibinfo {pages} {3697} (\bibinfo {year} {1983})}\BibitemShut {NoStop}%
\bibitem [{\citenamefont {Hunklinger}\ \emph {et~al.}(1972)\citenamefont {Hunklinger}, \citenamefont {Arnold}, \citenamefont {Stein}, \citenamefont {Nava},\ and\ \citenamefont {Dransfeld}}]{HUNKLINGER1972PLA}%
  \BibitemOpen
  \bibfield  {author} {\bibinfo {author} {\bibfnamefont {S.}~\bibnamefont {Hunklinger}}, \bibinfo {author} {\bibfnamefont {W.}~\bibnamefont {Arnold}}, \bibinfo {author} {\bibfnamefont {S.}~\bibnamefont {Stein}}, \bibinfo {author} {\bibfnamefont {R.}~\bibnamefont {Nava}}, \ and\ \bibinfo {author} {\bibfnamefont {K.}~\bibnamefont {Dransfeld}},\ }\href {\doibase https://doi.org/10.1016/0375-9601(72)90884-5} {\bibfield  {journal} {\bibinfo  {journal} {Physics Letters A}\ }\textbf {\bibinfo {volume} {42}},\ \bibinfo {pages} {253} (\bibinfo {year} {1972})}\BibitemShut {NoStop}%
\bibitem [{\citenamefont {{Von Schickfus}}\ and\ \citenamefont {Hunklinger}(1977)}]{SCHICKFUS1977PLA}%
  \BibitemOpen
  \bibfield  {author} {\bibinfo {author} {\bibfnamefont {M.}~\bibnamefont {{Von Schickfus}}}\ and\ \bibinfo {author} {\bibfnamefont {S.}~\bibnamefont {Hunklinger}},\ }\href {\doibase https://doi.org/10.1016/0375-9601(77)90558-8} {\bibfield  {journal} {\bibinfo  {journal} {Physics Letters A}\ }\textbf {\bibinfo {volume} {64}},\ \bibinfo {pages} {144} (\bibinfo {year} {1977})}\BibitemShut {NoStop}%
\bibitem [{\citenamefont {Pellé}\ and\ \citenamefont {Auzel}(2000)}]{PELLE2000131JAC}%
  \BibitemOpen
  \bibfield  {author} {\bibinfo {author} {\bibfnamefont {F.}~\bibnamefont {Pellé}}\ and\ \bibinfo {author} {\bibfnamefont {F.}~\bibnamefont {Auzel}},\ }\href {\doibase https://doi.org/10.1016/S0925-8388(99)00743-4} {\bibfield  {journal} {\bibinfo  {journal} {Journal of Alloys and Compounds}\ }\textbf {\bibinfo {volume} {300-301}},\ \bibinfo {pages} {131} (\bibinfo {year} {2000})}\BibitemShut {NoStop}%
\bibitem [{\citenamefont {Graebner}\ and\ \citenamefont {Golding}(1979)}]{Golding1979PRB}%
  \BibitemOpen
  \bibfield  {author} {\bibinfo {author} {\bibfnamefont {J.~E.}\ \bibnamefont {Graebner}}\ and\ \bibinfo {author} {\bibfnamefont {B.}~\bibnamefont {Golding}},\ }\href {\doibase 10.1103/PhysRevB.19.964} {\bibfield  {journal} {\bibinfo  {journal} {Phys. Rev. B}\ }\textbf {\bibinfo {volume} {19}},\ \bibinfo {pages} {964} (\bibinfo {year} {1979})}\BibitemShut {NoStop}%
\bibitem [{\citenamefont {Burin}\ \emph {et~al.}(2013)\citenamefont {Burin}, \citenamefont {Leveritt}, \citenamefont {Ruyters}, \citenamefont {Schötz}, \citenamefont {Bazrafshan}, \citenamefont {Fassl}, \citenamefont {von Schickfus}, \citenamefont {Fleischmann},\ and\ \citenamefont {Enss}}]{Burin2013EL}%
  \BibitemOpen
  \bibfield  {author} {\bibinfo {author} {\bibfnamefont {A.~L.}\ \bibnamefont {Burin}}, \bibinfo {author} {\bibfnamefont {J.~M.}\ \bibnamefont {Leveritt}}, \bibinfo {author} {\bibfnamefont {G.}~\bibnamefont {Ruyters}}, \bibinfo {author} {\bibfnamefont {C.}~\bibnamefont {Schötz}}, \bibinfo {author} {\bibfnamefont {M.}~\bibnamefont {Bazrafshan}}, \bibinfo {author} {\bibfnamefont {P.}~\bibnamefont {Fassl}}, \bibinfo {author} {\bibfnamefont {M.}~\bibnamefont {von Schickfus}}, \bibinfo {author} {\bibfnamefont {A.}~\bibnamefont {Fleischmann}}, \ and\ \bibinfo {author} {\bibfnamefont {C.}~\bibnamefont {Enss}},\ }\href {\doibase 10.1209/0295-5075/104/57006} {\bibfield  {journal} {\bibinfo  {journal} {Europhysics Letters}\ }\textbf {\bibinfo {volume} {104}},\ \bibinfo {pages} {57006} (\bibinfo {year} {2013})}\BibitemShut {NoStop}%
\bibitem [{\citenamefont {Hahn}(1950)}]{Hahn1950PR}%
  \BibitemOpen
  \bibfield  {author} {\bibinfo {author} {\bibfnamefont {E.~L.}\ \bibnamefont {Hahn}},\ }\href {\doibase 10.1103/PhysRev.80.580} {\bibfield  {journal} {\bibinfo  {journal} {Phys. Rev.}\ }\textbf {\bibinfo {volume} {80}},\ \bibinfo {pages} {580} (\bibinfo {year} {1950})}\BibitemShut {NoStop}%
\bibitem [{\citenamefont {Xin}\ \emph {et~al.}(2021)\citenamefont {Xin}, \citenamefont {Stolt}, \citenamefont {Song}, \citenamefont {Dai},\ and\ \citenamefont {Halperin}}]{Xin2021PRB}%
  \BibitemOpen
  \bibfield  {author} {\bibinfo {author} {\bibfnamefont {Y.}~\bibnamefont {Xin}}, \bibinfo {author} {\bibfnamefont {I.}~\bibnamefont {Stolt}}, \bibinfo {author} {\bibfnamefont {Y.}~\bibnamefont {Song}}, \bibinfo {author} {\bibfnamefont {P.}~\bibnamefont {Dai}}, \ and\ \bibinfo {author} {\bibfnamefont {W.~P.}\ \bibnamefont {Halperin}},\ }\href {\doibase 10.1103/PhysRevB.104.144421} {\bibfield  {journal} {\bibinfo  {journal} {Phys. Rev. B}\ }\textbf {\bibinfo {volume} {104}},\ \bibinfo {pages} {144421} (\bibinfo {year} {2021})}\BibitemShut {NoStop}%
\bibitem [{\citenamefont {Serbyn}\ and\ \citenamefont {Abanin}(2017)}]{Abanin2017PRB}%
  \BibitemOpen
  \bibfield  {author} {\bibinfo {author} {\bibfnamefont {M.}~\bibnamefont {Serbyn}}\ and\ \bibinfo {author} {\bibfnamefont {D.~A.}\ \bibnamefont {Abanin}},\ }\href {\doibase 10.1103/PhysRevB.96.014202} {\bibfield  {journal} {\bibinfo  {journal} {Phys. Rev. B}\ }\textbf {\bibinfo {volume} {96}},\ \bibinfo {pages} {014202} (\bibinfo {year} {2017})}\BibitemShut {NoStop}%
\bibitem [{\citenamefont {Kurnit}\ \emph {et~al.}(1964)\citenamefont {Kurnit}, \citenamefont {Abella},\ and\ \citenamefont {Hartmann}}]{Hartmann1964PRL}%
  \BibitemOpen
  \bibfield  {author} {\bibinfo {author} {\bibfnamefont {N.~A.}\ \bibnamefont {Kurnit}}, \bibinfo {author} {\bibfnamefont {I.~D.}\ \bibnamefont {Abella}}, \ and\ \bibinfo {author} {\bibfnamefont {S.~R.}\ \bibnamefont {Hartmann}},\ }\href {\doibase 10.1103/PhysRevLett.13.567} {\bibfield  {journal} {\bibinfo  {journal} {Phys. Rev. Lett.}\ }\textbf {\bibinfo {volume} {13}},\ \bibinfo {pages} {567} (\bibinfo {year} {1964})}\BibitemShut {NoStop}%
\bibitem [{\citenamefont {Chen}\ \emph {et~al.}(2020)\citenamefont {Chen}, \citenamefont {Zhang}, \citenamefont {Zheng}, \citenamefont {Wu},\ and\ \citenamefont {Zhai}}]{Zhai2020PRA}%
  \BibitemOpen
  \bibfield  {author} {\bibinfo {author} {\bibfnamefont {Y.-Y.}\ \bibnamefont {Chen}}, \bibinfo {author} {\bibfnamefont {P.}~\bibnamefont {Zhang}}, \bibinfo {author} {\bibfnamefont {W.}~\bibnamefont {Zheng}}, \bibinfo {author} {\bibfnamefont {Z.}~\bibnamefont {Wu}}, \ and\ \bibinfo {author} {\bibfnamefont {H.}~\bibnamefont {Zhai}},\ }\href {\doibase 10.1103/PhysRevA.102.011301} {\bibfield  {journal} {\bibinfo  {journal} {Phys. Rev. A}\ }\textbf {\bibinfo {volume} {102}},\ \bibinfo {pages} {011301} (\bibinfo {year} {2020})}\BibitemShut {NoStop}%
\bibitem [{\citenamefont {Mao}\ \emph {et~al.}(2025)\citenamefont {Mao}, \citenamefont {Tutunnikov}, \citenamefont {Krems},\ and\ \citenamefont {Averbukh}}]{Ilya2025PRA}%
  \BibitemOpen
  \bibfield  {author} {\bibinfo {author} {\bibfnamefont {Y.-W.}\ \bibnamefont {Mao}}, \bibinfo {author} {\bibfnamefont {I.}~\bibnamefont {Tutunnikov}}, \bibinfo {author} {\bibfnamefont {R.~V.}\ \bibnamefont {Krems}}, \ and\ \bibinfo {author} {\bibfnamefont {I.~S.}\ \bibnamefont {Averbukh}},\ }\href {\doibase 10.1103/bv4c-6q9k} {\bibfield  {journal} {\bibinfo  {journal} {Phys. Rev. A}\ }\textbf {\bibinfo {volume} {112}},\ \bibinfo {pages} {L040601} (\bibinfo {year} {2025})}\BibitemShut {NoStop}%
\bibitem [{\citenamefont {Phillips}(1972)}]{Phillips1972JLTP}%
  \BibitemOpen
  \bibfield  {author} {\bibinfo {author} {\bibfnamefont {W.~A.}\ \bibnamefont {Phillips}},\ }\href {\doibase 10.1007/BF00660072} {\bibfield  {journal} {\bibinfo  {journal} {Journal of Low Temperature Physics}\ }\textbf {\bibinfo {volume} {7}},\ \bibinfo {pages} {351} (\bibinfo {year} {1972})}\BibitemShut {NoStop}%
\bibitem [{\citenamefont {J{\"a}ckle}(1972)}]{Jackle1972ZP}%
  \BibitemOpen
  \bibfield  {author} {\bibinfo {author} {\bibfnamefont {J.}~\bibnamefont {J{\"a}ckle}},\ }\href {\doibase 10.1007/BF01401204} {\bibfield  {journal} {\bibinfo  {journal} {Zeitschrift f{\"u}r Physik}\ }\textbf {\bibinfo {volume} {257}},\ \bibinfo {pages} {212} (\bibinfo {year} {1972})}\BibitemShut {NoStop}%
\bibitem [{\citenamefont {Berret}\ and\ \citenamefont {Meißner}(1988)}]{Berret1988ZPBCM}%
  \BibitemOpen
  \bibfield  {author} {\bibinfo {author} {\bibfnamefont {J.~F.}\ \bibnamefont {Berret}}\ and\ \bibinfo {author} {\bibfnamefont {M.}~\bibnamefont {Meißner}},\ }\href {\doibase 10.1007/BF01320540} {\bibfield  {journal} {\bibinfo  {journal} {Zeitschrift für Physik B Condensed Matter}\ }\textbf {\bibinfo {volume} {70}},\ \bibinfo {pages} {65} (\bibinfo {year} {1988})}\BibitemShut {NoStop}%
\bibitem [{\citenamefont {Hunklinger}\ \emph {et~al.}(1975)\citenamefont {Hunklinger}, \citenamefont {Piché}, \citenamefont {Lasjaunias},\ and\ \citenamefont {Dransfeld}}]{Hunklinger1975JPCSSP}%
  \BibitemOpen
  \bibfield  {author} {\bibinfo {author} {\bibfnamefont {S.}~\bibnamefont {Hunklinger}}, \bibinfo {author} {\bibfnamefont {L.}~\bibnamefont {Piché}}, \bibinfo {author} {\bibfnamefont {J.~C.}\ \bibnamefont {Lasjaunias}}, \ and\ \bibinfo {author} {\bibfnamefont {K.}~\bibnamefont {Dransfeld}},\ }\href {\doibase 10.1088/0022-3719/8/21/001} {\bibfield  {journal} {\bibinfo  {journal} {Journal of Physics C: Solid State Physics}\ }\textbf {\bibinfo {volume} {8}},\ \bibinfo {pages} {L423} (\bibinfo {year} {1975})}\BibitemShut {NoStop}%
\bibitem [{\citenamefont {Smolyakov}\ and\ \citenamefont {Kha{\u\i}movich}(1982)}]{Smolyakov1982SPU}%
  \BibitemOpen
  \bibfield  {author} {\bibinfo {author} {\bibfnamefont {B.~P.}\ \bibnamefont {Smolyakov}}\ and\ \bibinfo {author} {\bibfnamefont {E.~P.}\ \bibnamefont {Kha{\u\i}movich}},\ }\href {\doibase 10.1070/PU1982v025n02ABEH004500} {\bibfield  {journal} {\bibinfo  {journal} {Soviet Physics Uspekhi}\ }\textbf {\bibinfo {volume} {25}},\ \bibinfo {pages} {102} (\bibinfo {year} {1982})}\BibitemShut {NoStop}%
\bibitem [{\citenamefont {Hunklinger}\ and\ \citenamefont {Raychaudhuri}(1986)}]{Hunklinger1986PLTP}%
  \BibitemOpen
  \bibfield  {author} {\bibinfo {author} {\bibfnamefont {S.}~\bibnamefont {Hunklinger}}\ and\ \bibinfo {author} {\bibfnamefont {A.~K.}\ \bibnamefont {Raychaudhuri}},\ }in\ \href@noop {} {\emph {\bibinfo {booktitle} {Progress in Low Temperature Physics}}},\ \bibinfo {series} {Progress in Low Temperature Physics}, Vol.~\bibinfo {volume} {9},\ \bibinfo {editor} {edited by\ \bibinfo {editor} {\bibfnamefont {D.~F.}\ \bibnamefont {Brewer}}}\ (\bibinfo  {publisher} {Elsevier Science Publishers B.V.},\ \bibinfo {address} {Amsterdam},\ \bibinfo {year} {1986})\ pp.\ \bibinfo {pages} {265--344}\BibitemShut {NoStop}%
\bibitem [{\citenamefont {Phillips}(1981)}]{Phillips1981}%
  \BibitemOpen
  \bibinfo {editor} {\bibfnamefont {W.~A.}\ \bibnamefont {Phillips}},\ ed.,\ \href {\doibase 10.1007/978-3-642-81563-8} {\emph {\bibinfo {title} {Amorphous Solids: Low-Temperature Properties}}},\ \bibinfo {series} {Topics in Current Physics}, Vol.~\bibinfo {volume} {24}\ (\bibinfo  {publisher} {Springer-Verlag Berlin Heidelberg},\ \bibinfo {year} {1981})\BibitemShut {NoStop}%
\bibitem [{\citenamefont {Charbonneau}\ \emph {et~al.}(2016{\natexlab{b}})\citenamefont {Charbonneau}, \citenamefont {Corwin}, \citenamefont {Parisi}, \citenamefont {Poncet},\ and\ \citenamefont {Zamponi}}]{Charbonneau2016PRL}%
  \BibitemOpen
  \bibfield  {author} {\bibinfo {author} {\bibfnamefont {P.}~\bibnamefont {Charbonneau}}, \bibinfo {author} {\bibfnamefont {E.~I.}\ \bibnamefont {Corwin}}, \bibinfo {author} {\bibfnamefont {G.}~\bibnamefont {Parisi}}, \bibinfo {author} {\bibfnamefont {A.}~\bibnamefont {Poncet}}, \ and\ \bibinfo {author} {\bibfnamefont {F.}~\bibnamefont {Zamponi}},\ }\href {\doibase 10.1103/PhysRevLett.117.045503} {\bibfield  {journal} {\bibinfo  {journal} {Phys. Rev. Lett.}\ }\textbf {\bibinfo {volume} {117}},\ \bibinfo {pages} {045503} (\bibinfo {year} {2016}{\natexlab{b}})}\BibitemShut {NoStop}%
\bibitem [{\citenamefont {Keim}\ \emph {et~al.}(2019)\citenamefont {Keim}, \citenamefont {Paulsen}, \citenamefont {Zeravcic}, \citenamefont {Sastry},\ and\ \citenamefont {Nagel}}]{Nagel2019RMP}%
  \BibitemOpen
  \bibfield  {author} {\bibinfo {author} {\bibfnamefont {N.~C.}\ \bibnamefont {Keim}}, \bibinfo {author} {\bibfnamefont {J.~D.}\ \bibnamefont {Paulsen}}, \bibinfo {author} {\bibfnamefont {Z.}~\bibnamefont {Zeravcic}}, \bibinfo {author} {\bibfnamefont {S.}~\bibnamefont {Sastry}}, \ and\ \bibinfo {author} {\bibfnamefont {S.~R.}\ \bibnamefont {Nagel}},\ }\href {\doibase 10.1103/RevModPhys.91.035002} {\bibfield  {journal} {\bibinfo  {journal} {Rev. Mod. Phys.}\ }\textbf {\bibinfo {volume} {91}},\ \bibinfo {pages} {035002} (\bibinfo {year} {2019})}\BibitemShut {NoStop}%
\bibitem [{\citenamefont {Kartashov}\ \emph {et~al.}(2011)\citenamefont {Kartashov}, \citenamefont {Malomed},\ and\ \citenamefont {Torner}}]{Torner2011RMP}%
  \BibitemOpen
  \bibfield  {author} {\bibinfo {author} {\bibfnamefont {Y.~V.}\ \bibnamefont {Kartashov}}, \bibinfo {author} {\bibfnamefont {B.~A.}\ \bibnamefont {Malomed}}, \ and\ \bibinfo {author} {\bibfnamefont {L.}~\bibnamefont {Torner}},\ }\href {\doibase 10.1103/RevModPhys.83.247} {\bibfield  {journal} {\bibinfo  {journal} {Rev. Mod. Phys.}\ }\textbf {\bibinfo {volume} {83}},\ \bibinfo {pages} {247} (\bibinfo {year} {2011})}\BibitemShut {NoStop}%
\bibitem [{\citenamefont {Ma}\ \emph {et~al.}(2023)\citenamefont {Ma}, \citenamefont {Tang}, \citenamefont {Shi}, \citenamefont {Wu}, \citenamefont {Yang}, \citenamefont {Zhou}, \citenamefont {Yao},\ and\ \citenamefont {Li}}]{ma2023PRL}%
  \BibitemOpen
  \bibfield  {author} {\bibinfo {author} {\bibfnamefont {F.}~\bibnamefont {Ma}}, \bibinfo {author} {\bibfnamefont {Z.}~\bibnamefont {Tang}}, \bibinfo {author} {\bibfnamefont {X.}~\bibnamefont {Shi}}, \bibinfo {author} {\bibfnamefont {Y.}~\bibnamefont {Wu}}, \bibinfo {author} {\bibfnamefont {J.}~\bibnamefont {Yang}}, \bibinfo {author} {\bibfnamefont {D.}~\bibnamefont {Zhou}}, \bibinfo {author} {\bibfnamefont {Y.}~\bibnamefont {Yao}}, \ and\ \bibinfo {author} {\bibfnamefont {F.}~\bibnamefont {Li}},\ }\href {\doibase 10.1103/PhysRevLett.131.046101} {\bibfield  {journal} {\bibinfo  {journal} {Phys. Rev. Lett.}\ }\textbf {\bibinfo {volume} {131}},\ \bibinfo {pages} {046101} (\bibinfo {year} {2023})}\BibitemShut {NoStop}%
\bibitem [{\citenamefont {Beltukov}\ and\ \citenamefont {Parshin}(2010)}]{BELTUKOV2010SSC}%
  \BibitemOpen
  \bibfield  {author} {\bibinfo {author} {\bibfnamefont {Y.}~\bibnamefont {Beltukov}}\ and\ \bibinfo {author} {\bibfnamefont {D.}~\bibnamefont {Parshin}},\ }\href {\doibase https://doi.org/10.1016/j.ssc.2009.11.026} {\bibfield  {journal} {\bibinfo  {journal} {Solid State Communications}\ }\textbf {\bibinfo {volume} {150}},\ \bibinfo {pages} {316} (\bibinfo {year} {2010})}\BibitemShut {NoStop}%
\bibitem [{\citenamefont {Khomenko}\ \emph {et~al.}(2020)\citenamefont {Khomenko}, \citenamefont {Scalliet}, \citenamefont {Berthier}, \citenamefont {Reichman},\ and\ \citenamefont {Zamponi}}]{Zamponi2020PRL}%
  \BibitemOpen
  \bibfield  {author} {\bibinfo {author} {\bibfnamefont {D.}~\bibnamefont {Khomenko}}, \bibinfo {author} {\bibfnamefont {C.}~\bibnamefont {Scalliet}}, \bibinfo {author} {\bibfnamefont {L.}~\bibnamefont {Berthier}}, \bibinfo {author} {\bibfnamefont {D.~R.}\ \bibnamefont {Reichman}}, \ and\ \bibinfo {author} {\bibfnamefont {F.}~\bibnamefont {Zamponi}},\ }\href {\doibase 10.1103/PhysRevLett.124.225901} {\bibfield  {journal} {\bibinfo  {journal} {Phys. Rev. Lett.}\ }\textbf {\bibinfo {volume} {124}},\ \bibinfo {pages} {225901} (\bibinfo {year} {2020})}\BibitemShut {NoStop}%
\bibitem [{\citenamefont {Schenzle}\ \emph {et~al.}(1976)\citenamefont {Schenzle}, \citenamefont {Grossman},\ and\ \citenamefont {Brewer}}]{Richard1976PRA}%
  \BibitemOpen
  \bibfield  {author} {\bibinfo {author} {\bibfnamefont {A.}~\bibnamefont {Schenzle}}, \bibinfo {author} {\bibfnamefont {S.}~\bibnamefont {Grossman}}, \ and\ \bibinfo {author} {\bibfnamefont {R.~G.}\ \bibnamefont {Brewer}},\ }\href {\doibase 10.1103/PhysRevA.13.1891} {\bibfield  {journal} {\bibinfo  {journal} {Phys. Rev. A}\ }\textbf {\bibinfo {volume} {13}},\ \bibinfo {pages} {1891} (\bibinfo {year} {1976})}\BibitemShut {NoStop}%
\bibitem [{\citenamefont {Burton}\ and\ \citenamefont {Nagel}(2016)}]{Burton2016PRE}%
  \BibitemOpen
  \bibfield  {author} {\bibinfo {author} {\bibfnamefont {J.~C.}\ \bibnamefont {Burton}}\ and\ \bibinfo {author} {\bibfnamefont {S.~R.}\ \bibnamefont {Nagel}},\ }\href {\doibase 10.1103/PhysRevE.93.032905} {\bibfield  {journal} {\bibinfo  {journal} {Phys. Rev. E}\ }\textbf {\bibinfo {volume} {93}},\ \bibinfo {pages} {032905} (\bibinfo {year} {2016})}\BibitemShut {NoStop}%
\bibitem [{\citenamefont {Li}\ \emph {et~al.}(2012)\citenamefont {Li}, \citenamefont {Ngo}, \citenamefont {Yang},\ and\ \citenamefont {Daraio}}]{li2012ARPL}%
  \BibitemOpen
  \bibfield  {author} {\bibinfo {author} {\bibfnamefont {F.}~\bibnamefont {Li}}, \bibinfo {author} {\bibfnamefont {D.}~\bibnamefont {Ngo}}, \bibinfo {author} {\bibfnamefont {J.}~\bibnamefont {Yang}}, \ and\ \bibinfo {author} {\bibfnamefont {C.}~\bibnamefont {Daraio}},\ }\href {\doibase 10.1063/1.4762832} {\bibfield  {journal} {\bibinfo  {journal} {Applied Physics Letters}\ }\textbf {\bibinfo {volume} {101}},\ \bibinfo {pages} {171903} (\bibinfo {year} {2012})}\BibitemShut {NoStop}%
\bibitem [{\citenamefont {Peskin}\ and\ \citenamefont {Schroeder}(2018)}]{peskin2018introduction}%
  \BibitemOpen
  \bibfield  {author} {\bibinfo {author} {\bibfnamefont {M.~E.}\ \bibnamefont {Peskin}}\ and\ \bibinfo {author} {\bibfnamefont {D.~V.}\ \bibnamefont {Schroeder}},\ }\href@noop {} {\emph {\bibinfo {title} {An Introduction To Quantum Field Theory}}},\ \bibinfo {edition} {1st}\ ed.\ (\bibinfo  {publisher} {CRC Press},\ \bibinfo {address} {Boca Raton, FL},\ \bibinfo {year} {2018})\BibitemShut {NoStop}%
\bibitem [{\citenamefont {{Joffrin, J.}}\ and\ \citenamefont {{Levelut, A.}}(1975)}]{Joffrin1975JDP}%
  \BibitemOpen
  \bibfield  {author} {\bibinfo {author} {\bibnamefont {{Joffrin, J.}}}\ and\ \bibinfo {author} {\bibnamefont {{Levelut, A.}}},\ }\href {\doibase 10.1051/jphys:01975003609081100} {\bibfield  {journal} {\bibinfo  {journal} {J. Phys. France}\ }\textbf {\bibinfo {volume} {36}},\ \bibinfo {pages} {811} (\bibinfo {year} {1975})}\BibitemShut {NoStop}%
\bibitem [{\citenamefont {Busiello}\ \emph {et~al.}(2010)\citenamefont {Busiello}, \citenamefont {Saburova}, \citenamefont {Gazeeva},\ and\ \citenamefont {Marenkov}}]{Busiello2010RPJ}%
  \BibitemOpen
  \bibfield  {author} {\bibinfo {author} {\bibfnamefont {G.}~\bibnamefont {Busiello}}, \bibinfo {author} {\bibfnamefont {R.~V.}\ \bibnamefont {Saburova}}, \bibinfo {author} {\bibfnamefont {E.~V.}\ \bibnamefont {Gazeeva}}, \ and\ \bibinfo {author} {\bibfnamefont {M.~S.}\ \bibnamefont {Marenkov}},\ }\href {\doibase 10.1007/s11182-010-9424-z} {\bibfield  {journal} {\bibinfo  {journal} {Russian Physics Journal}\ }\textbf {\bibinfo {volume} {53}},\ \bibinfo {pages} {336} (\bibinfo {year} {2010})}\BibitemShut {NoStop}%
\bibitem [{\citenamefont {Gramila}\ \emph {et~al.}(1993)\citenamefont {Gramila}, \citenamefont {Eisenstein}, \citenamefont {MacDonald}, \citenamefont {Pfeiffer},\ and\ \citenamefont {West}}]{West1993PRB}%
  \BibitemOpen
  \bibfield  {author} {\bibinfo {author} {\bibfnamefont {T.~J.}\ \bibnamefont {Gramila}}, \bibinfo {author} {\bibfnamefont {J.~P.}\ \bibnamefont {Eisenstein}}, \bibinfo {author} {\bibfnamefont {A.~H.}\ \bibnamefont {MacDonald}}, \bibinfo {author} {\bibfnamefont {L.~N.}\ \bibnamefont {Pfeiffer}}, \ and\ \bibinfo {author} {\bibfnamefont {K.~W.}\ \bibnamefont {West}},\ }\href {\doibase 10.1103/PhysRevB.47.12957} {\bibfield  {journal} {\bibinfo  {journal} {Phys. Rev. B}\ }\textbf {\bibinfo {volume} {47}},\ \bibinfo {pages} {12957} (\bibinfo {year} {1993})}\BibitemShut {NoStop}%
\bibitem [{\citenamefont {Burin}(2004)}]{Burin2004JLTP}%
  \BibitemOpen
  \bibfield  {author} {\bibinfo {author} {\bibfnamefont {A.}~\bibnamefont {Burin}},\ }\href {\doibase 10.1023/B:JOLT.0000049053.73309.96} {\bibfield  {journal} {\bibinfo  {journal} {Journal of Low Temperature Physics}\ }\textbf {\bibinfo {volume} {189}},\ \bibinfo {pages} {137} (\bibinfo {year} {2004})}\BibitemShut {NoStop}%
\bibitem [{\citenamefont {Van~Horn}\ and\ \citenamefont {Salpeter}(1967)}]{Van1967PR}%
  \BibitemOpen
  \bibfield  {author} {\bibinfo {author} {\bibfnamefont {H.~M.}\ \bibnamefont {Van~Horn}}\ and\ \bibinfo {author} {\bibfnamefont {E.~E.}\ \bibnamefont {Salpeter}},\ }\href {\doibase 10.1103/PhysRev.157.751} {\bibfield  {journal} {\bibinfo  {journal} {Phys. Rev.}\ }\textbf {\bibinfo {volume} {157}},\ \bibinfo {pages} {751} (\bibinfo {year} {1967})}\BibitemShut {NoStop}%
\bibitem [{\citenamefont {Vakakis}\ \emph {et~al.}(2001)\citenamefont {Vakakis}, \citenamefont {Manevitch}, \citenamefont {Mikhlin}, \citenamefont {Pilipchuk},\ and\ \citenamefont {Zevin}}]{vakakis2001normal}%
  \BibitemOpen
  \bibfield  {author} {\bibinfo {author} {\bibfnamefont {A.~F.}\ \bibnamefont {Vakakis}}, \bibinfo {author} {\bibfnamefont {L.~I.}\ \bibnamefont {Manevitch}}, \bibinfo {author} {\bibfnamefont {Y.~V.}\ \bibnamefont {Mikhlin}}, \bibinfo {author} {\bibfnamefont {V.~N.}\ \bibnamefont {Pilipchuk}}, \ and\ \bibinfo {author} {\bibfnamefont {A.~A.}\ \bibnamefont {Zevin}},\ }\href@noop {} {\emph {\bibinfo {title} {Normal modes and localization in nonlinear systems}}}\ (\bibinfo  {publisher} {Springer},\ \bibinfo {year} {2001})\BibitemShut {NoStop}%
\bibitem [{\citenamefont {Li}\ and\ \citenamefont {Wan}(2023)}]{Yuan2023PRB}%
  \BibitemOpen
  \bibfield  {author} {\bibinfo {author} {\bibfnamefont {Z.-L.}\ \bibnamefont {Li}}\ and\ \bibinfo {author} {\bibfnamefont {Y.}~\bibnamefont {Wan}},\ }\href {\doibase 10.1103/PhysRevB.108.165151} {\bibfield  {journal} {\bibinfo  {journal} {Phys. Rev. B}\ }\textbf {\bibinfo {volume} {108}},\ \bibinfo {pages} {165151} (\bibinfo {year} {2023})}\BibitemShut {NoStop}%
\bibitem [{\citenamefont {Landau}\ \emph {et~al.}(1986)\citenamefont {Landau}, \citenamefont {Pitaevskii}, \citenamefont {Kosevich},\ and\ \citenamefont {Lifshitz}}]{landau1986theory}%
  \BibitemOpen
  \bibfield  {author} {\bibinfo {author} {\bibfnamefont {L.~D.}\ \bibnamefont {Landau}}, \bibinfo {author} {\bibfnamefont {L.~P.}\ \bibnamefont {Pitaevskii}}, \bibinfo {author} {\bibfnamefont {A.~M.}\ \bibnamefont {Kosevich}}, \ and\ \bibinfo {author} {\bibfnamefont {E.~M.}\ \bibnamefont {Lifshitz}},\ }\href@noop {} {\emph {\bibinfo {title} {Theory of Elasticity: Volume 7}}},\ \bibinfo {edition} {3rd}\ ed.\ (\bibinfo  {publisher} {Butterworth-Heinemann},\ \bibinfo {year} {1986})\BibitemShut {NoStop}%
\bibitem [{\citenamefont {Kubo}(1962)}]{Kubo1962JPSJ}%
  \BibitemOpen
  \bibfield  {author} {\bibinfo {author} {\bibfnamefont {R.}~\bibnamefont {Kubo}},\ }\href {\doibase 10.1143/JPSJ.17.1100} {\bibfield  {journal} {\bibinfo  {journal} {Journal of the Physical Society of Japan}\ }\textbf {\bibinfo {volume} {17}},\ \bibinfo {pages} {1100} (\bibinfo {year} {1962})}\BibitemShut {NoStop}%
\bibitem [{\citenamefont {Zhao}\ \emph {et~al.}(2017)\citenamefont {Zhao}, \citenamefont {Tan}, \citenamefont {Yu}, \citenamefont {Zhu},\ and\ \citenamefont {Yu}}]{Yu2017PRA}%
  \BibitemOpen
  \bibfield  {author} {\bibinfo {author} {\bibfnamefont {P.}~\bibnamefont {Zhao}}, \bibinfo {author} {\bibfnamefont {X.}~\bibnamefont {Tan}}, \bibinfo {author} {\bibfnamefont {H.}~\bibnamefont {Yu}}, \bibinfo {author} {\bibfnamefont {S.-L.}\ \bibnamefont {Zhu}}, \ and\ \bibinfo {author} {\bibfnamefont {Y.}~\bibnamefont {Yu}},\ }\href {\doibase 10.1103/PhysRevA.96.043833} {\bibfield  {journal} {\bibinfo  {journal} {Phys. Rev. A}\ }\textbf {\bibinfo {volume} {96}},\ \bibinfo {pages} {043833} (\bibinfo {year} {2017})}\BibitemShut {NoStop}%
\bibitem [{\citenamefont {Zhou}(2019)}]{zhou2019universal}%
  \BibitemOpen
  \bibfield  {author} {\bibinfo {author} {\bibfnamefont {D.}~\bibnamefont {Zhou}},\ }\href@noop {} {\bibfield  {journal} {\bibinfo  {journal} {Journal of Physics: Condensed Matter}\ }\textbf {\bibinfo {volume} {32}},\ \bibinfo {pages} {055704} (\bibinfo {year} {2019})}\BibitemShut {NoStop}%
\bibitem [{\citenamefont {Albuquerque}\ \emph {et~al.}(2024)\citenamefont {Albuquerque}, \citenamefont {V\"olkel},\ and\ \citenamefont {Kokkotas}}]{PhysRevD.109.096014}%
  \BibitemOpen
  \bibfield  {author} {\bibinfo {author} {\bibfnamefont {S.}~\bibnamefont {Albuquerque}}, \bibinfo {author} {\bibfnamefont {S.~H.}\ \bibnamefont {V\"olkel}}, \ and\ \bibinfo {author} {\bibfnamefont {K.~D.}\ \bibnamefont {Kokkotas}},\ }\href {\doibase 10.1103/PhysRevD.109.096014} {\bibfield  {journal} {\bibinfo  {journal} {Phys. Rev. D}\ }\textbf {\bibinfo {volume} {109}},\ \bibinfo {pages} {096014} (\bibinfo {year} {2024})}\BibitemShut {NoStop}%
\bibitem [{\citenamefont {Fronk}\ and\ \citenamefont {Leamy}(2019)}]{Leamy2019PRE}%
  \BibitemOpen
  \bibfield  {author} {\bibinfo {author} {\bibfnamefont {M.~D.}\ \bibnamefont {Fronk}}\ and\ \bibinfo {author} {\bibfnamefont {M.~J.}\ \bibnamefont {Leamy}},\ }\href {\doibase 10.1103/PhysRevE.100.032213} {\bibfield  {journal} {\bibinfo  {journal} {Phys. Rev. E}\ }\textbf {\bibinfo {volume} {100}},\ \bibinfo {pages} {032213} (\bibinfo {year} {2019})}\BibitemShut {NoStop}%
\bibitem [{\citenamefont {{Zhou}}\ \emph {et~al.}(2022)\citenamefont {{Zhou}}, \citenamefont {{Zeb Rocklin}}, \citenamefont {{Leamy}},\ and\ \citenamefont {{Yao}}}]{Zhou2022NC}%
  \BibitemOpen
  \bibfield  {author} {\bibinfo {author} {\bibfnamefont {D.}~\bibnamefont {{Zhou}}}, \bibinfo {author} {\bibfnamefont {D.}~\bibnamefont {{Zeb Rocklin}}}, \bibinfo {author} {\bibfnamefont {M.}~\bibnamefont {{Leamy}}}, \ and\ \bibinfo {author} {\bibfnamefont {Y.}~\bibnamefont {{Yao}}},\ }\href {\doibase 10.1038/s41467-022-31084-y} {\bibfield  {journal} {\bibinfo  {journal} {Nature Communications}\ }\textbf {\bibinfo {volume} {13}},\ \bibinfo {pages} {3379} (\bibinfo {year} {2022})}\BibitemShut {NoStop}%
\bibitem [{\citenamefont {Rosa}\ \emph {et~al.}(2023)\citenamefont {Rosa}, \citenamefont {Leamy},\ and\ \citenamefont {Ruzzene}}]{Rosa2023NJP}%
  \BibitemOpen
  \bibfield  {author} {\bibinfo {author} {\bibfnamefont {M.~I.~N.}\ \bibnamefont {Rosa}}, \bibinfo {author} {\bibfnamefont {M.~J.}\ \bibnamefont {Leamy}}, \ and\ \bibinfo {author} {\bibfnamefont {M.}~\bibnamefont {Ruzzene}},\ }\href {\doibase 10.1088/1367-2630/ad016f} {\bibfield  {journal} {\bibinfo  {journal} {New Journal of Physics}\ }\textbf {\bibinfo {volume} {25}},\ \bibinfo {pages} {103053} (\bibinfo {year} {2023})}\BibitemShut {NoStop}%
\bibitem [{\citenamefont {Pal}\ \emph {et~al.}(2018)\citenamefont {Pal}, \citenamefont {Vila}, \citenamefont {Leamy},\ and\ \citenamefont {Ruzzene}}]{Ruzzene2018PRE}%
  \BibitemOpen
  \bibfield  {author} {\bibinfo {author} {\bibfnamefont {R.~K.}\ \bibnamefont {Pal}}, \bibinfo {author} {\bibfnamefont {J.}~\bibnamefont {Vila}}, \bibinfo {author} {\bibfnamefont {M.}~\bibnamefont {Leamy}}, \ and\ \bibinfo {author} {\bibfnamefont {M.}~\bibnamefont {Ruzzene}},\ }\href {\doibase 10.1103/PhysRevE.97.032209} {\bibfield  {journal} {\bibinfo  {journal} {Phys. Rev. E}\ }\textbf {\bibinfo {volume} {97}},\ \bibinfo {pages} {032209} (\bibinfo {year} {2018})}\BibitemShut {NoStop}%
\bibitem [{\citenamefont {Zhou}(2024)}]{Zhou2024NJP}%
  \BibitemOpen
  \bibfield  {author} {\bibinfo {author} {\bibfnamefont {D.}~\bibnamefont {Zhou}},\ }\href {\doibase 10.1088/1367-2630/ad5b14} {\bibfield  {journal} {\bibinfo  {journal} {New Journal of Physics}\ }\textbf {\bibinfo {volume} {26}},\ \bibinfo {pages} {073009} (\bibinfo {year} {2024})}\BibitemShut {NoStop}%
\bibitem [{\citenamefont {Romeo}\ and\ \citenamefont {Ruzzene}(2012)}]{Romeo2012}%
  \BibitemOpen
  \bibinfo {editor} {\bibfnamefont {F.}~\bibnamefont {Romeo}}\ and\ \bibinfo {editor} {\bibfnamefont {M.}~\bibnamefont {Ruzzene}},\ eds.,\ \href {\doibase 10.1007/978-3-7091-1309-7} {\emph {\bibinfo {title} {Wave Propagation in Linear and Nonlinear Periodic Media: Analysis and Applications}}},\ \bibinfo {series} {CISM International Centre for Mechanical Sciences}, Vol.\ \bibinfo {volume} {540}\ (\bibinfo  {publisher} {Springer Vienna},\ \bibinfo {year} {2012})\BibitemShut {NoStop}%
\bibitem [{\citenamefont {Zhou}\ \emph {et~al.}(2020)\citenamefont {Zhou}, \citenamefont {Ma}, \citenamefont {Sun}, \citenamefont {Gonella},\ and\ \citenamefont {Mao}}]{Zhou2020PRB}%
  \BibitemOpen
  \bibfield  {author} {\bibinfo {author} {\bibfnamefont {D.}~\bibnamefont {Zhou}}, \bibinfo {author} {\bibfnamefont {J.}~\bibnamefont {Ma}}, \bibinfo {author} {\bibfnamefont {K.}~\bibnamefont {Sun}}, \bibinfo {author} {\bibfnamefont {S.}~\bibnamefont {Gonella}}, \ and\ \bibinfo {author} {\bibfnamefont {X.}~\bibnamefont {Mao}},\ }\href {\doibase 10.1103/PhysRevB.101.104106} {\bibfield  {journal} {\bibinfo  {journal} {Phys. Rev. B}\ }\textbf {\bibinfo {volume} {101}},\ \bibinfo {pages} {104106} (\bibinfo {year} {2020})}\BibitemShut {NoStop}%
\bibitem [{\citenamefont {Zhou}\ and\ \citenamefont {Zhang}(2005)}]{ZhouHaiJun2005PRL}%
  \BibitemOpen
  \bibfield  {author} {\bibinfo {author} {\bibfnamefont {H.}~\bibnamefont {Zhou}}\ and\ \bibinfo {author} {\bibfnamefont {Y.}~\bibnamefont {Zhang}},\ }\href {\doibase 10.1103/PhysRevLett.94.028104} {\bibfield  {journal} {\bibinfo  {journal} {Phys. Rev. Lett.}\ }\textbf {\bibinfo {volume} {94}},\ \bibinfo {pages} {028104} (\bibinfo {year} {2005})}\BibitemShut {NoStop}%
\bibitem [{\citenamefont {Wu}\ \emph {et~al.}(2025)\citenamefont {Wu}, \citenamefont {Barrat},\ and\ \citenamefont {Kob}}]{Wu2025NC}%
  \BibitemOpen
  \bibfield  {author} {\bibinfo {author} {\bibfnamefont {Z.~W.}\ \bibnamefont {Wu}}, \bibinfo {author} {\bibfnamefont {J.-L.}\ \bibnamefont {Barrat}}, \ and\ \bibinfo {author} {\bibfnamefont {W.}~\bibnamefont {Kob}},\ }\href {\doibase 10.1038/s41467-025-66923-1} {\bibfield  {journal} {\bibinfo  {journal} {Nature Communications}\ }\textbf {\bibinfo {volume} {16}},\ \bibinfo {pages} {66923} (\bibinfo {year} {2025})}\BibitemShut {NoStop}%
\bibitem [{\citenamefont {Zhou}\ \emph {et~al.}(2018)\citenamefont {Zhou}, \citenamefont {Zhang},\ and\ \citenamefont {Mao}}]{Zhou2018PRL}%
  \BibitemOpen
  \bibfield  {author} {\bibinfo {author} {\bibfnamefont {D.}~\bibnamefont {Zhou}}, \bibinfo {author} {\bibfnamefont {L.}~\bibnamefont {Zhang}}, \ and\ \bibinfo {author} {\bibfnamefont {X.}~\bibnamefont {Mao}},\ }\href {\doibase 10.1103/PhysRevLett.120.068003} {\bibfield  {journal} {\bibinfo  {journal} {Phys. Rev. Lett.}\ }\textbf {\bibinfo {volume} {120}},\ \bibinfo {pages} {068003} (\bibinfo {year} {2018})}\BibitemShut {NoStop}%
\bibitem [{\citenamefont {Tang}\ \emph {et~al.}(2024)\citenamefont {Tang}, \citenamefont {Ma}, \citenamefont {Li}, \citenamefont {Yao},\ and\ \citenamefont {Zhou}}]{Tang2024PRL}%
  \BibitemOpen
  \bibfield  {author} {\bibinfo {author} {\bibfnamefont {Z.}~\bibnamefont {Tang}}, \bibinfo {author} {\bibfnamefont {F.}~\bibnamefont {Ma}}, \bibinfo {author} {\bibfnamefont {F.}~\bibnamefont {Li}}, \bibinfo {author} {\bibfnamefont {Y.}~\bibnamefont {Yao}}, \ and\ \bibinfo {author} {\bibfnamefont {D.}~\bibnamefont {Zhou}},\ }\href {\doibase 10.1103/PhysRevLett.133.106101} {\bibfield  {journal} {\bibinfo  {journal} {Phys. Rev. Lett.}\ }\textbf {\bibinfo {volume} {133}},\ \bibinfo {pages} {106101} (\bibinfo {year} {2024})}\BibitemShut {NoStop}%
\bibitem [{\citenamefont {Mims}(1966)}]{PhysRev.141.499}%
  \BibitemOpen
  \bibfield  {author} {\bibinfo {author} {\bibfnamefont {W.~B.}\ \bibnamefont {Mims}},\ }\href {\doibase 10.1103/PhysRev.141.499} {\bibfield  {journal} {\bibinfo  {journal} {Phys. Rev.}\ }\textbf {\bibinfo {volume} {141}},\ \bibinfo {pages} {499} (\bibinfo {year} {1966})}\BibitemShut {NoStop}%
\bibitem [{\citenamefont {Cooke}\ \emph {et~al.}(1982)\citenamefont {Cooke}, \citenamefont {Chaplin},\ and\ \citenamefont {Wilson}}]{Cooke1982JPCSSP}%
  \BibitemOpen
  \bibfield  {author} {\bibinfo {author} {\bibfnamefont {P.}~\bibnamefont {Cooke}}, \bibinfo {author} {\bibfnamefont {D.~H.}\ \bibnamefont {Chaplin}}, \ and\ \bibinfo {author} {\bibfnamefont {G.~V.~H.}\ \bibnamefont {Wilson}},\ }\href {\doibase 10.1088/0022-3719/15/34/015} {\bibfield  {journal} {\bibinfo  {journal} {Journal of Physics C: Solid State Physics}\ }\textbf {\bibinfo {volume} {15}},\ \bibinfo {pages} {7031} (\bibinfo {year} {1982})}\BibitemShut {NoStop}%
\bibitem [{\citenamefont {Cho}\ \emph {et~al.}(2019)\citenamefont {Cho} \emph {et~al.}}]{cho2019JOSA}%
  \BibitemOpen
  \bibfield  {author} {\bibinfo {author} {\bibfnamefont {M.}~\bibnamefont {Cho}} \emph {et~al.},\ }\href@noop {} {\bibfield  {journal} {\bibinfo  {journal} {Journal of the Optical Society of America B}\ }\textbf {\bibinfo {volume} {36}},\ \bibinfo {pages} {223} (\bibinfo {year} {2019})}\BibitemShut {NoStop}%
\bibitem [{\citenamefont {Piryatinski}\ \emph {et~al.}(2004)\citenamefont {Piryatinski}, \citenamefont {Tretiak}, \citenamefont {Fenimore}, \citenamefont {Saxena}, \citenamefont {Martin},\ and\ \citenamefont {Bishop}}]{Bishop2004PRB}%
  \BibitemOpen
  \bibfield  {author} {\bibinfo {author} {\bibfnamefont {A.}~\bibnamefont {Piryatinski}}, \bibinfo {author} {\bibfnamefont {S.}~\bibnamefont {Tretiak}}, \bibinfo {author} {\bibfnamefont {P.~W.}\ \bibnamefont {Fenimore}}, \bibinfo {author} {\bibfnamefont {A.}~\bibnamefont {Saxena}}, \bibinfo {author} {\bibfnamefont {R.~L.}\ \bibnamefont {Martin}}, \ and\ \bibinfo {author} {\bibfnamefont {A.~R.}\ \bibnamefont {Bishop}},\ }\href {\doibase 10.1103/PhysRevB.70.161404} {\bibfield  {journal} {\bibinfo  {journal} {Phys. Rev. B}\ }\textbf {\bibinfo {volume} {70}},\ \bibinfo {pages} {161404} (\bibinfo {year} {2004})}\BibitemShut {NoStop}%
\bibitem [{\citenamefont {Langbein}\ and\ \citenamefont {Patton}(2005)}]{Patton2005PRL}%
  \BibitemOpen
  \bibfield  {author} {\bibinfo {author} {\bibfnamefont {W.}~\bibnamefont {Langbein}}\ and\ \bibinfo {author} {\bibfnamefont {B.}~\bibnamefont {Patton}},\ }\href {\doibase 10.1103/PhysRevLett.95.017403} {\bibfield  {journal} {\bibinfo  {journal} {Phys. Rev. Lett.}\ }\textbf {\bibinfo {volume} {95}},\ \bibinfo {pages} {017403} (\bibinfo {year} {2005})}\BibitemShut {NoStop}%
\bibitem [{\citenamefont {Berthier}\ and\ \citenamefont {Reichman}(2023)}]{Berthier2023ModernCS}%
  \BibitemOpen
  \bibfield  {author} {\bibinfo {author} {\bibfnamefont {L.}~\bibnamefont {Berthier}}\ and\ \bibinfo {author} {\bibfnamefont {D.~R.}\ \bibnamefont {Reichman}},\ }\href {\doibase 10.1038/s42254-022-00548-x} {\bibfield  {journal} {\bibinfo  {journal} {Nature Reviews Physics}\ }\textbf {\bibinfo {volume} {5}},\ \bibinfo {pages} {102} (\bibinfo {year} {2023})}\BibitemShut {NoStop}%
\bibitem [{\citenamefont {Yanagishima}\ \emph {et~al.}(2021)\citenamefont {Yanagishima}, \citenamefont {Russo}, \citenamefont {Dullens},\ and\ \citenamefont {Tanaka}}]{PhysRevLett.127.215501}%
  \BibitemOpen
  \bibfield  {author} {\bibinfo {author} {\bibfnamefont {T.}~\bibnamefont {Yanagishima}}, \bibinfo {author} {\bibfnamefont {J.}~\bibnamefont {Russo}}, \bibinfo {author} {\bibfnamefont {R.~P.~A.}\ \bibnamefont {Dullens}}, \ and\ \bibinfo {author} {\bibfnamefont {H.}~\bibnamefont {Tanaka}},\ }\href {\doibase 10.1103/PhysRevLett.127.215501} {\bibfield  {journal} {\bibinfo  {journal} {Phys. Rev. Lett.}\ }\textbf {\bibinfo {volume} {127}},\ \bibinfo {pages} {215501} (\bibinfo {year} {2021})}\BibitemShut {NoStop}%
\bibitem [{\citenamefont {Das}\ \emph {et~al.}(2005)\citenamefont {Das}, \citenamefont {Tang}, \citenamefont {Kim}, \citenamefont {Theissmann}, \citenamefont {Baier}, \citenamefont {Wang},\ and\ \citenamefont {Eckert}}]{PhysRevLett.94.205501}%
  \BibitemOpen
  \bibfield  {author} {\bibinfo {author} {\bibfnamefont {J.}~\bibnamefont {Das}}, \bibinfo {author} {\bibfnamefont {M.~B.}\ \bibnamefont {Tang}}, \bibinfo {author} {\bibfnamefont {K.~B.}\ \bibnamefont {Kim}}, \bibinfo {author} {\bibfnamefont {R.}~\bibnamefont {Theissmann}}, \bibinfo {author} {\bibfnamefont {F.}~\bibnamefont {Baier}}, \bibinfo {author} {\bibfnamefont {W.~H.}\ \bibnamefont {Wang}}, \ and\ \bibinfo {author} {\bibfnamefont {J.}~\bibnamefont {Eckert}},\ }\href {\doibase 10.1103/PhysRevLett.94.205501} {\bibfield  {journal} {\bibinfo  {journal} {Phys. Rev. Lett.}\ }\textbf {\bibinfo {volume} {94}},\ \bibinfo {pages} {205501} (\bibinfo {year} {2005})}\BibitemShut {NoStop}%
\bibitem [{\citenamefont {Monnier}\ \emph {et~al.}(2021)\citenamefont {Monnier}, \citenamefont {Colmenero}, \citenamefont {Wolf},\ and\ \citenamefont {Cangialosi}}]{PhysRevLett.126.118004}%
  \BibitemOpen
  \bibfield  {author} {\bibinfo {author} {\bibfnamefont {X.}~\bibnamefont {Monnier}}, \bibinfo {author} {\bibfnamefont {J.}~\bibnamefont {Colmenero}}, \bibinfo {author} {\bibfnamefont {M.}~\bibnamefont {Wolf}}, \ and\ \bibinfo {author} {\bibfnamefont {D.}~\bibnamefont {Cangialosi}},\ }\href {\doibase 10.1103/PhysRevLett.126.118004} {\bibfield  {journal} {\bibinfo  {journal} {Phys. Rev. Lett.}\ }\textbf {\bibinfo {volume} {126}},\ \bibinfo {pages} {118004} (\bibinfo {year} {2021})}\BibitemShut {NoStop}%
\bibitem [{\citenamefont {Yeh}\ \emph {et~al.}(2020)\citenamefont {Yeh}, \citenamefont {Ozawa}, \citenamefont {Miyazaki}, \citenamefont {Kawasaki},\ and\ \citenamefont {Berthier}}]{PhysRevLett.124.225502}%
  \BibitemOpen
  \bibfield  {author} {\bibinfo {author} {\bibfnamefont {W.-T.}\ \bibnamefont {Yeh}}, \bibinfo {author} {\bibfnamefont {M.}~\bibnamefont {Ozawa}}, \bibinfo {author} {\bibfnamefont {K.}~\bibnamefont {Miyazaki}}, \bibinfo {author} {\bibfnamefont {T.}~\bibnamefont {Kawasaki}}, \ and\ \bibinfo {author} {\bibfnamefont {L.}~\bibnamefont {Berthier}},\ }\href {\doibase 10.1103/PhysRevLett.124.225502} {\bibfield  {journal} {\bibinfo  {journal} {Phys. Rev. Lett.}\ }\textbf {\bibinfo {volume} {124}},\ \bibinfo {pages} {225502} (\bibinfo {year} {2020})}\BibitemShut {NoStop}%
\bibitem [{\citenamefont {Sun}\ \emph {et~al.}(2019)\citenamefont {Sun}, \citenamefont {Peng}, \citenamefont {Yang}, \citenamefont {Zhang}, \citenamefont {Yang}, \citenamefont {Wang}, \citenamefont {Ho},\ and\ \citenamefont {Yu}}]{PhysRevLett.123.105701}%
  \BibitemOpen
  \bibfield  {author} {\bibinfo {author} {\bibfnamefont {Y.}~\bibnamefont {Sun}}, \bibinfo {author} {\bibfnamefont {S.-X.}\ \bibnamefont {Peng}}, \bibinfo {author} {\bibfnamefont {Q.}~\bibnamefont {Yang}}, \bibinfo {author} {\bibfnamefont {F.}~\bibnamefont {Zhang}}, \bibinfo {author} {\bibfnamefont {M.-H.}\ \bibnamefont {Yang}}, \bibinfo {author} {\bibfnamefont {C.-Z.}\ \bibnamefont {Wang}}, \bibinfo {author} {\bibfnamefont {K.-M.}\ \bibnamefont {Ho}}, \ and\ \bibinfo {author} {\bibfnamefont {H.-B.}\ \bibnamefont {Yu}},\ }\href {\doibase 10.1103/PhysRevLett.123.105701} {\bibfield  {journal} {\bibinfo  {journal} {Phys. Rev. Lett.}\ }\textbf {\bibinfo {volume} {123}},\ \bibinfo {pages} {105701} (\bibinfo {year} {2019})}\BibitemShut {NoStop}%
\bibitem [{\citenamefont {Shen}(1976)}]{RevModPhys.48.1}%
  \BibitemOpen
  \bibfield  {author} {\bibinfo {author} {\bibfnamefont {Y.~R.}\ \bibnamefont {Shen}},\ }\href {\doibase 10.1103/RevModPhys.48.1} {\bibfield  {journal} {\bibinfo  {journal} {Rev. Mod. Phys.}\ }\textbf {\bibinfo {volume} {48}},\ \bibinfo {pages} {1} (\bibinfo {year} {1976})}\BibitemShut {NoStop}%
\bibitem [{\citenamefont {Franken}\ \emph {et~al.}(1961)\citenamefont {Franken}, \citenamefont {Hill}, \citenamefont {Peters},\ and\ \citenamefont {Weinreich}}]{PhysRevLett.7.118}%
  \BibitemOpen
  \bibfield  {author} {\bibinfo {author} {\bibfnamefont {P.~A.}\ \bibnamefont {Franken}}, \bibinfo {author} {\bibfnamefont {A.~E.}\ \bibnamefont {Hill}}, \bibinfo {author} {\bibfnamefont {C.~W.}\ \bibnamefont {Peters}}, \ and\ \bibinfo {author} {\bibfnamefont {G.}~\bibnamefont {Weinreich}},\ }\href {\doibase 10.1103/PhysRevLett.7.118} {\bibfield  {journal} {\bibinfo  {journal} {Phys. Rev. Lett.}\ }\textbf {\bibinfo {volume} {7}},\ \bibinfo {pages} {118} (\bibinfo {year} {1961})}\BibitemShut {NoStop}%
\bibitem [{\citenamefont {Xu}\ \emph {et~al.}(2019)\citenamefont {Xu}, \citenamefont {Morimoto},\ and\ \citenamefont {Moore}}]{PhysRevB.100.220501}%
  \BibitemOpen
  \bibfield  {author} {\bibinfo {author} {\bibfnamefont {T.}~\bibnamefont {Xu}}, \bibinfo {author} {\bibfnamefont {T.}~\bibnamefont {Morimoto}}, \ and\ \bibinfo {author} {\bibfnamefont {J.~E.}\ \bibnamefont {Moore}},\ }\href {\doibase 10.1103/PhysRevB.100.220501} {\bibfield  {journal} {\bibinfo  {journal} {Phys. Rev. B}\ }\textbf {\bibinfo {volume} {100}},\ \bibinfo {pages} {220501} (\bibinfo {year} {2019})}\BibitemShut {NoStop}%
\bibitem [{\citenamefont {Taghinejad}\ \emph {et~al.}(2020)\citenamefont {Taghinejad}, \citenamefont {Xu}, \citenamefont {Lee}, \citenamefont {Lian},\ and\ \citenamefont {Cai}}]{PhysRevLett.124.013901}%
  \BibitemOpen
  \bibfield  {author} {\bibinfo {author} {\bibfnamefont {M.}~\bibnamefont {Taghinejad}}, \bibinfo {author} {\bibfnamefont {Z.}~\bibnamefont {Xu}}, \bibinfo {author} {\bibfnamefont {K.-T.}\ \bibnamefont {Lee}}, \bibinfo {author} {\bibfnamefont {T.}~\bibnamefont {Lian}}, \ and\ \bibinfo {author} {\bibfnamefont {W.}~\bibnamefont {Cai}},\ }\href {\doibase 10.1103/PhysRevLett.124.013901} {\bibfield  {journal} {\bibinfo  {journal} {Phys. Rev. Lett.}\ }\textbf {\bibinfo {volume} {124}},\ \bibinfo {pages} {013901} (\bibinfo {year} {2020})}\BibitemShut {NoStop}%
\bibitem [{\citenamefont {Tang}\ \emph {et~al.}(2023)\citenamefont {Tang}, \citenamefont {Ma}, \citenamefont {Li}, \citenamefont {Guo},\ and\ \citenamefont {Zhou}}]{Tang2023FOP}%
  \BibitemOpen
  \bibfield  {author} {\bibinfo {author} {\bibfnamefont {J.}~\bibnamefont {Tang}}, \bibinfo {author} {\bibfnamefont {F.}~\bibnamefont {Ma}}, \bibinfo {author} {\bibfnamefont {F.}~\bibnamefont {Li}}, \bibinfo {author} {\bibfnamefont {H.}~\bibnamefont {Guo}}, \ and\ \bibinfo {author} {\bibfnamefont {D.}~\bibnamefont {Zhou}},\ }\href@noop {} {\bibfield  {journal} {\bibinfo  {journal} {Frontiers of Physics}\ }\textbf {\bibinfo {volume} {18}},\ \bibinfo {pages} {33311} (\bibinfo {year} {2023})}\BibitemShut {NoStop}%
\bibitem [{\citenamefont {Tempelman}\ \emph {et~al.}(2021)\citenamefont {Tempelman}, \citenamefont {Matlack},\ and\ \citenamefont {Vakakis}}]{Tempelman2021PRB}%
  \BibitemOpen
  \bibfield  {author} {\bibinfo {author} {\bibfnamefont {J.~R.}\ \bibnamefont {Tempelman}}, \bibinfo {author} {\bibfnamefont {K.~H.}\ \bibnamefont {Matlack}}, \ and\ \bibinfo {author} {\bibfnamefont {A.~F.}\ \bibnamefont {Vakakis}},\ }\href {\doibase 10.1103/PhysRevB.104.174306} {\bibfield  {journal} {\bibinfo  {journal} {Phys. Rev. B}\ }\textbf {\bibinfo {volume} {104}},\ \bibinfo {pages} {174306} (\bibinfo {year} {2021})}\BibitemShut {NoStop}%
\bibitem [{\citenamefont {Lu}\ \emph {et~al.}(2021)\citenamefont {Lu}, \citenamefont {He}, \citenamefont {Addison}, \citenamefont {Mele},\ and\ \citenamefont {Zhen}}]{PhysRevLett.126.113901}%
  \BibitemOpen
  \bibfield  {author} {\bibinfo {author} {\bibfnamefont {J.}~\bibnamefont {Lu}}, \bibinfo {author} {\bibfnamefont {L.}~\bibnamefont {He}}, \bibinfo {author} {\bibfnamefont {Z.}~\bibnamefont {Addison}}, \bibinfo {author} {\bibfnamefont {E.~J.}\ \bibnamefont {Mele}}, \ and\ \bibinfo {author} {\bibfnamefont {B.}~\bibnamefont {Zhen}},\ }\href {\doibase 10.1103/PhysRevLett.126.113901} {\bibfield  {journal} {\bibinfo  {journal} {Phys. Rev. Lett.}\ }\textbf {\bibinfo {volume} {126}},\ \bibinfo {pages} {113901} (\bibinfo {year} {2021})}\BibitemShut {NoStop}%
\bibitem [{\citenamefont {Lumer}\ \emph {et~al.}(2013)\citenamefont {Lumer}, \citenamefont {Plotnik}, \citenamefont {Rechtsman},\ and\ \citenamefont {Segev}}]{PhysRevLett.111.263901}%
  \BibitemOpen
  \bibfield  {author} {\bibinfo {author} {\bibfnamefont {Y.}~\bibnamefont {Lumer}}, \bibinfo {author} {\bibfnamefont {Y.}~\bibnamefont {Plotnik}}, \bibinfo {author} {\bibfnamefont {M.~C.}\ \bibnamefont {Rechtsman}}, \ and\ \bibinfo {author} {\bibfnamefont {M.}~\bibnamefont {Segev}},\ }\href {\doibase 10.1103/PhysRevLett.111.263901} {\bibfield  {journal} {\bibinfo  {journal} {Phys. Rev. Lett.}\ }\textbf {\bibinfo {volume} {111}},\ \bibinfo {pages} {263901} (\bibinfo {year} {2013})}\BibitemShut {NoStop}%
\bibitem [{\citenamefont {Shalchi}(2011)}]{Shalchi2011PRE}%
  \BibitemOpen
  \bibfield  {author} {\bibinfo {author} {\bibfnamefont {A.}~\bibnamefont {Shalchi}},\ }\href {\doibase 10.1103/PhysRevE.83.046402} {\bibfield  {journal} {\bibinfo  {journal} {Phys. Rev. E}\ }\textbf {\bibinfo {volume} {83}},\ \bibinfo {pages} {046402} (\bibinfo {year} {2011})}\BibitemShut {NoStop}%
\bibitem [{\citenamefont {Runge}\ \emph {et~al.}(2012)\citenamefont {Runge}, \citenamefont {Heitzig}, \citenamefont {Petoukhov},\ and\ \citenamefont {Kurths}}]{PhysRevLett.108.258701}%
  \BibitemOpen
  \bibfield  {author} {\bibinfo {author} {\bibfnamefont {J.}~\bibnamefont {Runge}}, \bibinfo {author} {\bibfnamefont {J.}~\bibnamefont {Heitzig}}, \bibinfo {author} {\bibfnamefont {V.}~\bibnamefont {Petoukhov}}, \ and\ \bibinfo {author} {\bibfnamefont {J.}~\bibnamefont {Kurths}},\ }\href {\doibase 10.1103/PhysRevLett.108.258701} {\bibfield  {journal} {\bibinfo  {journal} {Phys. Rev. Lett.}\ }\textbf {\bibinfo {volume} {108}},\ \bibinfo {pages} {258701} (\bibinfo {year} {2012})}\BibitemShut {NoStop}%
\bibitem [{\citenamefont {Wu}\ and\ \citenamefont {Yang}(2007)}]{PhysRevLett.98.013601}%
  \BibitemOpen
  \bibfield  {author} {\bibinfo {author} {\bibfnamefont {Y.}~\bibnamefont {Wu}}\ and\ \bibinfo {author} {\bibfnamefont {X.}~\bibnamefont {Yang}},\ }\href {\doibase 10.1103/PhysRevLett.98.013601} {\bibfield  {journal} {\bibinfo  {journal} {Phys. Rev. Lett.}\ }\textbf {\bibinfo {volume} {98}},\ \bibinfo {pages} {013601} (\bibinfo {year} {2007})}\BibitemShut {NoStop}%
\bibitem [{\citenamefont {Paulose}\ \emph {et~al.}(2015)\citenamefont {Paulose}, \citenamefont {Chen},\ and\ \citenamefont {Vitelli}}]{paulose2015topological}%
  \BibitemOpen
  \bibfield  {author} {\bibinfo {author} {\bibfnamefont {J.}~\bibnamefont {Paulose}}, \bibinfo {author} {\bibfnamefont {B.~G.-g.}\ \bibnamefont {Chen}}, \ and\ \bibinfo {author} {\bibfnamefont {V.}~\bibnamefont {Vitelli}},\ }\href@noop {} {\bibfield  {journal} {\bibinfo  {journal} {Nature Physics}\ }\textbf {\bibinfo {volume} {11}},\ \bibinfo {pages} {153} (\bibinfo {year} {2015})}\BibitemShut {NoStop}%
\bibitem [{\citenamefont {Zhou}\ \emph {et~al.}(2019)\citenamefont {Zhou}, \citenamefont {Zhang},\ and\ \citenamefont {Mao}}]{Zhou2019PRX}%
  \BibitemOpen
  \bibfield  {author} {\bibinfo {author} {\bibfnamefont {D.}~\bibnamefont {Zhou}}, \bibinfo {author} {\bibfnamefont {L.}~\bibnamefont {Zhang}}, \ and\ \bibinfo {author} {\bibfnamefont {X.}~\bibnamefont {Mao}},\ }\href {\doibase 10.1103/PhysRevX.9.021054} {\bibfield  {journal} {\bibinfo  {journal} {Phys. Rev. X}\ }\textbf {\bibinfo {volume} {9}},\ \bibinfo {pages} {021054} (\bibinfo {year} {2019})}\BibitemShut {NoStop}%
\bibitem [{\citenamefont {Levine}\ and\ \citenamefont {Steinhardt}(1986)}]{PhysRevB.34.596}%
  \BibitemOpen
  \bibfield  {author} {\bibinfo {author} {\bibfnamefont {D.}~\bibnamefont {Levine}}\ and\ \bibinfo {author} {\bibfnamefont {P.~J.}\ \bibnamefont {Steinhardt}},\ }\href {\doibase 10.1103/PhysRevB.34.596} {\bibfield  {journal} {\bibinfo  {journal} {Phys. Rev. B}\ }\textbf {\bibinfo {volume} {34}},\ \bibinfo {pages} {596} (\bibinfo {year} {1986})}\BibitemShut {NoStop}%
\bibitem [{\citenamefont {Socolar}\ and\ \citenamefont {Steinhardt}(1984)}]{PhysRevB.34.617}%
  \BibitemOpen
  \bibfield  {author} {\bibinfo {author} {\bibfnamefont {J.~E.~S.}\ \bibnamefont {Socolar}}\ and\ \bibinfo {author} {\bibfnamefont {P.~J.}\ \bibnamefont {Steinhardt}},\ }\href {\doibase 10.1103/PhysRevB.34.617} {\bibfield  {journal} {\bibinfo  {journal} {Phys. Rev. B}\ }\textbf {\bibinfo {volume} {34}},\ \bibinfo {pages} {617} (\bibinfo {year} {1984})}\BibitemShut {NoStop}%
\bibitem [{\citenamefont {Levine}\ and\ \citenamefont {Steinhardt}(1984)}]{PhysRevLett.53.2477}%
  \BibitemOpen
  \bibfield  {author} {\bibinfo {author} {\bibfnamefont {D.}~\bibnamefont {Levine}}\ and\ \bibinfo {author} {\bibfnamefont {P.~J.}\ \bibnamefont {Steinhardt}},\ }\href {\doibase 10.1103/PhysRevLett.53.2477} {\bibfield  {journal} {\bibinfo  {journal} {Phys. Rev. Lett.}\ }\textbf {\bibinfo {volume} {53}},\ \bibinfo {pages} {2477} (\bibinfo {year} {1984})}\BibitemShut {NoStop}%
\bibitem [{\citenamefont {Duncan}\ \emph {et~al.}(2020)\citenamefont {Duncan}, \citenamefont {Manna},\ and\ \citenamefont {Nielsen}}]{PhysRevB.101.115413}%
  \BibitemOpen
  \bibfield  {author} {\bibinfo {author} {\bibfnamefont {C.~W.}\ \bibnamefont {Duncan}}, \bibinfo {author} {\bibfnamefont {S.}~\bibnamefont {Manna}}, \ and\ \bibinfo {author} {\bibfnamefont {A.~E.~B.}\ \bibnamefont {Nielsen}},\ }\href {\doibase 10.1103/PhysRevB.101.115413} {\bibfield  {journal} {\bibinfo  {journal} {Phys. Rev. B}\ }\textbf {\bibinfo {volume} {101}},\ \bibinfo {pages} {115413} (\bibinfo {year} {2020})}\BibitemShut {NoStop}%
\bibitem [{\citenamefont {Scheibner}\ \emph {et~al.}(2020)\citenamefont {Scheibner}, \citenamefont {Irvine},\ and\ \citenamefont {Vitelli}}]{PhysRevLett.125.118001}%
  \BibitemOpen
  \bibfield  {author} {\bibinfo {author} {\bibfnamefont {C.}~\bibnamefont {Scheibner}}, \bibinfo {author} {\bibfnamefont {W.~T.~M.}\ \bibnamefont {Irvine}}, \ and\ \bibinfo {author} {\bibfnamefont {V.}~\bibnamefont {Vitelli}},\ }\href {\doibase 10.1103/PhysRevLett.125.118001} {\bibfield  {journal} {\bibinfo  {journal} {Phys. Rev. Lett.}\ }\textbf {\bibinfo {volume} {125}},\ \bibinfo {pages} {118001} (\bibinfo {year} {2020})}\BibitemShut {NoStop}%
\bibitem [{\citenamefont {Sone}\ and\ \citenamefont {Ashida}(2019)}]{Sone2019PRL}%
  \BibitemOpen
  \bibfield  {author} {\bibinfo {author} {\bibfnamefont {K.}~\bibnamefont {Sone}}\ and\ \bibinfo {author} {\bibfnamefont {Y.}~\bibnamefont {Ashida}},\ }\href {\doibase 10.1103/PhysRevLett.123.205502} {\bibfield  {journal} {\bibinfo  {journal} {Phys. Rev. Lett.}\ }\textbf {\bibinfo {volume} {123}},\ \bibinfo {pages} {205502} (\bibinfo {year} {2019})}\BibitemShut {NoStop}%
\bibitem [{\citenamefont {Zhou}\ and\ \citenamefont {Zhang}(2020)}]{Zhou2020PRR}%
  \BibitemOpen
  \bibfield  {author} {\bibinfo {author} {\bibfnamefont {D.}~\bibnamefont {Zhou}}\ and\ \bibinfo {author} {\bibfnamefont {J.}~\bibnamefont {Zhang}},\ }\href {\doibase 10.1103/PhysRevResearch.2.023173} {\bibfield  {journal} {\bibinfo  {journal} {Phys. Rev. Research}\ }\textbf {\bibinfo {volume} {2}},\ \bibinfo {pages} {023173} (\bibinfo {year} {2020})}\BibitemShut {NoStop}%
\bibitem [{\citenamefont {Yang}\ \emph {et~al.}(2021)\citenamefont {Yang}, \citenamefont {Zhu}, \citenamefont {Liu}, \citenamefont {Liu}, \citenamefont {Shi}, \citenamefont {Chen}, \citenamefont {Zheng}, \citenamefont {Ye},\ and\ \citenamefont {Yang}}]{PhysRevLett.126.198001}%
  \BibitemOpen
  \bibfield  {author} {\bibinfo {author} {\bibfnamefont {Q.}~\bibnamefont {Yang}}, \bibinfo {author} {\bibfnamefont {H.}~\bibnamefont {Zhu}}, \bibinfo {author} {\bibfnamefont {P.}~\bibnamefont {Liu}}, \bibinfo {author} {\bibfnamefont {R.}~\bibnamefont {Liu}}, \bibinfo {author} {\bibfnamefont {Q.}~\bibnamefont {Shi}}, \bibinfo {author} {\bibfnamefont {K.}~\bibnamefont {Chen}}, \bibinfo {author} {\bibfnamefont {N.}~\bibnamefont {Zheng}}, \bibinfo {author} {\bibfnamefont {F.}~\bibnamefont {Ye}}, \ and\ \bibinfo {author} {\bibfnamefont {M.}~\bibnamefont {Yang}},\ }\href {\doibase 10.1103/PhysRevLett.126.198001} {\bibfield  {journal} {\bibinfo  {journal} {Phys. Rev. Lett.}\ }\textbf {\bibinfo {volume} {126}},\ \bibinfo {pages} {198001} (\bibinfo {year} {2021})}\BibitemShut {NoStop}%
\bibitem [{\citenamefont {Ma}\ \emph {et~al.}(2025)\citenamefont {Ma}, \citenamefont {Feng}, \citenamefont {Li}, \citenamefont {Wu},\ and\ \citenamefont {Zhou}}]{Ma2025PRB}%
  \BibitemOpen
  \bibfield  {author} {\bibinfo {author} {\bibfnamefont {F.}~\bibnamefont {Ma}}, \bibinfo {author} {\bibfnamefont {J.}~\bibnamefont {Feng}}, \bibinfo {author} {\bibfnamefont {F.}~\bibnamefont {Li}}, \bibinfo {author} {\bibfnamefont {Y.}~\bibnamefont {Wu}}, \ and\ \bibinfo {author} {\bibfnamefont {D.}~\bibnamefont {Zhou}},\ }\href {\doibase 10.1103/PhysRevB.111.134301} {\bibfield  {journal} {\bibinfo  {journal} {Phys. Rev. B}\ }\textbf {\bibinfo {volume} {111}},\ \bibinfo {pages} {134301} (\bibinfo {year} {2025})}\BibitemShut {NoStop}%
\bibitem [{\citenamefont {Souslov}\ \emph {et~al.}(2017)\citenamefont {Souslov}, \citenamefont {Van~Zuiden}, \citenamefont {Bartolo},\ and\ \citenamefont {Vitelli}}]{souslov2017topological}%
  \BibitemOpen
  \bibfield  {author} {\bibinfo {author} {\bibfnamefont {A.}~\bibnamefont {Souslov}}, \bibinfo {author} {\bibfnamefont {B.~C.}\ \bibnamefont {Van~Zuiden}}, \bibinfo {author} {\bibfnamefont {D.}~\bibnamefont {Bartolo}}, \ and\ \bibinfo {author} {\bibfnamefont {V.}~\bibnamefont {Vitelli}},\ }\href@noop {} {\bibfield  {journal} {\bibinfo  {journal} {Nature Physics}\ }\textbf {\bibinfo {volume} {13}},\ \bibinfo {pages} {1091} (\bibinfo {year} {2017})}\BibitemShut {NoStop}%
\bibitem [{\citenamefont {Kunst}\ \emph {et~al.}(2018)\citenamefont {Kunst}, \citenamefont {Edvardsson}, \citenamefont {Budich},\ and\ \citenamefont {Bergholtz}}]{PhysRevLett.121.026808}%
  \BibitemOpen
  \bibfield  {author} {\bibinfo {author} {\bibfnamefont {F.~K.}\ \bibnamefont {Kunst}}, \bibinfo {author} {\bibfnamefont {E.}~\bibnamefont {Edvardsson}}, \bibinfo {author} {\bibfnamefont {J.~C.}\ \bibnamefont {Budich}}, \ and\ \bibinfo {author} {\bibfnamefont {E.~J.}\ \bibnamefont {Bergholtz}},\ }\href {\doibase 10.1103/PhysRevLett.121.026808} {\bibfield  {journal} {\bibinfo  {journal} {Phys. Rev. Lett.}\ }\textbf {\bibinfo {volume} {121}},\ \bibinfo {pages} {026808} (\bibinfo {year} {2018})}\BibitemShut {NoStop}%
\bibitem [{\citenamefont {Yao}\ and\ \citenamefont {Wang}(2018)}]{PhysRevLett.121.086803}%
  \BibitemOpen
  \bibfield  {author} {\bibinfo {author} {\bibfnamefont {S.}~\bibnamefont {Yao}}\ and\ \bibinfo {author} {\bibfnamefont {Z.}~\bibnamefont {Wang}},\ }\href {\doibase 10.1103/PhysRevLett.121.086803} {\bibfield  {journal} {\bibinfo  {journal} {Phys. Rev. Lett.}\ }\textbf {\bibinfo {volume} {121}},\ \bibinfo {pages} {086803} (\bibinfo {year} {2018})}\BibitemShut {NoStop}%
\end{thebibliography}
\end{document}